\documentclass[twocolumn,aps,floatfix,superscriptaddress,longbibliography]{revtex4-1}
\pdfoutput=1 
\usepackage{amsmath,amssymb,eucal,graphicx,float,epstopdf}
\usepackage{epsfig}
\usepackage[utf8]{inputenc}

\usepackage[colorlinks=true, urlcolor=blue, anchorcolor=blue, citecolor=blue,filecolor=blue,linkcolor=blue,menucolor=blue]{hyperref}

\usepackage{subfigure}

\newcommand{\br}{\boldsymbol{r}}

\begin{document}
\title{Dynamic Space Packing}

\author{Rahul Dandekar}
\affiliation{Institut de Physique Th\'eorique, CEA/Saclay, F-91191 Gif-sur-Yvette Cedex, France}
\author{P. L. Krapivsky}
\affiliation{Department of Physics, Boston University, Boston, Massachusetts 02215, USA}
\affiliation{Santa Fe Institute, Santa Fe, New Mexico 87501, USA}

\begin{abstract}
Dynamic space packing (DSP) is a random process with sequential addition and removal of identical objects into space. In the lattice version, objects are particles occupying single lattice sites, and adding a particle to a lattice site leads to the removal of particles on neighboring sites. We show that the model is solvable and determine the steady-state occupancy, correlation functions, desorption probabilities, and other statistical features for the DSP of hyper-cubic lattices. We also solve a continuous DSP of balls into $\mathbb{R}^d$.  
\end{abstract}

\maketitle

\section{Introduction}

Packing of space by non-overlapping objects is an old and fascinating subject. Packings by identical spheres are especially popular. The densest sphere packings in three dimensions were `discovered' when people stacked oranges or cannonballs. Confirming this ancient observation required a complicated computer-assisted proof \cite{Hales05}. Optimal sphere packings in dimensions 8 and 24 were guessed a while ago \cite{Conway}. Proving that they are the densest \cite{Maryna17a,Maryna17b} is much simpler than setting the three-dimensional situation; $d=1,2,3,8,24$ are the only dimensions in which the sphere packing problem is solved.

Despite much work \cite{Minkowski05,Blichfeldt29,Rogers,KL78,Cohn03,Cohn02,Vardy,Vance11,Venkatesh,Parisi08,Cohn14,Cohn17,Cohn16,Perkins,Cohn22a,
Cohn23,Courcy23, Elser23}, little is known about dense high-dimensional sphere packings. Such packings are essential for communication over noisy channels \cite{Shannon48a,Shannon48b} and combinatorial optimization \cite{Optimization,Belitz13}. Kissing numbers defined as the greatest numbers of non-overlapping spheres that can touch a sphere \cite{VDW,Leech,Levenshtein79,Odlyzko79,Ziegler,Musin08,Min-energy15,Kiss18,Hardin} also have applications to coding theory \cite{Shannon59,Wyner}. 

Interacting spheres underlie many physical systems. The analysis greatly simplifies in high spatial dimensions, and emerging results shed light on liquids, granular matter, strongly correlated electrons, jamming, glass transitions, crystallization, active matter, etc. \cite{Frisch85,Frisch86,Frisch87,Frisch88,Frisch89, Mehta,Georges96, Frisch99,Parisi00,TS10,Parisi10,Schilling10,Charbonneau11,Kallus13,Kurchan16a,Kurchan16b,Szamel17,Charbonneau17,Biroli18,Torquato18, Parisi20,Fred19,Fred21,Charbonneau21,Rastelli19, Cohn20,Biroli22}.  

Packings generated by algorithms arise in applications to coding theory and material science. Here we analyze dynamic space packing (DSP), a sequential deposition process in which each deposition of an object is accompanied by the removal of the previously deposited objects overlapping with a newly added one. The system quickly reaches a steady state. 

In Secs.~\ref{sec:lattice}--\ref{sec:lattice-d}, we analyze the DSP of the hyper-cubic lattices $\mathbb{Z}^d$ by particles occupying single lattice sites. Particles are added randomly with unit rate per site, and after each deposition event $2d$ neighboring sites are emptied. Hence there is at most one particle on each pair of neighboring sites. In the analysis, we treat a deposition into an occupied site as replacing the particle at the site.

The maximal allowed density is $\frac{1}{2}$ since the lattices $\mathbb{Z}^d$ are bipartite. The actual steady-state density is 
\begin{equation}
\label{rho-d}
\rho = \frac{1}{2d+1}
\end{equation}
as we show in Sec.~\ref{sec:lattice}. Our results are readily extended to {\em regular} graphs in which each vertex has the same number of neighbors. In an infinite $q-$regular graph (each vertex has $q$ neighbors), the steady-state density is 
\begin{equation}
\label{rho-q}
\rho = \frac{1}{q+1}
\end{equation}

\begin{figure}
  \begin{center}
\includegraphics[width=2.48in]{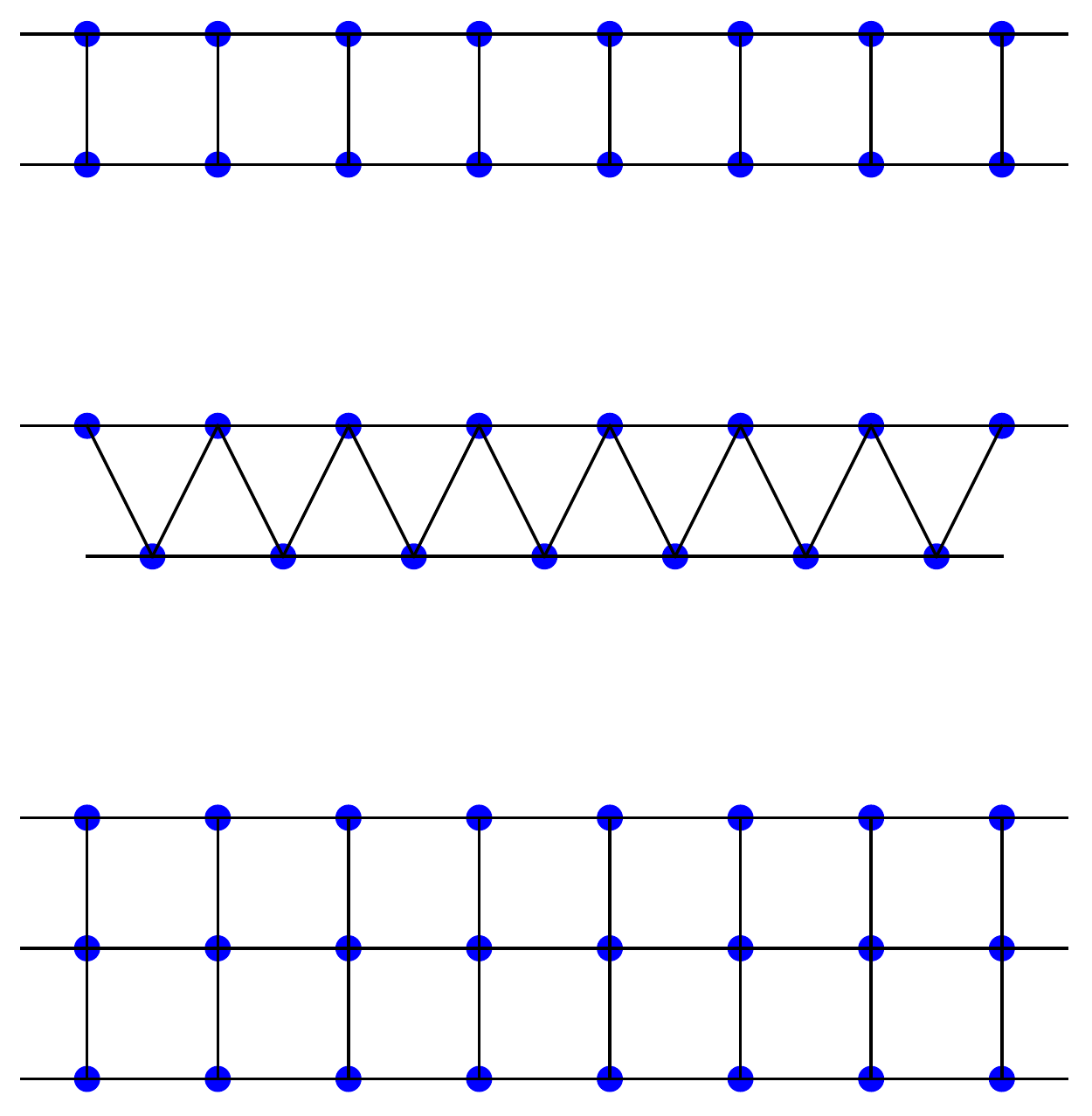}
 \caption{The ladder at the top is a $3-$regular graph. The ladder in the middle is a $4-$regular graph. The ladder at the bottom is not a regular graph. }
\label{Fig:Ladders}
  \end{center}
\end{figure}

For the DSP on the one-dimensional lattice, $q=2$ and $\rho=\frac{1}{3}$. This model was investigated in \cite{KR-birds22}, and in Sec.~\ref{sec:lattice-1d} we further explore it, e.g., we determine the empty interval distribution and its counterpart, the distribution of most congested intervals. More generally, we study the occupancy distribution, and for large intervals, we determine all cumulants. Some results can be extended to the DSP on quasi-one-dimensional lattices, e.g., ladders shown in Fig.~\ref{Fig:Ladders}. The ladder at the top with $w=2$ wires is a $3-$regular graph, so the steady-state density is $\rho=\frac{1}{4}$. The ladder in the middle of Fig.~\ref{Fig:Ladders} is a $4-$regular graph, so the steady-state density is $\rho=\frac{1}{5}$. The ladder at the bottom in Fig.~\ref{Fig:Ladders} with $w=3$ wires is not a regular graph, albeit its steady-state density $\rho=\frac{7}{30}$ can be established using the methods developed below, namely by noting that the density in the bulk (the middle wire) is $\rho_\text{bulk}=\frac{1}{5}$ while the density on the edges (the border wires) is $\rho_\text{edge}=\frac{1}{4}$. For similar ladders with $w\geq 2$ wires, the average density is $\rho= \frac{1}{5}+\frac{1}{10 w}$. 

The chief feature of the DSP is complete de-correlation beyond a finite distance. We call sites $i$ and $j$ well-separated if they are not neighbors and do not have common neighbors, i.e., the Manhattan distance \footnote{For $\mathbb{Z}^d$, we use the Manhattan norm $|i|=|i_1|+\ldots+|i_d|$.} between sites satisfies $|i-j|>2$. For any set $S\subset \mathbb{Z}^d$ composed of mutually well-separated sites
\begin{equation}
\label{decorr}
\left\langle \prod_{k\in S} \eta_k\right\rangle  =  \prod_{k\in S} \langle \eta_k\rangle = \rho^{|S|}
\end{equation}
where the angle brackets denote an average in the steady-state distribution and $\eta_k$ is the occupancy of site $k$:
\begin{equation}
\label{eta:def}
\eta_k =
\begin{cases} 
1 &  \text{if $k$ is occupied}\\
0  & \text{if $k$ is empty}
\end{cases}
\end{equation}
Equation \eqref{decorr} and more general results applicable to well-separated sets are established in Sec.~\ref{sec:lattice}. Short-distance correlations remain, e.g., the pair correlation function $\langle \eta_i \eta_j\rangle$ is non-trivial when $1<|i-j|\leq 2$; for neighboring sites, the pair correlation function  vanishes by the definition of the DSP. Complete de-correlation is the reason behind the solvability of the DSP processes. In more complicated systems, de-correlation often occurs in high spatial dimensions and simplifies the analysis of such systems. Examples range from some sphere packing processes \cite{Charbonneau10,Torquato16} and equilibrium hard spheres \cite{Santos07} to Gaussian core model \cite{Cohn08,Torquato08a,Cohn18} and point processes \cite{Torquato08b}.

Apart from the density and correlations, we analyze the probabilities $p_n$ that $n$ objects are removed after a deposition event. These `desorption' probabilities satisfy 
\begin{subequations}
\begin{align}
\label{norm:p}
\sum_{n\geq 0}    p_n & =1\\
\label{SS:p}
\sum_{n\geq 0} n p_n & =1
\end{align}
\end{subequations}
The first constraint expresses normalization. The second reflects that in the steady state, one object is on average removed after a deposition event. On the one-dimensional lattice
\begin{equation}
\label{p012}
(p_0,p_1,p_2) =\left(\tfrac{1}{5}, \tfrac{3}{5}, \tfrac{1}{5}\right)
\end{equation}
in the steady state \cite{KR-birds22}. On the square lattice 
\begin{equation}
\label{p01234}
(p_0,p_1,p_2,p_3,p_4)=\big(\tfrac{119}{396}, \tfrac{5}{11}, \tfrac{7}{36}, \tfrac{1}{22},\tfrac{1}{198}\big)
\end{equation}
in the steady state. We derive \eqref{p01234} in Sec.~\ref{sec:lattice-2d} where we also determine an approach to the steady state, viz., $p_n(t)$ for all $n=0,\ldots,4$. 

The non-vanishing desorption probabilities for the DSP on $\mathbb{Z}^d$ are $p_0(d),\ldots,p_{2d}(d)$. The calculation of these probabilities becomes cumbersome as the dimension increases, albeit it is in principle possible to determine the steady-state values in arbitrary dimension as we show in Sec.~\ref{sec:lattice-d}. The answers come from computer-assisted exact calculations using an algorithm described in Appendix~\ref{ap:algorithm}. 


\begin{table}[h!]
\centering
\renewcommand{\arraystretch}{1.5}{
\begin{tabular}{| c | c | c | c |}
\hline
$d$         &  $3$                                                            & $4$  & $5$ \\ 
\hline
$p_0$   & $\frac{415\,986\,817}{1\,280\,916\,000}$   & $\frac{5\,281\,278\,681\,782\,515}{15\,709\,851\,298\,132\,992}$     
   & $\frac{2375275880412462072499}{6928755085630538334720}$ \\ 
\hline
$p_1$   & $\frac{271\,348\,349}{640\,458\,000}$     & $\frac{89\,237\,426\,291\,711}{218\,192\,379\,140\,736}$   
& $\frac{695914927071909500851309}{1737962733978993365625600}$  \\ 
\hline
$p_2$   & $\frac{6\,069\,977}{32\,022\,900}$          & $\frac{2\,950\,539\,568\,090\,175}{15\,709\,851\,298\,132\,992}$ 
 & $\frac{974439468820087031128267}{5213888201936980096876800}$ \\ 
\hline
$p_3$    & $\frac{144\,541}{2\,784\,600}$                & $\frac{18\,624\,482\,925\,199}{341\,518\,506\,481\,152}$     
 & $\frac{36121813090436723231}{644963904247523515200}$     \\  
\hline
$p_4$   & $\frac{2\,326\,741}{256\,183\,200}$      & $\frac{4\,362\,095\,202\,317}{402\,816\,699\,952\,128}$  
& $\frac{8795432899903151773471}{744841171705282870982400}$       \\  
\hline
$p_5$    & $\frac{622\,649}{640\,458\,000}$                 &$\frac{ 5\,919\,397\,345\,913}{3\,927\,462\,824\,533\,248}$ 
 & $\frac{887714927717509909}{485554870733561193600}$    \\  
\hline
$p_6$   & $\frac{32\,771}{640\,458\,000}$                   &$\frac{815\,681\,663}{5\,644\,933\,991\,424}$   
  & $\frac{39286987966716032503}{186210292926320717745600}$           \\ 
\hline
$p_7$    & $0$                                                               &$\frac{190\,904\,219}{21\,638\,913\,633\,792}$ 
& $\frac{3351387587134093093}{186210292926320717745600}$                 \\ 
\hline
$p_8$    & $0$                                                               &$\frac{17\,354\,929}{64\,916\,740\,901\,376}$ 
& $\frac{11396860157332670819}{10427776403873960193753600}$                 \\
\hline
$p_9$     & $0$       &$0$      &    $\frac{74776774478832653}{1737962733978993365625600}$  \\
\hline
$p_{10}$  & $0$     &$0$      &  $\frac{4398633792872509}{5213888201936980096876800}$ \\
\hline
\end{tabular}
}
\caption{The probabilities $p_n$ for $d=3,4,5$.} 
\label{Table:345}
\end{table}

\begin{figure}[ht]
\begin{center}
\includegraphics[width=0.44\textwidth]{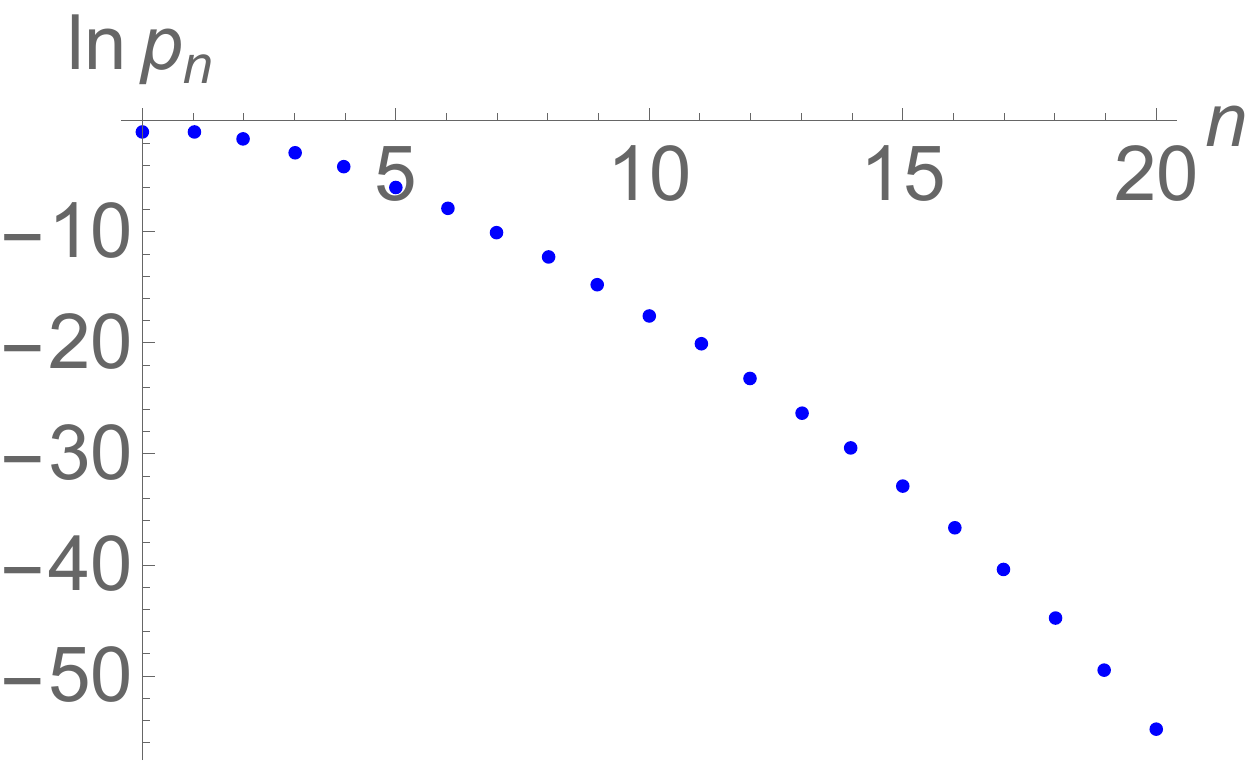}
\caption{Desorption probabilities for $d=10$. These probabilities are positive when $n\leq 2d=20$. The smallest probability is $p_{20}(10)\approx 1.84\times 10^{-24}$.}
\label{fig:pn-10}
  \end{center}
\end{figure}

General analytical expressions for $p_n(d)$ remain unknown. The results in $d=3,4,5$ shown in Table~\ref{Table:345}, and exact probabilities that we computed in $d\leq 11$ dimensions, hint that the existence of such expressions in arbitrary dimension is unlikely. We discovered (Sec.~\ref{sec:lattice-d}) two exact results for the ratios 
\begin{subequations}
\label{ratio}
\begin{align}
\label{ratio-1}
\frac{p_{2d-1}(d)}{p_{2d}(d)} &= 2d^2 + 1\\
\label{ratio-2}
2\,\frac{p_{2d-2}(d)}{p_{2d}(d)} &= 4d^4 + 4d^2 -2d + 1
\end{align}
\end{subequations}
The first ratio holds in all dimensions, while Eq.~\eqref{ratio-2} is applicable for $d\geq 2$. 

Results shown in Table~\ref{Table:345} and Fig.~\ref{fig:pn-10} suggest that $p_1(d)$ is maximal among $p_0(d),\ldots,p_{2d}(d)$. This feature agrees with the exact results in dimensions $d\leq 11$ computed using the algorithm described in Appendix~\ref{ap:algorithm}. Exact rational expressions for $p_n(d)$ are huge, e.g., $p_{22}(11)$ has the denominator close to $6.3\times10^{122}$. We applied the Richardson extrapolation \cite{Bender} to the exact results for $p_0(d)$ to accelerate the convergence to $p_0(\infty)$. Using this method, we extracted a numerical value of $p_0(\infty)$ close to $e^{-1}$ supporting our theoretical prediction  (Appendix~\ref{ap:algorithm}). 

In Sec.~\ref{sec:lattice-d}, we argue that 
\begin{equation}
\label{pn:inf}
p_n(\infty) \equiv \lim_{d\to\infty}p_n(d)=\frac{e^{-1}}{n!} 
\end{equation}
There is no disagreement between \eqref{pn:inf} and \eqref{ratio} since \eqref{pn:inf} is valid when $n=O(1)$ and $d\to\infty$. Thus if $d\gg 1$, we cannot use \eqref{pn:inf} to estimate $p_{2d-1}$ and $p_{2d}$.  

In Sec.~\ref{sec:cont}, we analyze the continuous DSP by spheres and show that the steady-state volume fraction equals $2^{-d}$. The same volume fraction appears in the realm of dense sphere packings. Indeed, any {\em saturated} sphere packing \footnote{A sphere packing is saturated if no other sphere can be added without overlap.} has volume fraction exceeding $2^{-d}$. Thus, $2^{-d}$ is the lower bound (known as Minkowski's bound) for the densest sphere packing. The proof of the lower bound is elementary \cite{Cohn16}, yet the exponential factor has not changed since it was first written down in 1905; algebraic improvements like $d\cdot 2^{-d}$ are the best achievements \cite{Rogers,Vardy,Vance11,Venkatesh,Perkins}. The emergence of the $2^{-d}$ volume fraction in the realm of the DSP is intriguing. Indeed, Minkowski's bound is non-constructive, and in large dimensions, there are no algorithmically described packings with volume fraction $2^{-d}$  or higher. Thus, the continuous DSP provides a randomized algorithm generating packings fulfilling Minkowski's bound. Closely related processes \cite{Viot97,TS06b,TS06a,TS06c} also provide randomized algorithms generating packings fulfilling Minkowski's bound. The multilayer adsorption  \cite{Viot97} and ghost random sequential adsorption \cite{TS06b,TS06a} are highly irreversible processes. Still, many features of the jammed states in these processes are identical to the steady states of the continuous DSP.

For the continuous DSP, the desorption probabilities are known \cite{KR-birds22} in one dimension. In the steady state
\begin{equation}
\label{p012:cont}
p_0 = p_2 = -1+4\ln(4/3), \quad  p_1 =3-8\ln(4/3)
\end{equation}
In Sec.~\ref{sec:cont}, we express the steady-state desorption probabilities in arbitrary dimensions via certain integrals. The multiplicity of these integrals  quickly increases with $d$, and analytical computations appear impossible already in two spatial dimensions. The $d\to\infty$ limit is tractable, and we show that the steady-state desorption probabilities $p_n(\infty)$ are given by the same Poisson distribution \eqref{pn:inf} as in the lattice case. In Sec.~\ref{sec:remarks}, we compare the DSP with an irreversible random deposition algorithm known as random sequential adsorption.  

The algorithm for computing desorption probabilities for the lattice DSP is described in Appendix~\ref{ap:algorithm}. Derivations of several assertions are relegated to Appendices~\ref{ap:induction}, \ref{ap:empty}, \ref{ap:var}. In Appendix \ref{ap:perm}, we describe the connection between the DSP on an arbitrary graph, the statistics of maxima of a random surface (with independent heights) on the same graph, and permutations.

\section{DSP of $\mathbb{Z}^d$: Basic results}
\label{sec:lattice}

Here and in Sects.~\ref{sec:lattice-1d}--\ref{sec:lattice-d}, we consider the DSP of the hyper-cubic lattices. In studying the relaxation to the steady state, we always begin with an empty lattice and set the deposition rate to unity.

Let $\eta_i$ be the occupancy of site $i\in \mathbb{Z}^d$, i.e., $\eta_i=1$ if $i$ is occupied, $\eta_i=0$ otherwise. There are $2d+1$ deposition events resulting in the desorption of a particle from the site $i$ (if occupied) and one attempt of filling the site $i$. The rate equation for the average occupancy $\langle \eta_i \rangle$ reads
\begin{equation}
\label{ni:eq}
\frac{d}{dt}\,\langle \eta_i \rangle = 1-(2d+1)\langle \eta_i \rangle
\end{equation}
Using translational invariance we conclude that $\langle \eta_i \rangle=\rho$, and thus the density of occupied sites satisfies 
\begin{equation}
\label{evol:lattice}
\frac{d \rho}{dt} = 1 - (2d+1)\rho
\end{equation}
Solving Eq.~\eqref{evol:lattice} subject to $\rho(0)=0$ gives 
\begin{equation}
\label{rho-d:sol}
\rho(t) = \frac{1-e^{-(2d+1)t}}{2d+1}
\end{equation}
Thus the density exhibits a purely exponential relaxation to the steady-state density \eqref{rho-d}. 

Now consider the pair correlation function $ \langle \eta_i \eta_j\rangle$. This correlation function vanishes if $i$ and $j$ are neighboring sites. Otherwise, it satisfies the rate equation 
\begin{equation}
\label{nij:eq}
\frac{d}{dt}\,\langle \eta_i \eta_j\rangle = \langle \eta_i \rangle + \langle \eta_j \rangle -V_{i,j}\langle \eta_i \eta_j\rangle
\end{equation}
The gain terms on the right-hand side of \eqref{nij:eq} are due to the deposition to site $j$ and $i$, respectively. The loss term accounts for the deposition to the joined neighborhood $\mathcal{V}_{i,j}$ of sites $i$ and $j$. This neighborhood contains all sites on distance $\leq 1$ from either $i$ or $j$, and $V_{i,j}=|\mathcal{V}_{i,j}|$ is the total number of sites in this neighborhood. 

If sites $j$ and $i$ are well-separated, $|i-j|>2$, we have $V_{i,j}=2(2d+1)$. Equation \eqref{nij:eq} in this situation becomes
\begin{equation}
\label{nij:far-eq}
\frac{d}{dt}\,\langle \eta_i \eta_j\rangle = \langle \eta_i \rangle + \langle \eta_j \rangle  -  2(2d+1)\langle \eta_i \eta_j\rangle
\end{equation}
In the steady state the time derivative is zero, so
\begin{equation}
\label{nij:far}
\langle \eta_i \eta_j\rangle = \langle \eta_i \rangle \langle \eta_j \rangle = \rho^2
\end{equation}
where we have also used $\langle \eta_j \rangle = \rho$ due to translational invariance. Equation \eqref{nij:far} is the simplest manifestation of complete de-correlation. 

De-correlation is the chief feature of the DSP. To generalize de-correlation from two sites to two sets of sites we begin by defining the distance between two arbitrary non-empty sets $S_1, S_2\subset \mathbb{Z}^d$ via:
\begin{equation}
\label{dist}
d(S_1,S_2)=\min_{i\in S_1, j\in S_2}|i-j|
\end{equation}
We call sets $S_1$ and $S_2$ well-separated if $d(S_1,S_2)>2$. (The distance \eqref{dist} is a pseudo-metric \cite{pseudo} as the strict positiveness does not hold: Vanishing of the distance between two sets does not imply that these sets are equal, the distance between overlapping sets, $S_1\cap S_2\ne  \emptyset$, vanishes, $d(S_1, S_2)=0$.)
 
We then define the correlation function
\begin{equation}
\label{CS:def}
C(S)=\left\langle \prod_{k\in S} \eta_k\right\rangle
\end{equation}
for any set $S\subset \mathbb{Z}^d$. De-correlation is the statement that the correlation function of the union of two arbitrary well-separated sets factorizes into the product of the correlation functions of the sets:
\begin{equation}
\label{2:far}
C(S_1\cup S_2)=C(S_1)C(S_2) \quad\text{if} \quad d(S_1,S_2)>2
\end{equation}

Equation \eqref{2:far} reduces to \eqref{nij:far} when $S_1$ and $S_2$ are one-element sets. To prove \eqref{2:far} in the general case, one writes equations for the correlation functions similar to \eqref{nij:far-eq} and employs induction in the size of sets (see Appendix~\ref{ap:induction}).

De-correlation holds for more than two well-separated sets: 
\begin{equation}
\label{ell:far}
C\left(S_1\cup S_2\cup\cdots\cup S_\ell\right)=\prod_{a=1}^\ell C(S_a) 
\end{equation}
when $d(S_a,S_b)>2$ for all $a\ne b$. For one-element sets, $|S_a|=\ldots=|S_\ell|=1$, Eq.~\eqref{ell:far} reduces to Eq.~\eqref{decorr}.

One derives \eqref{ell:far} using \eqref{2:far} and induction in the number of sets $\ell$.

\section{One-dimensional lattice}
\label{sec:lattice-1d}

In one dimension, the density is
\begin{equation}
\label{rho:sol}
\rho(t) = \frac{1-e^{-3t}}{3}
\end{equation}
The void distribution ($\circ$ denotes an empty site, $\bullet$ denotes an occupied site)
\begin{equation}
\label{Vk:def}
V_k = \text{Prob}\big[\bullet\underbrace{\circ\,\ldots\,\circ}_k\bullet\big]
\end{equation}
has been also studied in \cite{KR-birds22}. In Secs.~\ref{subsec:void}--\ref{subsec:CF}, we recapitulate some results of \cite{KR-birds22} for the void distribution and desorption probabilities and present a direct derivation of de-correlation in the case of one-element sets. We then probe the characteristics that have not been investigated in \cite{KR-birds22}. We determine the probabilities of the least congested and most congested configurations, explore occupation probabilities, and compute all cumulants of the occupation number of long intervals (Secs.~\ref{subsec:empty}--\ref{subsec:max}).

\subsection{Voids}
\label{subsec:void}

The void distribution is given by \cite{KR-birds22}
\begin{equation}
\label{Vk:sol}
V_k  = 2^{k+1}\,\frac{k(k+3)}{(k+4)!}
\end{equation}
in the steady state. The time-dependent void densities $V_k(t)$ can be obtained by recurrently solving 
\begin{equation}
\label{Vk:eq}
\frac{d V_k}{d t} = -(k+4)V_k - 2\sum_{j=1}^{k-2}V_j + 2\rho
\end{equation}
One gets
\begin{equation}
\label{Vt}
\begin{split}
V_1 &= \tfrac{1}{15}\left(2-5e^{-3t}+3e^{-5t}\right)\\
V_2 &= \tfrac{1}{9}\left(1-e^{-3t}\right)^2\\
V_3 &= \tfrac{1}{35}\left(2-7e^{-5t}+5e^{-7t}\right)\\
V_4 &= \tfrac{1}{45}\left(1+4e^{-3t}-6e^{-5t}-5e^{-6t}+6e^{-8t}\right)\\
V_5 &= \tfrac{1}{567}\left(4 + 42 e^{-3t} - 42 e^{-6t} - 81 e^{-7t}+77 e^{-9t}\right)
\end{split}
\end{equation}
etc. The exact expressions for $V_k(t)$ become more and more cumbersome as $k$ increases, see Eqs.~\eqref{Vt}. These exact expressions do not reveal a pattern. One can still find the time-dependent solution as we explain in Sec.~\ref{subsec:empty} and Appendix~\ref{ap:empty}. Before going into this analysis, we discuss desorption probabilities $p_n$ which can be determined using the density $V_1$ of  the shortest voids. 

In the one-dimensional model, $p_0, p_1, p_2$ are non-vanishing. (For the hyper-cubic lattice $\mathbb{Z}^d$, the probabilities $p_n$ are non-vanishing for $n=0,1,\ldots, 2d$.) To derive the desorption probabilities \eqref{p012} in the steady state we use the sum rules \eqref{norm:p}--\eqref{SS:p}. In one dimension, these sum rules reduce to 
\begin{subequations}
\begin{align}
\label{norm}
p_0+p_1+p_2 &= 1\\
\label{SS}
p_1+2p_2 &=1
\end{align}
\end{subequations}
Equation \eqref{norm} applies throughout the evolution while Eq.~\eqref{SS} is valid only in the steady state. Equations \eqref{norm}--\eqref{SS} imply $p_0=p_2$ in the steady state. 

\begin{figure}
\centering
\includegraphics[width=7.89cm]{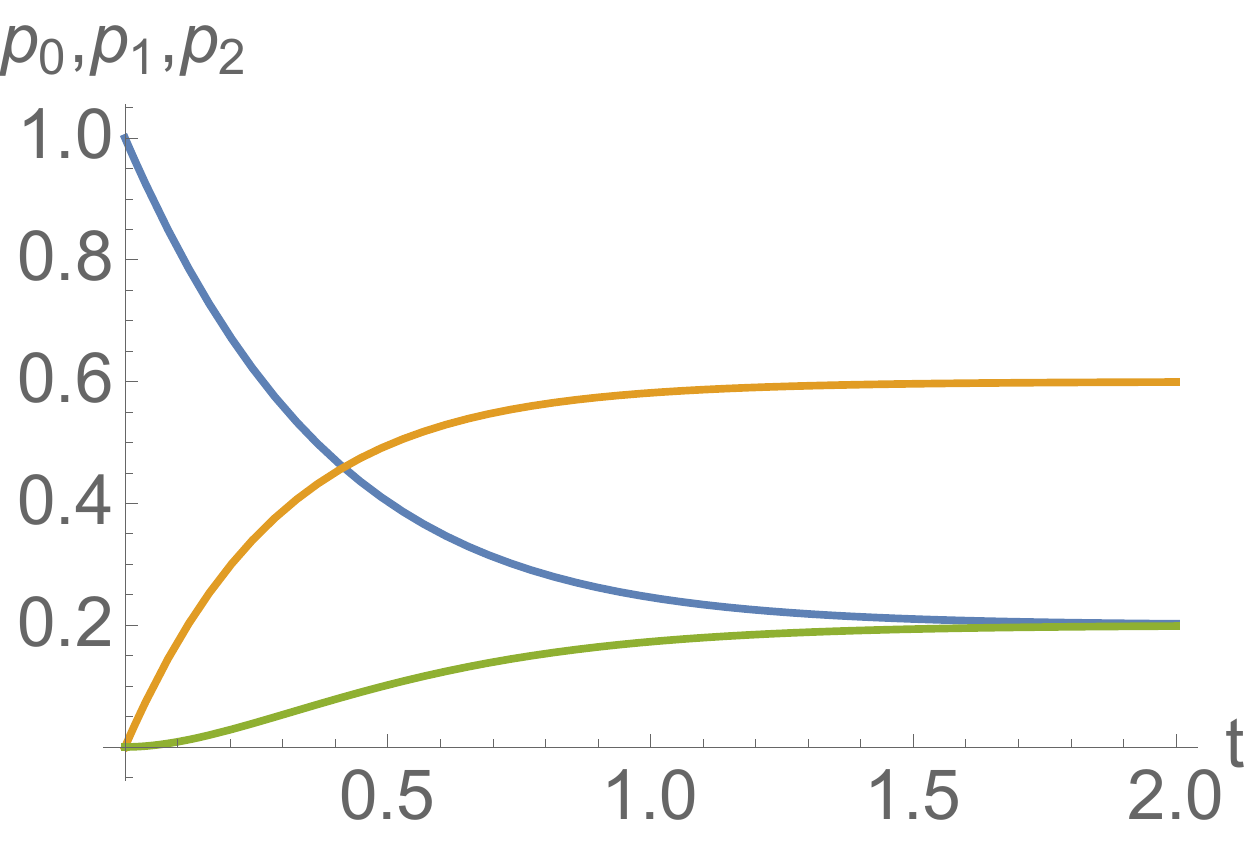}
\caption{The evolution of the desorption probabilities $p_0, p_1, p_2$ (top to bottom at small $t$) on the one-dimensional lattice. }
\label{Fig:1d-probs}
\end{figure}

The probability $p_2$ coincides with the void density $V_1$ up to the $(1-\rho)^{-1}$ factor accounting for successful deposition events
\begin{subequations}
\begin{equation}
\label{p2:eq}
p_2= \frac{V_1}{1-\rho}
\end{equation}
In the steady state, \eqref{p2:eq} gives $p_2=1/5$ which in conjunction with \eqref{norm}--\eqref{SS} lead to \eqref{p012}. 

To derive time-dependent desorption probabilities we rely on \eqref{p2:eq} and
\begin{equation}
\label{p1:eq}
p_1 = \frac{2}{1-\rho}\sum_{k\geq 2} V_k = 2\,\frac{\rho-V_1}{1-\rho}
\end{equation}
\end{subequations}
(The last expression in \eqref{p1:eq} follows from the sum rule $\sum_{k\geq 1} V_k=\rho$.) Using \eqref{p2:eq}--\eqref{p1:eq}, the normalization condition \eqref{norm} and the density given by \eqref{rho:sol} together with $V_1(t)$ appearing in \eqref{Vt} we obtain 
\begin{subequations}
\begin{align}
\label{p0}
p_0 &= \frac{1}{5}\,\frac{2+ 10e^{-3t}+3e^{-5t}}{2+e^{-3t}}\\
\label{p1}
p_1 &=\frac{6}{5}\,\frac{1-e^{-5t}}{2+e^{-3t}}\\
\label{p2}
p_2 &= \frac{1}{5}\,\frac{2-5e^{-3t}+3e^{-5t}}{2+e^{-3t}}
\end{align}
\end{subequations}
These desorption probabilities are shown in  Fig.~\ref{Fig:1d-probs}.

\subsection{Correlation functions}
\label{subsec:CF}

The void distribution \eqref{Vk:def} can be alternatively written as a correlation function 
\begin{equation}
V_k = \langle \eta_0 (1-\eta_1)\ldots (1-\eta_k)\eta_{k+1}\rangle
\end{equation}
Denote by $C_j = \langle \eta_0 \eta_j\rangle$ the pair correlation function. We have $C_0 = \langle \eta_0^2\rangle=\langle \eta_0\rangle=\rho$, $C_1=0$, $C_2=V_1$ and $C_3=V_2$. Thus 
\begin{subequations}
\begin{align}
\label{C0}
C_0 &= \rho = \tfrac{1}{3}\left(1-e^{-3t}\right)\\
\label{C2}
C_2 &= V_1 = \tfrac{1}{15}\left(2-5e^{-3t}+3e^{-5t}\right)\\
\label{C3}
C_3 &=V_2 = \tfrac{1}{9}\left(1-e^{-3t}\right)^2
\end{align}
\end{subequations}
Needless to say, $C_3=\rho^2$ in agreement with the general relation \eqref{nij:far} for sufficiently well-separated sites. More generally,
\begin{equation}
\label{Cj}
C_j = \rho^2 = \tfrac{1}{9}\left(1-e^{-3t}\right)^2 \qquad (j\geq 3)
\end{equation}
follows from \eqref{nij:far} in one dimension.

It is useful to re-derive $C_2= \langle \eta_0 \eta_2\rangle$ from Eq.~\eqref{nij:eq}. The neighborhood of sites $0$ and $2$ consists of five sites $-1,0,1,2,3$. Thus $\mathcal{N}_{0,2}=5$ and \eqref{nij:eq} becomes 
\begin{equation}
\label{C2:eq}
\frac{d C_2}{dt}=  -5C_2 + 2\rho
\end{equation}
Solving \eqref{C2:eq} with $\rho$ given by \eqref{C0} one re-derives \eqref{C2}. 

The correlation function $F_\ell=\langle \eta_{i_1}\ldots \eta_{i_\ell}\rangle$ with sufficiently separated sites, $i_2-i_1\geq 3, \ldots, i_\ell-i_{\ell-1}\geq 3$, is universal (independent on $i_1,\ldots,i_\ell$), so we use the shorthand notation $F_\ell$ for this correlated function. The rate equation for $F_\ell$ is derived similarly to \eqref{nij:far-eq} to yield
\begin{equation}
\label{Fp:eq}
\frac{d F_\ell}{d t} = -3\ell F_\ell + \ell F_{\ell-1}
\end{equation}
We already know $F_1=\rho$ and $F_2=\rho^2$. Solving \eqref{Fp:eq} recurrently we obtain
\begin{equation}
\label{Fp:sol}
F_\ell = \rho^\ell
\end{equation}
Needless to say, the general result \eqref{ell:far} reduces to \eqref{Fp:sol} in the case of one-element sets: $S_a=\{i_a\}$. 

\subsection{Empty intervals}
\label{subsec:empty}

The empty interval distribution is defined via 
\begin{equation}
\label{Ek:def}
E_k = \text{Prob}\big[\underbrace{\circ\,\ldots\,\circ}_k\big]
\end{equation}
The void distribution can be extracted from the empty interval distribution through the general formula \cite{KRB}
\begin{equation}
\label{VEk}
V_k = E_k - 2 E_{k+1}+E_{k+2}
\end{equation}

The empty interval distribution evolves according to
\begin{equation}
\label{Ek:eq}
\frac{d E_k}{d t} = -(k+2)E_k +2E_{k-1}
\end{equation}
valid for all $k\geq 1$ if we set $E_0\equiv 1$. In the steady state $(k+2)E_k = 2E_{k-1}$ which is iterated to give
\begin{equation}
\label{Ek:sol}
E_k  = \frac{2^{k+1}}{(k+2)!}
\end{equation}
By inserting \eqref{Ek:sol} to \eqref{VEk} we recover \eqref{Vk:sol}. 

In the steady state, the empty interval distribution \eqref{Ek:sol}  is simpler than the void distribution \eqref{Vk:sol}.  The evolution equations \eqref{Ek:eq} are also simpler than Eqs.~\eqref{Vk:eq}, so one could hope to solve Eqs.~\eqref{Ek:eq} and then find the void distribution by using \eqref{VEk}. To solve Eqs.~\eqref{Ek:eq} we introduce the generating function 
\begin{equation}
\label{GF}
\mathcal{E}(t,z) = \sum_{k\geq 1} E_k(t)\, e^{(k+2)z}
\end{equation}
An infinite set of recurrent ordinary differential equations \eqref{Ek:eq} reduces to a single partial differential equation for the generating function
\begin{equation}
\label{E:PDE}
\left(\partial_t+ \partial_z\right)\mathcal{E} = 2 e^z\left[\mathcal{E}+ e^{2z}\right]
\end{equation}
which we solve in Appendix~\ref{ap:empty}. From the generating function we extract the solution 
\begin{eqnarray}
\label{Ekt}
E_k  &=&2^{k+1}\,\frac{(1-e^{-t})^{k+2}}{(k+2)!} \nonumber\\
&+&e^{2(e^t-1)-(k+2)t}\,\frac{\Gamma[k+2,2(e^t-1)]}{\Gamma(k+2)}
\end{eqnarray}
where
\begin{equation*}
\Gamma[k+2,Y] = \int_{Y}^\infty dy\,y^{k+1} e^{-y}
\end{equation*}
is the incomplete gamma function \cite{NIST}. 

For small  $k$, we can either use the general solution \eqref{Ekt} or solve Eqs.~\eqref{Ek:eq} recurrently. In particular
\begin{equation}
\label{E15}
\begin{split}
E_1 &= \frac{2+e^{-3t}}{3}   \\
E_2 &= \frac{1+2e^{-3t}}{3}\\
E_3 &= \frac{2+10e^{-3t}+3 e^{-5t}}{15}\\
E_4 &= \frac{2+20e^{-3t}+ 18 e^{-5t} +5e^{-6t}}{45} \\
E_5 &= \frac{4+70 e^{-3t} + 126 e^{-5t} +70 e^{-6t} +45 e^{-7t}}{315}
\end{split}
\end{equation}
suggesting that the empty interval distribution relaxes to the steady state according to 
\begin{equation}
\label{Ekt-1}
E_k(t) = \frac{2^{k+1}}{(k+2)!} +  A_k\,e^{-3t}+B_k\,e^{-5t}+O\big(e^{-6t}\big)
\end{equation}
Rather than deriving \eqref{Ekt-1} from the exact solution \eqref{Ekt} it is easier to prove this relaxation law by induction. 

To determine the amplitudes $A_k$ and $B_k$ we plug \eqref{Ekt-1} into \eqref{Ek:eq} and arrive at the recurrences $(k-1)A_k = 2A_{k-1}$ and $(k-3)B_k = 2B_{k-1}$ fixing the amplitudes:
\begin{subequations}
\label{ABk:sol}
\begin{align}
\label{Ak:sol}
A_k  &= \frac{2^{k-1}}{3(k-1)!} ~\quad  (k\geq 1)\\
\label{Bk:sol}
B_k  &= \frac{2^{k-1}}{20(k-3)!}   \quad  (k\geq 3)
\end{align}
\end{subequations}
and $B_1=B_2=0$. Combining \eqref{VEk} with \eqref{Ekt-1}--\eqref{ABk:sol} leads to the relaxation law of the void distribution 
\begin{equation*}
V_k(t)  = 2^{k+1}\,\frac{k(k+3)}{(k+4)!}+a_k\,e^{-3t}+b_k\,e^{-5t}+O\big(e^{-6t}\big)
\end{equation*}
with $b_1=0$ and other amplitudes
\begin{subequations}
\begin{align}
\label{ak:sol}
a_k  &= \frac{2^{k-1}\,k(k-3)}{3(k+1)!} ~\qquad\quad  (k\geq 1)\\
\label{bk:sol}
b_k &= \frac{2^{k-1}\,(k-2)(k-5)}{20(k-1)!}   \quad  (k\geq 2)
\end{align}
\end{subequations}

\subsection{Occupation probabilities}
\label{subsec:pnk}

Let $p_{n,k}$ be the probability that an interval of $n$ sites is occupied by $k$ particles. For each $n$, we introduce the generating function
\begin{equation}
\label{Pn:def}
\mathcal{P}_n(x) = \sum_{k=0}^{n} p_{n,k} x^k
\end{equation}
We are primarily interested in the steady-state results. However, to appreciate the governing equations for the generating functions, it is easier to consider the evolution and then set $t=\infty$. To avoid cluttering the formulae, we shortly write $\mathcal{P}_n(x)$ instead of $\mathcal{P}_n(x, t)$.

\begin{figure}
\centering
\includegraphics[width=7.89cm]{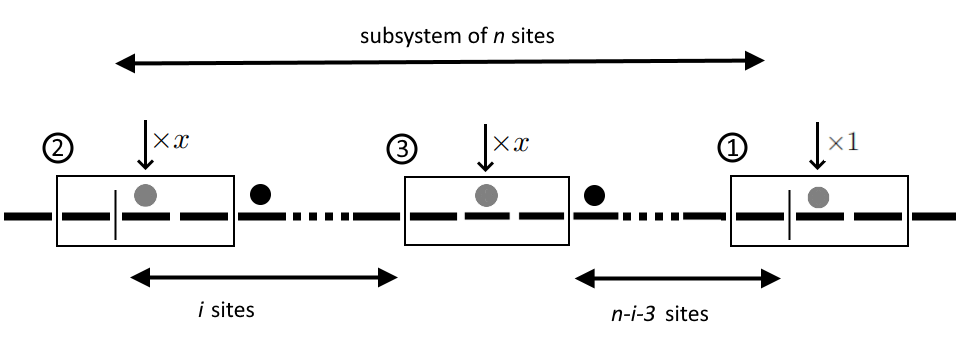}
\caption{Three types of updates affect an interval of $n$ sites. Black circles show some already present particles; gray circles designate newly added particles. Updates of the types (1) and (2) involve subsystems of sizes $n-1$ and $n-2$ and occur at both ends of the subsystem. Type (3) updates occur in bulk and involve the coalescence of two subsystems of sizes $i$ and $n-i-3$ where $i$ is between $0$ and $n-3$. Type (3) updates occur only for subsystems of length $n > 2$. Updates of types (2) and (3) involve adding a particle inside the subsystem and hence come with an additional weight $x$, while type (1) updates add a particle outside the subsystem and hence come with a factor $1$. }
\label{Fig:illustr}
\end{figure}

Consider first the evolution equation for $\mathcal{P}_1(x)$ corresponding to the single site. There are three updates involving this site, two deleting the particle on the site and one adding a particle (see Fig.~\ref{Fig:illustr}). Hence the gain term is $x+2$, and the evolution equation reads
\begin{subequations}
\begin{equation}
\label{P1}
\frac{d \mathcal{P}_1(x)}{dt} = -3 \mathcal{P}_1(x) + x+2
\end{equation}
Similarly, the evolution equation for $\mathcal{P}_2(x)$ is
\begin{equation}
\label{P2}
\frac{d \mathcal{P}_2(x)}{dt} = -4 \mathcal{P}_2(x) + 2\mathcal{P}_1(x) + 2x
\end{equation}
\end{subequations}

The evolution equation for $\mathcal{P}_n(x)$ with $n\geq 3$ has linear terms as in Eqs.~\eqref{P1}--\eqref{P2} coming from updates at the boundaries. There are also coalescence terms coming from updates that occur inside the interval. Each bulk update divides the interval into two subintervals of the lengths $i$ and $n-i-3$ and deposits a particle on the site $i+2$ (see Fig.~\ref{Fig:illustr}). The evolution equations read
\begin{eqnarray}
\label{Pn}
\frac{d \mathcal{P}_n(x)}{dt} &=& -(n+2) \mathcal{P}_n(x) + 2  \mathcal{P}_{n-1}(x) + 2  x \mathcal{P}_{n-2}(x) \nonumber\\
&+ &  x \sum_{i=0}^{n-3} \mathcal{P}_i(x) \mathcal{P}_{n-i-3}(x) 
\end{eqnarray}
The coalescence terms factorize because in the DSP all correlation functions on well-separated sets factorize, see Eq.~\eqref{2:far}. Setting
\begin{equation}
\label{P012}
\mathcal{P}_{-2}(x) = x^{-1}, \quad \mathcal{P}_{-1}(x) = \mathcal{P}_0(x)=1
\end{equation}
we incorporate the second and third terms on the right-hand side of Eq.~\eqref{Pn} into the coalescence term and obtain
\begin{equation}
\label{Pn:SS}
(n+2) \mathcal{P}_n(x) = x \sum_{i=-2}^{n-1} \mathcal{P}_i(x) \mathcal{P}_{n-i-3}(x)
\end{equation}
in the steady state. These equations hold for $n=1$ and $n=2$, namely, they reduce to the steady-state versions of \eqref{P1} and \eqref{P2} after simplifying the coalescence terms with the help of the boundary conditions \eqref{P012}. 

The recurrent nature of Eqs.~\eqref{Pn:SS} suggest to combine them into a single equation for the generating function
\begin{eqnarray}
G(x,y) &=& x \sum_{n\geq 0} y^n \mathcal{P}_{n-2}(x)  \nonumber \\
 &=& x y^2 \sum_{n\geq 0} \sum_{k=0}^{n} p_{n,k} x^k y^n + x y + 1
\end{eqnarray}
Using Eqs.~\eqref{Pn:SS} we deduce the governing equation for this grand canonical generating function
\begin{equation}
\label{G-eq}
\frac{d G}{d y} = G^2 + x -1
\end{equation}
Solving \eqref{G-eq} subject to $G(x,0)=1$ gives
\begin{equation}
\label{Gxy}
G(x,y) = \frac{\sqrt{1-x}-(1-x)\tanh[\sqrt{1-x} \,y]}{\sqrt{1-x}-\tanh[\sqrt{1-x}\,y]}
\end{equation}
Subtracting the contributions of $\mathcal{P}_{-2}(x)$ and $\mathcal{P}_{-1}(x)$ we obtain
\begin{equation}
\label{Fxy}
F(x,y) \equiv \sum_{n\geq 0} \sum_{k=0}^{n} p_{n,k}\, x^k y^n =\frac{G(x,y)-xy-1}{x y^2} 
\end{equation}

Noting that
\begin{equation}
\label{E:GF}
F(0,y) = \sum_{n\geq 0} p_{n,0}\, y^n = \sum_{n\geq 0} E_n y^n
\end{equation}
let us recover \eqref{Ek:sol} from \eqref{Gxy} and \eqref{Fxy} thereby providing a consistency check.  To find $F(0,y)$ we first expand $G(x,y)$ given by \eqref{Gxy} around $x=0$:
\begin{equation}
\label{Gxy:0}
G(x,y)  = 1 + \frac{e^{2 y} -1}{2}\,x+ O(x^2)
\end{equation}
Combining \eqref{Fxy} and \eqref{Gxy:0} we obtain 
\begin{equation*}
F(0,y) = \frac{e^{2y}-2 y - 1}{2 y^2} = \sum_{n\geq  0} \frac{2^{n+1}}{(2+n)!} y^n
\end{equation*}
leading indeed to \eqref{Ek:sol}.

\subsection{Cumulant generating function}
\label{subsec:cumulants}

The grand canonical generating function \eqref{Gxy} encapsulates the occupancy distribution for finite intervals. The emerging distribution is cumbersome, but
simplification occurs for very long intervals. In particular, the cumulants of the number of occupied sites scale linearly with the interval length $n$.  

The large deviation function defined via
\begin{equation}
\label{mu:def}
\mu(\lambda) = \lim_{n \rightarrow \infty} n^{-1} \ln \left\langle e^{\lambda k} \right\rangle
\equiv  \lim_{n \rightarrow \infty} n^{-1}  \ln \mathcal{P}_n\big(e^{\lambda}\big)
\end{equation}
encodes all cumulants. Since $F(x,y) = \sum_{n\geq 0} y^n \mathcal{P}_n(x)$, we have 
\begin{equation}
\mathcal{P}_n(x)  \sim \frac{1}{[y_*(x)]^n}
\end{equation}
for $n\gg  1$, where $y_*(x)$ is the pole of $F(x,y)$ closest to $y=0$. The large deviation function is then given by
\begin{subequations}
\label{mu:sol}
\begin{equation}
\label{mu-y}
\mu(\lambda) = -\ln{y_*\big(e^{\lambda}\big)}
\end{equation}
Using \eqref{Gxy} and \eqref{Fxy} we find 
\begin{equation}
\label{y-eq}
y_*\big(e^{\lambda}\big)  = \frac{\tanh^{-1}\!\big(\sqrt{1-e^{\lambda}}\big)}{\sqrt{1-e^{\lambda}}}
\end{equation}
\end{subequations}
Expanding \eqref{mu:sol} into the Taylor series one can extract all cumulants from the general formula
\begin{equation}
\label{mu:exp}
\mu(\lambda) = \sum_{p\geq 1} \frac{\lambda^p}{p!}\,\frac{\langle k^p\rangle_c}{n}
\end{equation}

The first cumulant is $\frac{\langle k\rangle}{n}=\frac{1}{3}$ in  agreement with already known steady-state density. The second cumulant gives the variance  of the occupation number:
\begin{equation}
\label{var}
\frac{\langle k^2\rangle_c}{n}\equiv \frac{\langle k^2\rangle - \langle k\rangle^2}{n} = \frac{2}{45}
\end{equation}
The so-called Mandel $Q$ parameter \cite{Mandel79} 
\begin{equation}
\label{Mandel:def}
Q=\frac{\langle k^2\rangle_c}{\langle k\rangle}-1
\end{equation}
is a basic measure characterizing the deviation from the Poisson statistics. The values  $-1\leq Q<\infty$ are permissible. For the Poisson statistics, $Q=0$ and the range $-1\leq Q<0$ is sub-Poissonian. Since
\begin{equation}
\label{Q:DSP}
Q = -\frac{13}{15}
\end{equation}
the occupation statistics is strongly sub-Poissonian. 

The ratios $\Phi_n=\langle k^n\rangle_c/\langle k\rangle$ of cumulants to the average are known as Fano factors \cite{Fano}. For the  Poisson random variable, all Fano factors are equal to unity. Here are a few Fano factors for the DSP of the one-dimensional lattice which we extracted from \eqref{mu:sol}--\eqref{mu:exp}:
\begin{equation}
\label{Fano}
\begin{split}
\Phi_2 & = \tfrac{2}{15}\\
\Phi_3 & = -\tfrac{2}{315}\\
\Phi_4 & =  -\tfrac{22}{1575}\\
\Phi_5 & =  \tfrac{2}{1485}\\
\Phi_6 & =  \tfrac{94\,442}{14\,189\,175}\\
\Phi_7 & =  -\tfrac{1622}{2\,027\,025}\\
\Phi_8 & =  -\tfrac{3\,581\,702}{516\,891\,375}\\
\Phi_9 & =  \tfrac{196\,599\,626}{206\,239\,658\,625}\\
\Phi_{10} & =  \tfrac{47\,221\,599\,182}{3\,781\,060\,408\,125} \\
\Phi_{11} & =  -\tfrac{532\,489\,978}{279\,030\,126\,375} \\
\Phi_{12} & =  -\tfrac{223\,496\,668\,545\,998}{6\,474\,894\,082\,531\,875}
\end{split}
\end{equation}

These Fano factors differ by $2\cdot (-1)^n$ from the Fano factors in the problem of random sequential covering of the one-dimensional lattice by dimers \cite{PK23}. In the covering problem, the system falls into a jammed state while the DSP process continues forever. Moreover, the steady state in the DSP is universal, while jammed states in the covering problem depend on the initial condition. (Jammed states formed in an initially empty system were studied in \cite{PK23}.) Still, the relation between the Fano factors hints at a connection between the statistics of the occupancy in the jammed state in the dimer covering problem and steady states in the DSP. 

The large deviation function \eqref{mu:sol} also describes the distribution of the number of peaks (or valleys) of a random surface whose base is a one-dimensional lattice, and the heights are independent identically distributed random variables. (The height distribution is irrelevant as long as it does not contain delta functions.) The large deviation function \eqref{mu:sol} was computed in \cite{Derrida-SG86} in the context of the one-dimensional spin glass, and then by a different method in \cite{peaks-hivert} where the distribution $p_{n, k}$ was additionally derived; see \cite{peaks-Billera,Majumdar06,perm-Oshanin,peaks-carmi,peaks-Dani,peaks-JML} for other work about peaks in uncorrelated landscapes. In Appendix \ref{ap:perm}, we explain the connection between steady states in the DSP, peaks in random surfaces, and permutations.

\subsection{Maximum occupancy}
\label{subsec:max}

We have already established the probabilities to observe empty intervals. Here we look at another extreme, viz., the maximally congested configurations. Using \eqref{mu:sol} one can deduce the asymptotic behavior 
\begin{equation}
\mu(\lambda)\to\frac{\lambda}{2}-\ln(\pi/2)
\end{equation}
of the large deviation function in the $\lambda\to\infty$ limit, from which one can extract the dominant behavior of the probabilities to observe maximally congested configurations. As for empty intervals, a direct treatment that we now present gives more precise results for the probabilities to observe maximally congested configurations.

For intervals with $n$ sites, the maximal occupancy is $n/2$ if $n$ is even and $(n+1)/2$ if $n$ is odd. Hence we define two sequences depending on the parity of $n$:
\begin{subequations}
\begin{equation}
\label{Q:def}
Q_n = 
\begin{cases}
	p_{n,(n+1)/2} &\mbox{$n$ is odd}\\
	0 &\mbox{$n$ is even}
\end{cases}
\end{equation}
and 
\begin{equation}
\label{R:def}
R_n = 
\begin{cases}
	0              &\mbox{$n$ is odd}\\
	p_{n,n/2} &\mbox{$n$ is even}
\end{cases}
\end{equation}
\end{subequations}

The evolution equations for $Q_n$ can be written similarly to Eqs.~\eqref{Pn}, except that updates that reduce the total number of particles do not contribute, and the coalescing sub-intervals must have the maximum occupancy. The evolution equation for the shortest intervals is
\begin{subequations}
\label{Q:eq}
\begin{align}
\label{Q1}
\frac{d Q_1}{dt}  = - 3 Q_1 +  1
\end{align}
while for $n>1$ we have
\begin{align}
\label{Qn}
\frac{d Q_n}{dt}  =  -(n+2) Q_n + 2 Q_{n-2} + \sum_{i=0}^{n-3} Q_i Q_{n-i-3}
\end{align}
\end{subequations}

The evolution equations for $R_n$ involve $Q_j$ with $j<n$. The evolution equation for the shortest even intervals is
\begin{subequations}
\label{R:eq}
\begin{equation}
\label{R2}
\frac{d R_2}{dt}  = - 4 R_2 +  2 Q_1 + 2
\end{equation}
The evolution equations for $R_n$ with $n>2$ involve two types of coalescence terms:
\begin{eqnarray}
\label{Rn}
\frac{d R_n}{dt}  &=&   -(n+2) R_n + 2 Q_{n-1} + 2 R_{n-2} \nonumber\\
& + &\sum_{i=0}^{n-3} Q_i R_{n-i-3} + \sum_{i=0}^{n-3} R_i Q_{n-i-3}
\end{eqnarray}
\end{subequations}

Setting
\begin{equation}
\label{Q:BC}
Q_{-2} = 0, \qquad  Q_{-1} = 1
\end{equation}
and introducing the generating function 
\begin{equation}
\label{Q:gf}
\mathcal{Q}(y) = \sum_{n\geq 0} Q_{n-2} y^n
\end{equation}
we reduce an infinite system \eqref{Q:eq} to a single partial differential equation which becomes
\begin{equation}
\label{Q-ODE}
\frac{d \mathcal{Q}}{d y} = 1+\mathcal{Q}^2
\end{equation}
in the steady state. Setting
\begin{equation}
\label{R:BC}
R_{-2} = 1, \qquad  R_{-1} = 0
\end{equation}
and introducing the generating function 
\begin{equation}
\label{R:gf}
\mathcal{R}(y) = \sum_{n\geq 0} R_{n-2} y^n
\end{equation}
we reduce an infinite system \eqref{R:eq} to a single partial differential equation which becomes 
\begin{equation}
\label{R-ODE}
\frac{d \mathcal{R}}{d y} = 2\mathcal{Q}\mathcal{R}
\end{equation}
in the steady state. 

Solving \eqref{Q-ODE} subject to $\mathcal{Q}(0)=Q_{-2}=0$ we obtain
\begin{equation}
\label{Q:sol}
\mathcal{Q}(y) = \tan y
\end{equation}
Solving \eqref{R-ODE} subject to $\mathcal{R}(0)=R_{-2} =1$ gives
\begin{equation}
\label{R:sol}
\mathcal{R}(y) = \frac{1}{\cos^2 y}
\end{equation}
Using \eqref{Q:sol}--\eqref{R:sol} we deduce $\frac{d \mathcal{Q}}{d y}=\mathcal{R}(y)$ from which
\begin{equation}
\label{RQ:n}
R_{2n}=(2n+3)Q_{2n+1} 
\end{equation}

The asymptotic of $Q_n$ for $n\gg  1$ can be obtained by looking at the poles at $\tan(y)$ closest to the origin. These are the simple poles at $y = \pm \frac{\pi}{2}$. One finds
\begin{subequations}
\begin{equation}
\label{Q:asymp}
Q_{2n+1} \simeq 2 \left(\frac{2}{\pi}\right)^{2n+4} 
\end{equation}
Using \eqref{RQ:n} and \eqref{Q:asymp} we establish the asymptotic of $R_n$ 
\begin{equation}
\label{R:asymp}
R_{2n} \simeq 2(2n+3) \left(\frac{2}{\pi}\right)^{2n+4} 
\end{equation}
\end{subequations}

The connection with permutations (Appendix \ref{ap:perm}) implies the similarity between maximally congested configurations and alternating permutations that is a classical subject going back to work of Andr\'{e} \cite{Andre1879,Andre1881} in the 19th century, see \cite{Arnold92,Stanley} for review. Different definitions give qualitatively similar predictions. For instance, it is customary to consider alternating up/down permutations which, e.g., for $n=4$ give the unique $\circ\bullet\circ\bullet$ configuration, while there are two more maximally congested configurations:  $\bullet\circ\bullet\circ$ and $\bullet\circ\circ\bullet$. In the even case, the generating function \eqref{R:sol} differs from the generating function $(\cos y)^{-1}$ describing corresponding alternating up/down permutations \cite{Andre1879,Andre1881,Arnold92,Stanley}. Still, the dominant exponential $(2/\pi)^{2n}$ asymptotic is the same.

\section{Square lattice}
\label{sec:lattice-2d}

The pair correlation function $C({\bf a}) = \langle \eta_{\bf 0} \eta_{\bf a}\rangle$ equals the density when ${\bf a}={\bf 0}$:
\begin{equation}
\label{rho-2d}
C({\bf 0}) =  \langle \eta_{\bf 0}^2\rangle = \langle \eta_{\bf 0}\rangle=\rho = \tfrac{1}{5}\left(1-e^{-5t}\right)
\end{equation}
If $a=|{\bf a}|=1$, i.e. ${\bf a}=(\pm 1, 0)$ or ${\bf a}=(0,\pm 1)$, the pair correlation function vanishes. When $a=|{\bf a}|>2$, 
\begin{equation}
\label{F2:sol}
C({\bf a}) = F_2 = \rho^2
\end{equation}

\begin{figure}
\centering
\includegraphics[width=7.89cm]{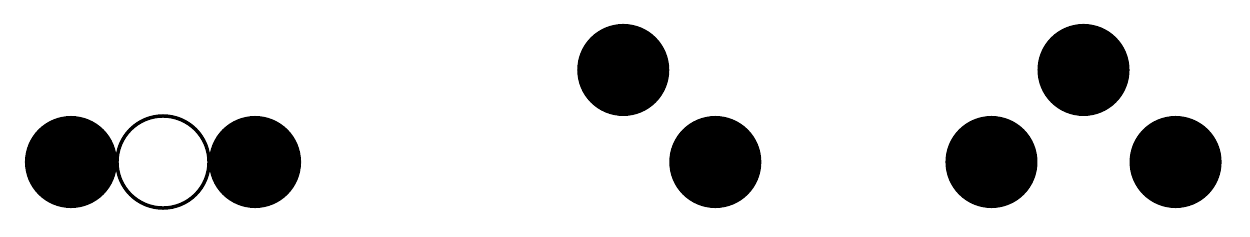}
\caption{The `horizontal' configuration on the left occurs with probability $H$. The open circle emphasizes that the site adjacent to two occupied sites (shown by disks) is empty; this and other necessarily empty sites are not shown in other configurations in this figure and Fig.~\ref{Fig:R1234}. There is also a  vertical configuration similar to the displayed horizontal configuration, it occurs with the same probability $H$. The configuration in the middle occurs with probability $D$. Three similar diagonal configurations differ by orientation and occur with the same probability. The configuration on the right occurs with probability $C(\therefore)$ defined by \eqref{C3:def}. }
\label{Fig:HDC}
\end{figure}

The pair correlation functions $C({\bf a})$ with ${\bf a}=(\pm 2, 0)$ or ${\bf a}=(0,\pm 2)$ are equal. Denote any such function by $H$, see Fig.~\ref{Fig:HDC}, and observe that it satisfies
\begin{equation}
\label{H:eq}
\frac{d H}{d t} = -9H + 2\rho
\end{equation}
This equation is analogous to \eqref{C2:eq} and again follows from \eqref{nij:eq} after noticing that $\mathcal{N}_{{\bf 0},{\bf a}}=9$ when  ${\bf a}=(\pm 2, 0)$ or ${\bf a}=(0,\pm 2)$. 
Solving \eqref{H:eq} we find
\begin{equation}
\label{H:sol}
H = \tfrac{1}{90}\left(4-9e^{-5t}+5 e^{-9t}\right)
\end{equation}

\begin{figure}
\centering
\includegraphics[width=7.89cm]{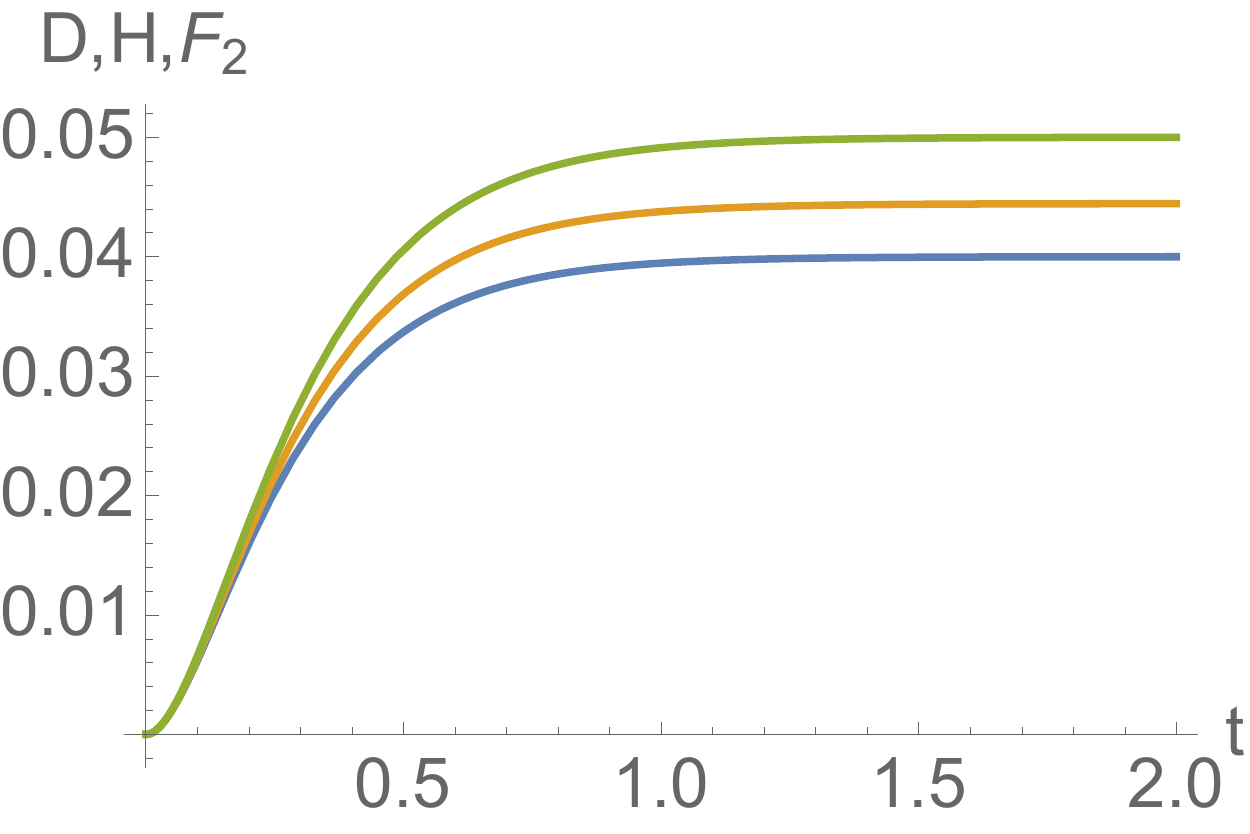}
\caption{The pair correlation functions $D, H, F_2$ (from top to bottom).  }
\label{Fig:DH}
\end{figure}

The pair correlation functions $C({\bf a})$ with ${\bf a}=(\pm 1, \pm 1)$ are also all equal. Denoting any such function by $D$, see Fig.~\ref{Fig:HDC}, and observing that it satisfies
\begin{equation}
\label{D:eq}
\frac{d D}{d t} = -8D + 2\rho
\end{equation}
we find
\begin{equation}
\label{D:sol}
D = \tfrac{1}{60}\left(3-8e^{-5t}+5 e^{-8t}\right)
\end{equation}
The correlation functions $F_2$, $H$ and $D$ given by \eqref{F2:sol}, \eqref{H:sol} and \eqref{D:sol} are plotted in Fig.~\ref{Fig:DH}. This plot suggests that these correlation functions exhibit the same asymptotic behavior in the $t\to 0$ limit. From Eqs.~\eqref{F2:sol}, \eqref{H:sol} and \eqref{D:sol} one finds that, as expected, all three correlation functions grow as $t^2$ when $t\ll 1$. 

Higher-order correlation functions $\langle \eta_{{\bf i}_1}\ldots \eta_{{\bf i}_\ell}\rangle$ are also universal, i.e. depend only on the order $\ell$, if the spatial positions are well separated: $|{\bf i}_m-{\bf i}_n|>2$ for all $m\ne n$. Denoting these correlation functions by $F_\ell$ we derive
\begin{equation}
\label{Fp-2:eq}
\frac{d F_\ell}{d  t} = -5\ell F_\ell + \ell F_{\ell-1}
\end{equation}
Solving \eqref{Fp-2:eq} recurrently we obtain the same solution \eqref{Fp:sol} as in one dimension, only the density is dimension-dependent. This is also the consequence of the general result \eqref{ell:far} specialized to the case of one-element sets. This strong de-correlation phenomenon occurs on all lattices $\mathbb{Z}^d$: For well-separated positions, the correlation functions are given by \eqref{Fp:sol} and depend only on the number of particles $\ell$. 

We now return to the square lattice and illustrate the computation of correlation functions with particles occupying nearby sites. The correlation function (see Fig.~\ref{Fig:HDC})
\begin{equation}
\label{C3:def}
C(\therefore) = \langle \eta_{-1,0} \eta_{0,1} \eta_{1,0}\rangle
\end{equation}
evolves according to
\begin{equation}
\label{C3:eq}
\left(\tfrac{d}{dt}+11\right) C(\therefore) =  H + 2D
\end{equation}
with $H$ given by \eqref{H:sol} and $D$ given by \eqref{D:sol}. Solving \eqref{C3:eq} we obtain
\begin{equation}
\label{C3:sol}
C(\therefore) = \tfrac{26-121e^{-5t}+110 e^{-8t} + 55 e^{-9t} - 70 e^{-11t}}{1980}
\end{equation}
The correlation function 
\begin{equation}
C_\diamond = \langle \eta_{1,0} \eta_{0,1} \eta_{-1,0}  \eta_{0,-1}\rangle
\end{equation}
evolves according to
\begin{equation}
\label{C-diamond:eq}
\frac{d C_\diamond}{d t} = -13 C_\diamond + 4 C(\therefore) 
\end{equation}
from which
\begin{equation}
\label{C-diamond:sol}
C_\diamond  = \tfrac{16-121e^{-5t}+176 e^{-8t} + 110 e^{-9t} - 280 e^{-11t} + 99 e^{-13t}}{3960}
\end{equation}

\begin{figure}
\centering
\includegraphics[width=7.89cm]{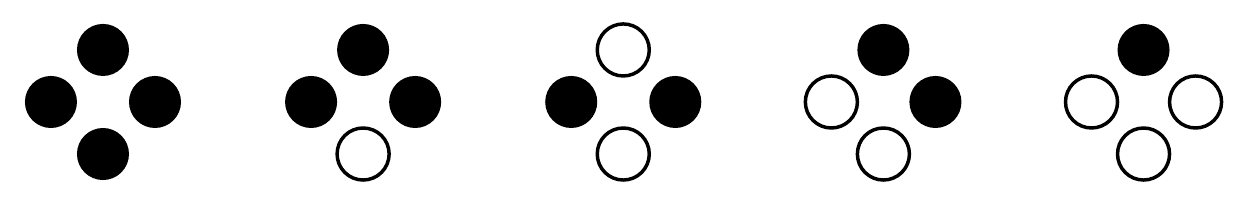}
\caption{Shown configurations (from left to right) occur with probabilities $R_4, R_3, H_2, D_2, R_1$. }
\label{Fig:R1234}
\end{figure}

To determine $p_n$ with $n=0,\ldots, 4$ we consider the occupation of the basic rhombus surrounding an empty site. If ${\bf 0}=(0,0)$ is the chosen empty site, the rhombus contains sites $(1,0), ~(0,1), ~(-1,0), ~(0,-1)$ each of which can be occupied or empty. We shall use the following notation: $R_4$ is the probability that all four sites are occupied; $R_3$ is the probability that only three sites $(1,0), ~(0,1), ~(-1,0)$ are occupied; $D_2$ is the probability that only two `diagonal' sites $(1,0), ~(0,1)$ are occupied; $H_2$ is the probability that only two `horizontal' sites $(1,0), ~(-1,0)$ are occupied; $R_1$ is the probability that only site $(1,0)$ is occupied. The corresponding configurations are shown in Fig.~\ref{Fig:R1234}. 

Accounting for different possible orientations of the occupied sites on the rhombus one finds four events with probabilities $R_3$; four probabilities $D_2$; two probabilities $H_2$; four probabilities $R_1$. 

It is possible to express $R_1, R_3, R_4$ and $D_2, H_2$ via already known correlation functions. By definition
\label{R-all}
\begin{equation}
\label{R4}
R_4 = C_\diamond 
\end{equation}
Further, $R_3+R_4=C(\therefore)$ as both left and right-hand sides give the probability that sites $(1,0), ~(0,1), ~(-1,0)$ are occupied. Therefore
\begin{equation}
\label{R3}
R_3 = C(\therefore) - R_4
\end{equation}
Similarly $D_2+2R_3+R_4=D$ and $H_2+2R_3+R_4=H$ which together with \eqref{R4} and \eqref{R3} lead to 
\begin{subequations}
\label{DH2}
\begin{align}
\label{D2}
D_2 &= D-2C(\therefore) + R_4\\
\label{H2}
H_2 &= H-2C(\therefore) + R_4
\end{align}
\end{subequations}
Finally, $R_1+2D_2+H_2+3R_3+R_4=\rho$ leading to
\begin{equation}
\label{R1}
R_1 = \rho - 2D-H + 3C(\therefore) - R_4
\end{equation}
Using \eqref{R4}--\eqref{R1} and accounting for the multiplicities we determine four desorption probabilities
\begin{equation}
\label{p1234}
(p_1,p_2,p_3,p_4)=\frac{(4R_1,4D_2+2H_2,4R_3,R_4)}{1-\rho}
\end{equation}
The probability of no desorption, $p_0=1-p_1-p_2-p_3-p_4$, is fixed by normalization. 

\begin{figure}
\centering
\includegraphics[width=7.89cm]{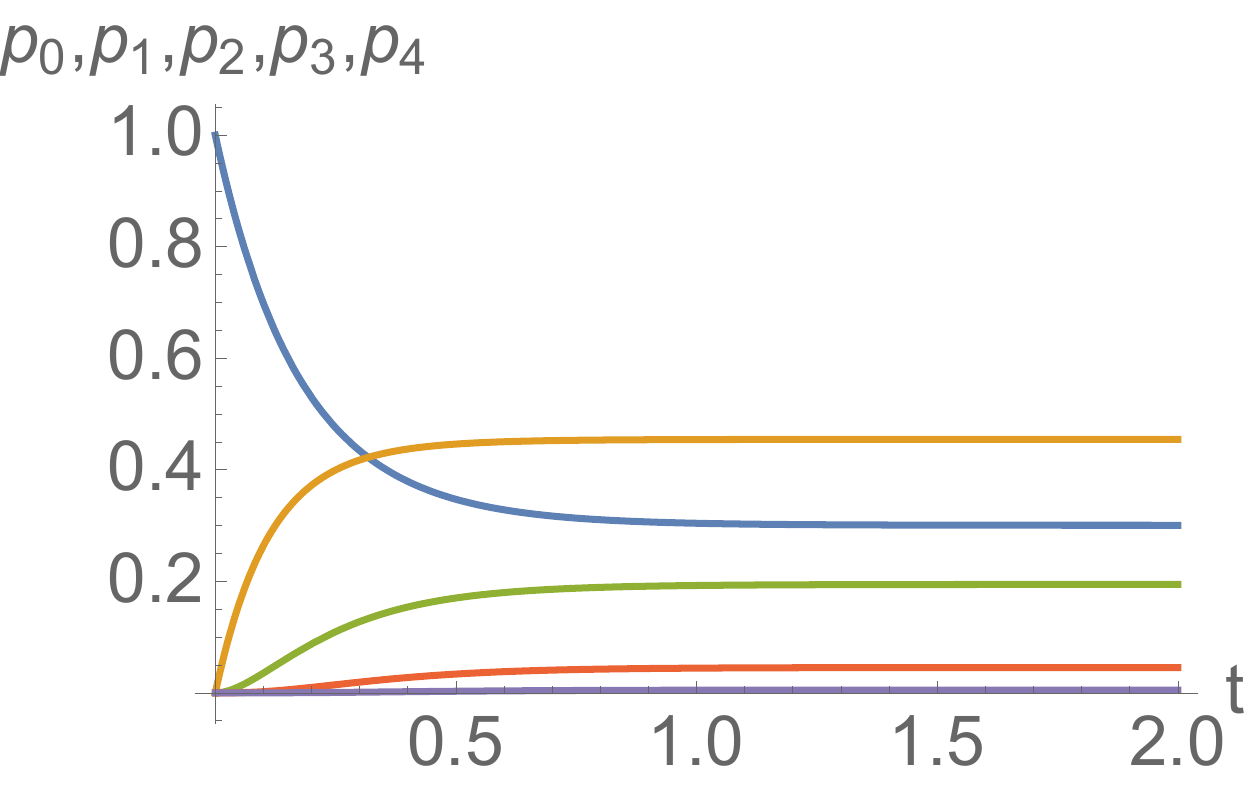}
\caption{The evolution of the desorption probabilities on the square lattice.  Throughout the evolution $p_1>p_2>p_3>p_4$; the first two probabilities satisfy $p_0>p_1$ when $0\leq t<t_*$ and $p_0<p_1$ when $t>t_*$ with $t_*\approx 0.317507542$. }
\label{Fig:2d-probs}
\end{figure}

The time-dependent solution for the desorption probabilities is thus given by \eqref{p1234} with $\rho$ determined by \eqref{rho-2d},  $H$ by \eqref{H:sol}, $D$ by \eqref{D:sol}, $C(\therefore)$ by \eqref{C3:sol}, and  $C_\diamond$ by  \eqref{C-diamond:sol}: 
\begin{equation*}
\begin{split}
p_0 & = \tfrac{952+1903 e^{-5t} +616 e^{-8t}  + 110 e^{-9t} +280 e^{-11t} + 99 e^{-13t}}{792(4+e^{-5t})}\\
p_1 & = \tfrac{360+55e^{-5t} -176 e^{-8t} - 140 e^{-11t} - 99 e^{-13t}}{198(4+e^{-5t})}\\
p_2 & = \tfrac{28-33e^{-5t}-12 e^{-8t} - 10 e^{-9t} +27 e^{-13t}}{36(4+e^{-5t})}\\
p_3 & = \tfrac{36-121e^{-5t}+44 e^{-8t} + 140 e^{-11t} - 99 e^{-13t}}{198(4+e^{-5t})}\\
p_4 & = \tfrac{16-121e^{-5t}+176 e^{-8t} + 110 e^{-9t} - 280 e^{-11t} + 99 e^{-13t}}{792(4+e^{-5t})}
\end{split}
\end{equation*}
Using these formulae one can verify the expected small time behavior: 
\begin{equation}
p_n=\binom{4}{n}t^n + O(t^{n+1})
\end{equation}
In the steady state, the desorption probabilities are given by the announced expressions \eqref{p01234}.

\section{Arbitrary dimension}
\label{sec:lattice-d}

For the hyper-cubic lattice $\mathbb{Z}^d$, the density is given by \eqref{rho-d:sol}. The correlation functions with well-separated sites are again universal and represented by \eqref{Fp:sol}. The computation of the desorption probabilities requires knowledge of non-trivial higher-order correlation functions involving sites that are not well-separated. The computation of these correlation functions quickly becomes very cumbersome as $d$ increases. Hence to gain insight, we begin with a mean-field approximation. 

\subsection{Mean-field approximation}
\label{subsec:MFA}

Neglecting correlations, we write a rate equation for the density
\begin{eqnarray}
\label{rho:MFT}
\frac{d\rho^\text{MF}}{dt} &=&- \sum_{n=0}^{2d}(n-1)\binom{2d}{n} (1-\rho^\text{MF}\big)^{2d-n+1}\big(\rho^\text{MF}\big)^{n} \nonumber \\
                         &=& (1-\rho^\text{MF}\big) \big(1-2 d \rho^\text{MF}\big)
\end{eqnarray}
The gain term ($n=0$) corresponds to the case when the landing site and all $2d$ neighbors are empty. The loss terms describe the situations when $n\geq 2$ neighbors are occupied. The mean-field prediction $\rho^\text{MF}=(2d)^{-1}$ for steady-state density approaches the exact result \eqref{rho-d} as $d\to\infty$.

The mean-field expressions for the steady-state desorption probabilities are
\begin{eqnarray}
\label{pn:MFT}
p_{n}^\text{MF} &=& \binom{2d}{n}(1-\rho^\text{MF}\big)^{2d-n}\big(\rho^\text{MF}\big)^{n}\nonumber\\
&=&  \binom{2d}{n}\frac{(2d-1)^{2d-n}}{(2d)^{2d}}
\end{eqnarray}
In one dimension, $\big(p_0^\text{MF}, p_1^\text{MF}, p_2^\text{MF}\big)=(\frac{1}{4},\frac{1}{2},\frac{1}{4})$ in the steady-state; the exact desorption probabilities are given by \eqref{p012}. The mean-field predictions \eqref{pn:MFT} become
\begin{equation}
\label{pn:MFT-2d}
\big(\tfrac{81}{256}, \tfrac{27}{64}, \tfrac{27}{128}, \tfrac{3}{64},\tfrac{1}{256}\big)
\end{equation}
for the square lattice and 
\begin{equation}
\label{pn:MFT-3d}
\big(\tfrac{15\,625}{46\,656}, \tfrac{3\,125}{7\,776}, \tfrac{3\,125}{15\,552}, \tfrac{625}{11\,664}, \tfrac{125}{15\,552}, \tfrac{5}{7\,776}, \tfrac{1}{46\,656}\big)
\end{equation}
for the cubic lattice.

Comparing \eqref{pn:MFT-2d} and exact results \eqref{p01234} we observe that the largest relative discrepancy is for $p_4$ when $d=2$. 
Comparing \eqref{pn:MFT-3d} and exact results [shown in Table~\ref{Table:345}] we observe that the largest relative discrepancy is for $p_6$ when $d=3$. This seemingly holds in all $d\geq 2$: the ratio $p_n(d)/p_{n}^\text{MF}(d)$ reaches maximum at $n=2d$.

\subsection{Derivation of Eqs.~\eqref{ratio}}

Exact results for the (steady-state) desorption probabilities in $d\leq 11$ dimensions indicate that for any $d$, the discrete distribution $p_k(d)$ is bell-shaped with the maximum at $k=1$. In other words
\begin{subequations}
\label{ineq}
\begin{equation}
\label{bell}
p_1(d)>p_0(d), \quad p_1(d)>p_2(d)>\ldots >p_{2d}(d)
\end{equation}
for $d\geq 1$. Additionally
\begin{equation}
\label{02:d}
p_0(d)>p_2(d)\qquad \text{for}\quad  d\geq 2.
\end{equation}
\end{subequations}

We have  not found a proof of the inequalities \eqref{ineq}. We do not have a guess for $p_0(d)$ in arbitrary $d$ or other similar infinite sequences $p_k(d)$ with fixed $k$. One could hope to guess $p_{2d}(d)$ as the mean-field prediction for this smallest probability is particularly simple: $p_{2d}^\text{MF}(d)=(2d)^{-2d}$. However, the probabilities $p_{2d}(d)$ with $d\leq 5$ shown in Table~\ref{Table:345}, and exact values that we computed up to $d=11$, make  questionable the existence of the general  formula as both the numerator and denominator of $p_{2d}(d)$ rapidly increase with dimension, e.g.,
\begin{equation*}
p_{12}(6)  = \frac{557507697256123784009641}{313826430834079851331840708512000000}
\end{equation*}

The desorption probabilities are rational numbers as we explain in Sec.~\ref{subsec:DP} and Appendix~\ref{ap:algorithm}. The numerators and denominators of $p_n(d)$ very quickly grow with $d$, e.g., the largest denominator increases roughly as $10^{d^2}$. To our surprise, the ratios 
\begin{equation}
\label{ratios:def}
\frac{p_{2d-1}(d)}{p_{2d}(d)} \qquad \text{and} \qquad \frac{2p_{2d-2}(d)}{p_{2d}(d)}
\end{equation}
are integer according to exact results for $d\leq 11$. Our data indicated that the ratios grow algebraically with $d$. For instance, $p_{2d-1}(d)/p_{2d}(d)$ are
\begin{equation*}
\label{rd:data}
3, 19, 33, 51, 73, 99, 129, 201, 243
\end{equation*}
for $d=1,\ldots,11$. Using these results we guessed \eqref{ratio-1}. The ratios $2p_{2d-2}(d)/p_{2d}(d)$ are
\begin{equation*}
\label{rd:hat-data}
77, 355, 1081, 2591,  5317,  9787,  16625,  26551,40381, 59027
\end{equation*}
for $d=2,\ldots,11$ which led us to \eqref{ratio-2} valid $d\geq 2$. Using \eqref{pn:MFT}  we deduce the mean-field predictions for these ratios 
\begin{equation*}
\begin{split}
 \frac{p_{2d-1}(d)}{p_{2d}(d)}\Big|_\text{MFT}   &= 2d(2d-1)\\
 \frac{2p_{2d-2}(d)}{p_{2d}(d)}\Big|_\text{MFT} & = 2d(2d-1)^3
 \end{split}
\end{equation*}
The second ratio is always even according to the mean-field theory, while the exact value is odd for $d\geq 2$. 

To derive \eqref{ratio-1}, we use two correlation functions. One correlation function
\begin{equation}
\label{Rd}
R_{2d} = \left\langle \prod \eta_{\bf i} \right\rangle
\end{equation}
with product taken over all $|{\bf i}|=1$ gives the probability that a $d-$dimensional rhombus around the selected site ${\bf 0}$ is fully occupied. Another correlation function is 
\begin{equation}
\label{Cd:1}
C_{2d-1} = \left\langle \prod \eta_{\bf i}\right\rangle, \qquad {\bf i}\ne (1,0,\ldots,0)
\end{equation}
with product taken over all $2d$ sites with $|{\bf i}|=1$ except one site. The correlation function \eqref{Cd:1} gives the probability that all neighbors of the selected site ${\bf 0}$,  with possible exception of site $(1,0,\ldots,0)$, are occupied. 

The correlation function \eqref{Rd} evolves according to the rate equation 
\begin{equation}
\label{Rd:eq}
\frac{d R_{2d} }{d t} = - A_{2d} R_{2d}  + 2d C_{2d-1}
\end{equation}
generalizing \eqref{C-diamond:eq} for the square lattice. The gain term in Eq.~\eqref{Rd:eq} is obvious. The amplitude $A_{2d}$ in the loss term counts the number of sites ${\bf i}$ of the following types:
\begin{enumerate}
\item Central site ${\bf i}= {\bf 0}$.
\item Sites $\pm \hat{j}$ on unit distance from ${\bf 0}$, i.e., neighbors of ${\bf 0}$. There are $2d$ such sites.
\item Sites $\pm 2\hat{j}$ on distance 2 from ${\bf 0}$. There are $2d$ such sites.
\item Sites $\pm \hat{j} \pm \hat{k}$ with $j\ne k$. There are $4 \binom{d}{2}=2d(d-1)$ such sites. 
\end{enumerate}
Summing the numbers of sites listed above yields
\begin{equation}
\label{Ad}
A_{2d}=1+2d+2d+2d(d-1)=1+2d+2d^2
\end{equation}

Let $R_{2d-1}$ be the probability that $(1,0,\ldots,0)$ is empty while remaining $2d-1$ sites of the $d-$dimensional rhombus around ${\bf 0}$ are occupied. We have
\begin{equation}
\label{Rd:1}
R_{2d-1} = C_{2d-1} - R_{2d}
\end{equation}
[cf. with analogous Eq.~\eqref{R3} for the square lattice]. The desorption probabilities in the ratio \eqref{ratio-1} 
\begin{align}
\label{pd1}
p_{2d}(d) = \frac{R_{2d}}{1-\rho}\,, \qquad p_{2d-1}(d) =\frac{2d R_{2d-1}}{1-\rho}
\end{align}
from which
\begin{equation}
\label{rd:eq}
\frac{p_{2d-1}(d)}{p_{2d}(d)} = 2d\,\frac{R_{2d-1}}{R_{2d}}= \frac{2d C_{2d-1}}{R_{2d}}-2d
\end{equation}
In the steady state we deduce $A_{2d} = 2d C_{2d-1}/R_{2d}$ from Eq.~\eqref{Rd:eq}. Thus \eqref{rd:eq} simplifies to
\begin{equation*}
\frac{p_{2d-1}(d)}{p_{2d}(d)} = A_{2d}-2d = 2d^2+1
\end{equation*}
in the steady state giving the announced result \eqref{ratio-1}. 

To derive \eqref{ratio-2}, we need correlation functions $\left\langle \prod \eta_{\bf i}\right\rangle$, with product taken over $2d-2$ neighbors of ${\bf 0}$. If the omitted neighbors are $\{+\hat{j},-\hat{j}\}$, we denote by $C_{2d-2}^{(1)}$ the corresponding correlation function; if  the omitted neighbors are $\{\pm \hat{j}, \pm \hat{k}\}$ with $j\ne k$, we denote by $C_{2d-2}^{(2)}$ the corresponding correlation function. 

Take a set of $2d-1$ sites that are neighbors of ${\bf 0}$. One can exclude a site from this set in $2d-1$ ways. One of these ways leads to a set of the first kind, so the appropriate correlation function is $C_{2d-2}^{(1)}$; other $2d-2$ exclusions lead to sets of the second kind with appropriate correlation function $C_{2d-2}^{(2)}$. Therefore
\begin{equation}
\label{CCA}
C_{2d-1} = \frac{C_{2d-2}^{(1)} + 2 (d-1) C_{2d-2}^{(2)}}{A_{2d-1}} 
\end{equation}
where $A_{2d-1}$ is the total volume of the set of the $2d-1$ sites and their neighbors. This set is a subset of the set which was counted in $A_{2d}$ above. If the excluded site is $\hat{i}$, only the sites $\hat{i}$ and $2 \hat{i}$ are not neighbors of the remaining $2d-1$ sites. Thus
\begin{equation}
\label{Ad:1}
A_{2d-1} = A_{2d}-2 = 2 d^2 + 2 d -1
\end{equation}

The probabilities that $2d-2$ neighbors of ${\bf 0}$ are occupied and the other two are empty also come in two types, $R_{2d-2}^{(1)}$ and $R_{2d-2}^{(2)}$. Expressing these probabilities via the correlation functions $C_{2d-2}^{(1)}, C_{2d-2}^{(2)}, C_{2d-1}, C_{2d}$ we arrive at equations
\begin{subequations}
\label{Rd:2}
\begin{align}
\label{Rd:21}
R_{2d-2}^{(1)}  &= C_{2d-2}^{(1)} -2 C_{2d-1} + R_{2d}\\
\label{Rd:22}
R_{2d-2}^{(2)} &= C_{2d-2}^{(2)} -2 C_{2d-1} + R_{2d}
\end{align}
\end{subequations}
generalizing Eqs.~\eqref{D2}--\eqref{H2} for the square lattice.  Similarly to Eqs.~\eqref{pd1}, we express $p_{2d-2}(d)$ through the  probabilities $R_{2d-2}^{(1)}$ and $R_{2d-2}^{(2)}$:
\begin{equation}
\label{pd2}
p_{2d-2}(d) = d\,\frac{R_{2d-2}^{(1)} + 2(d-1) R_{2d-2}^{(2)}}{1-\rho}
\end{equation}
Combining formulae \eqref{pd1} and \eqref{pd2} yields
\begin{equation}
\label{pd2d}
\frac{p_{2d-2}(d)}{p_{2d}(d)} = d\,\frac{R_{2d-2}^{(1)}  + 2(d-1) R_{2d-2}^{(2)}}{R_{2d}}
\end{equation}
We substitute \eqref{Rd:2} into the right-hand side of \eqref{pd2d}, and then use \eqref{CCA} and $A_{2d} = 2d C_{2d-1}/R_{2d}$, and finally \eqref{Ad} and \eqref{Ad:1} yielding 
\begin{eqnarray*}
\frac{p_{2d-2}(d)}{p_{2d}(d)} &=&  d\,\frac{C_{2d-2}^{(1)} +  2(d-1) C_{2d-2}^{(2)}}{R_{2d}} \\
&-& (2d-1)\frac{2d C_{2d-1}}{R_{2d}} + d(2d-1) \\
&=& \frac{1}{2}\,A_{2d-1}A_{2d}-(2d-1)A_{2d}+ d(2d-1) \\
&=&  2d^4 + 2 d^2 -d +\frac{1}{2}
\end{eqnarray*}
which is the announced result \eqref{ratio-2}. 

Given the striking simplicity of $p_{2d-1}(d)/p_{2d}(d)$ and $p_{2d-2}(d)/p_{2d}(d)$, one could hope that $p_{2d-3}(d)/p_{2d}(d)$ is rational with small denominators for all $d$. This is not so, the denominators  rapidly grow with $d$. For instance
\begin{equation*}
\frac{p_7(5)}{p_{10}(5)} = \frac{93\,838\,852\,439\,754\,606\,604}{4\,398\,633\,792\,872\,509}
\end{equation*}

\subsection{Desorption probabilities in arbitrary dimension}
\label{subsec:DP}

By definition, $p_k$ is the probability that $k$ neighbors of a `central'  site on $\mathbb{Z}^d$ are occupied, conditioned that the site itself is empty. Here we derive a formula for the generating function encapsulating $p_k$ in terms of correlation functions involving neighboring sites. The generating function is defined via
\begin{equation}
\label{GFP:def}
\mathcal{P}(x) = \sum_{k=0}^{2d} p_k x^k
\end{equation}
The desorption probabilities $p_k$ for $k \ge 1$ can be written in terms of correlation functions of the neighboring sites, while $p_0$ is determined by the normalization condition \eqref{norm:p}, which becomes $\mathcal{P}(1) = 1$.

Consider clusters of $k$ sites that are the neighbors of the central site. The set of such clusters ${\bf  S}_k=\{S_k\}$ contains $\binom{2d}{k}$ clusters. Recalling the definition of the desorption probabilities we re-write the generating function \eqref{GFP:def} as 
\begin{subequations}
\begin{equation}
\label{GF:P-av}
\mathcal{P}(x) = P_0 + \frac{\langle P(x)\rangle}{1-\rho} 
\end{equation}
where $P_0$ similarly to $p_0$ is fixed by normalization, and 
\begin{equation}
\label{P:def}
P(x) = \sum_{k=0}^{2d} \sum_{S_k\in {\bf  S}_k} \prod_{i \in S_k} x\eta_i \prod_{j \notin S_k} (1-\eta_j)
\end{equation}
\end{subequations}
The first product in the right-hand side of Eq.~\eqref{P:def} is over the $k$ sites in the cluster $S_k$, and the second product is over the $2d-k$ neighboring sites not in the cluster.  Expanding the right-hand side of \eqref{P:def} into the product of $x\eta_i$ and $(1-\eta_j)$ factors we re-write $P(x)$ as
\begin{equation*}
\label{P:prod}
P(x) = \prod_{\ell\in \mathcal{N}} \left[x \eta_\ell+(1-\eta_\ell)\right] = \prod_{\ell\in \mathcal{N}} \left[1+(x-1) \eta_\ell\right]
\end{equation*}
where $\mathcal{N}$ denotes the set of all the nearest neighbors of the central site. Expanding the second product yields
\begin{equation}
\label{P:final}
	P(x)  = \sum_{k=0}^{2d} (x-1)^k \sum_{S_k \in \boldsymbol{S}_k} \prod_{i \in S_k} \eta_i
\end{equation}
Averaging \eqref{P:final} we reduce \eqref{GF:P-av} to
\begin{eqnarray}
\label{GFP}
\mathcal{P}(x)  &=& P_0 + \frac{1}{1-\rho}\sum_{k=0}^{2d} (x-1)^k \sum_{S_k \in \boldsymbol{S}_k} C(S_k)  \nonumber\\
	                 &=&1 + \frac{1}{1-\rho} \sum_{k=1}^{2d} (x-1)^k \sum_{S_k \in \boldsymbol{S}_k} C(S_k)
\end{eqnarray}
where the correlation functions $C(S_k)$ are generally defined by \eqref{CS:def} and the second line in \eqref{GFP} follows from the first after accounting for the normalization $\mathcal{P}(1)=1$.

The correlation function $C(S_k)$ can be expressed via correlation functions of clusters with $k-1$ sites in $S_k$. Denote by $V[S_k]$ the `volume' of the neighborhood of $S_k$. This volume is equal to $k=|S_k|$ plus the number of neighboring sites on distance 1 from $S_k$. The evolution equation for $C(S_k)$ reads 
\begin{equation}
\label{CSk:eq}
\frac{d C(S_k)}{d t} = - V[S_k] C(S_k) + \sum_{j \in S_k} C(S_k - \{j\})
\end{equation}
In the steady state
\begin{equation}
\label{CSk}
V[S_k] C(S_k) = \sum_{j \in S_k} C(S_k - \{j\})
\end{equation}

In Appendix~\ref{ap:algorithm}, we rely on Eqs.~\eqref{GFP} and \eqref{CSk} to devise an algorithm computing desorption probabilities in arbitrary dimension.

\subsection{Infinite-dimensional limit}
\label{subsec:inf}

One anticipates that the mean-field predictions \eqref{pn:MFT} for desorption probabilities become exact when $d=\infty$. Taking the $d\to\infty$ limit of Eqs.~\eqref{pn:MFT} we arrive at the announced Poisson distribution \eqref{pn:inf}. 

A more rigorous derivation of \eqref{pn:inf} relies on Eqs.~\eqref{GFP} and \eqref{CSk} valid in arbitrary $d$. These equations greatly simplify when the spatial dimension diverges, namely, when $d\to\infty$ with $k$ kept fixed.  First, we notice that the volume is maximal, $V[S_k]=(2d+1)k$, for almost all clusters if $d\gg 1$, namely, the fraction of clusters with smaller volume, $V[S_k]<(2d+1)k$, vanishes in the $d\to\infty$ limit. Similarly $C(S_k)=C_k$ for almost all clusters, and hence \eqref{CSk} gives $C_k=(2d+1)^{-1}C_{k-1}$ from which $C_k=(2d+1)^{-k}$. With the same precision $|{\bf  S}_k|=\binom{2d}{k}\to (2d)^k/k!$, and $(1-\rho)^{-1}=\frac{2d+1}{2d}\to 1$. Therefore in the $d\to\infty$ limit, Eq.~\eqref{GFP} simplifies to
\begin{equation}
\label{Poisson}
 \sum_{k\geq  0} p_k x^k = 1+\sum_{k\geq 1} \frac{(x-1)^k}{k!}=e^{x-1}
\end{equation}
leading to the Poisson distribution \eqref{pn:inf}.

\subsection{Fluctuations of the occupation number}
\label{subsec:fluct}

In one dimension, we have computed cumulants of the occupation number of connected sets (intervals) and the extremal probabilities of observing empty and maximally congested configurations (Secs.~\ref{subsec:empty}--\ref{subsec:max}). In higher dimensions, connected sets are much more diverse than in one spatial dimension. Fortunately, for large sufficiently round connected sets, the leading asymptotic behaviors depend only on the number of sites, $N\gg 1$. (In a sufficiently round connected large set, linear dimensions in every direction are comparable.) The shape of the set affects only sub-leading terms. Thus we shortly denote by $E_{d,N}$ the probability that the set is empty and by  $M_{d,N}$ the probability that the set is maximally congested, i.e., occupied by $N/2$ particles. Recall that in one dimension
\begin{equation}
\label{EM:1}
E_{1,N} = \frac{2^{N+1}}{(N+2)!}\,, \qquad M_{1,N}\asymp \left(\frac{2}{\pi}\right)^N
\end{equation}
Here $A \asymp B$ denotes the asymptotic equality of the logarithms of $A$ and $B$, viz., $\lim_{N\to\infty} \frac{\ln A}{\ln B}=1$. 

Qualitatively similar behaviors are expected in higher dimensions:
\begin{align}
\label{EM-d}
E_{d,N} \asymp \frac{\alpha_d^N}{N!}\,, \qquad  M_{d,N} \asymp \kappa_d^N
\end{align}

The exponent $\kappa_d$ was computed numerically \cite{Majumdar06} in the range $2\leq d\leq 5$. Simulations suggest that $\kappa_d$ is a decreasing function of $d$. The exponent varies in the range
\begin{equation}
\label{bounds}
\frac{2}{\pi} \geq \kappa_d \geq \frac{1}{2}
\end{equation}
The upper bound is just $\kappa_1$, while the lower bound is the infinite-dimensional limit $\kappa_\infty$. This latter result can be deduced from the exact value \cite{Majumdar06} of the exponent on the Cayley tree with $K$ branches:
\begin{equation}
\label{kappa}
\kappa(K)=\frac{K}{B(K^{-1},K^{-1})}
\end{equation}
where $B(x,y)=\Gamma(x)\Gamma(y)/\Gamma(x+y)$ is the Euler's beta function. For $K=2$, we recover $\kappa_1=\frac{2}{\pi}$ from \eqref{kappa}, while when $K\gg  1$ we obtain $\kappa(K)\to \frac{1}{2}$ leading to $\kappa_\infty = \frac{1}{2}$ as the higher-dimensional hyper-cubic lattice is asymptotically the Cayley tree with $K=2d$. 

The lower bound in \eqref{bounds} is very natural as we now demonstrate. Take a finite bipartite graph $G=G_{-}\cup G_{+}$ with parts of equal size: $|G_{-}|=|G_{+}|=n$. Consider permutations of $\{1,\ldots,2n\}$ such that the numbers $\{1,\ldots,n\}$ are on the vertices of $G_{-}$ and the numbers $\{n+1,\ldots,2n\}$ are on the vertices of $G_{+}$. In this case, the maxima are on the subgraph $G_{+}$. The probability of such specific maximally congested configurations is
\begin{equation}
\label{binom}
\frac{n!n!}{(2n)!}\asymp 2^{-2n}
\end{equation}
provides the lower bound for the probability $M(G)$ to observe a maximally congested configuration with half vertices occupied. Comparing $M(G)\asymp[\kappa(G)]^{2n}$ with \eqref{binom} we arrive at the lower bound $\kappa(G)\geq \frac{1}{2}$ for all large finite bipartite graphs $G$ with parts of equal size. 

The hyper-cubic lattice is an infinite bipartite graph: $\mathbb{Z}^d= \mathbb{Z}^d_\text{even}\cup \mathbb{Z}^d_\text{odd}$ with sub-lattices
\begin{equation}
\label{Z+-}
\begin{split}
\mathbb{Z}^d_\text{even} &=\{{\bf x}\in \mathbb{Z}^d\,| x_1+\ldots+x_d \equiv 0 \mod 2\} \\
\mathbb{Z}^d_\text{odd} &=\{{\bf x}\in \mathbb{Z}^d\,| x_1+\ldots+x_d \equiv 1 \mod 2\} 
\end{split}
\end{equation}
For  a large sufficiently round connected set with $N$ sites, close to $N/2$ sites belong to each sub-lattice, and the above argument leads to $\kappa_d\geq \frac{1}{2}$.

The emptiness probability $E_{d,N}$ has not been studied so far in $d\geq 2$ dimensions. Heuristic arguments suggest an inverse factorial decay, albeit the asymptotic \eqref{EM-d} may be too bold, a more general $E_{d,N} \asymp \alpha_d^N/\Gamma(\beta_d N)$ asymptotic is a safer guess. We only know $\alpha_1=2$ and $\beta_1=1$. 

Finally, we mention the exact result for the variance of the occupation number:
\begin{equation}
\label{v-d}
v_d = \lim_{N\to\infty}\frac{\langle k^2\rangle-\langle k\rangle^2}{N}=\frac{4 d^2 - d - 1}{(2d+1)^2 (4d+1)}
\end{equation}
In one dimension, we recover \eqref{var}. This result appears in diverse contexts ranging from metastable states in one-dimensional spin glass models \cite{Derrida-SG86} to the occupation time of a one-dimensional non-Markovian sequence \cite{Majumdar02}. The variance on the square lattice, $v_2= \frac{113}{225}$, was also known \cite{peaks-hivert}. The Mandel $Q$ parameter
\begin{equation}
\label{Q-d}
Q_d = \frac{v_d}{\langle k\rangle} -1=-\frac{1}{2}-\frac{1}{2+4d} -\frac{1}{1+4d}
\end{equation}
remains negative, $Q_d>-\frac{1}{2}$, implying sub-Poisson statistics of the occupation number in all spatial dimensions. 

The derivation of \eqref{v-d} is presented in Appendix~\ref{ap:var}.

\section{Continuous Version}
\label{sec:cont}

The space is filled with balls of radii $\frac{1}{2}$ according to the following rule: After each deposition event, we remove the balls with centers on distance $\leq 1$ from the center of the added ball. The density $\rho_d(t)$ satisfies the evolution equation
\begin{equation}
\label{evol:cont}
\frac{d \rho_d}{d t} = 1 - V_d\rho_d
\end{equation}
where $V_d$ is the volume of the ball of unit radius:
\begin{equation}
\label{Vd}
V_d \equiv V_d(1) = \frac{\pi^{d/2}}{\Gamma(1+d/2)}
\end{equation}
Equation \eqref{evol:cont} gives 
\begin{equation}
\label{rho:cont}
\rho_d = \frac{1-e^{-V_d t}}{V_d}
\end{equation}
and in the steady state $\rho_d = 1/V_d$. The volume fraction occupied by balls in the steady state is $\phi_d = \rho_d V_d(\frac{1}{2})$, from which
\begin{equation}
\label{volume:cont}
\phi_d = \frac{V_d(\frac{1}{2})}{V_d(1)}= 2^{-d}
\end{equation}

The volume fraction of a {\em saturated} sphere packing exceeds $2^{-d}$. This `greedy' bound admits a one-sentence proof \cite{Minkowski05} and appears weak,  but even a guess for an anticipated exponential improvement is lacking. The first modest improvement,  the $2\zeta(d)/2^d$ lower bound where $\zeta(\cdot)$ is the zeta function, was also established by Minkowski \cite{Minkowski05}; linear improvements, i.e., bounds growing like $d/2^d$, were found in \cite{Rogers,Vance11}; a slightly super-linear $d\, \ln(\ln d)/2^d$ bound is proven \cite{Venkatesh} for a sparse sequence of dimensions. 

Thus non-saturated sphere packings generated by the DSP in $\mathbb{R}^d$ have the volume fraction $2^{-d}$ coinciding with the classical Minkowski's lower bound for the volume fractions of saturated sphere packings. 

\subsection{One dimension}

In one dimension, the density is
\begin{equation}
\label{rho:1-cont}
\rho_1 = \frac{1-e^{-2t}}{2}
\end{equation}
The density $V(x,t)$ of voids of length $x$ has been also computed in \cite{KR-birds22}. The results are more cumbersome than in the lattice version as one must recurrently determine $V(x,t)$ for each interval $n<x<n+1$ using $V(x,t)$ for $x<n$. For instance, one finds \cite{KR-birds22}
\begin{equation}
\label{Vx12:sol}
V(x,t)= \frac{1-e^{-(2+x)t}}{2+x}-\frac{e^{-2t}- e^{-(2+x)t}}{x}
\end{equation}
when $1<x<2$. The void density $V(x,t)$ in the following interval $2<x<3$ satisfies 
\begin{eqnarray}
\label{Vx2-rec}
\frac{d V(x,t)}{dt} &=& -(2+x)V(x,t)-2\int_1^{x-1}dy\,V(y,t)\nonumber \\
&+&1-e^{-2t}
\end{eqnarray}
with $V(y,t)$ in the integral taken from \eqref{Vx12:sol}.

In the steady state
\begin{subequations}
\begin{align}
\label{V:1-2}
V(x) &=\frac{1}{2+x} ~~\qquad\qquad 1<x<2 \\
\label{V:2-3}
V(x) &=\frac{1-2\ln\frac{x+1}{3}}{2+x}\qquad 2<x<3
\end{align}
In the following interval $3<x<4$ 
\begin{eqnarray}
\label{V:3-4}
(2+x)V(x)  &=& 1-\ln(4/9)  -2\text{Li}_2(-3)+2\text{Li}_2(-x) \nonumber\\
&-&(1+ 2\ln 3 - 2\ln x)\ln(1+x)
\end{eqnarray}
\end{subequations}
where $\text{Li}_2(-x) = \sum_{j\geq 1}(-x)^j/j^2$ is the dilogarithm function \cite{Zagier,NIST}. In the $3<x<4$ range relevant for  $\text{Li}_2(-x)$ appearing in \eqref{V:3-4}, and actually for all $x>1$, it is necessary to perform an analytical continuation of the infinite sum defining the dilogarithm. A convenient expression is given by an integral representation
\begin{equation}
\text{Li}_2(-x) =  -  \int_0^\infty  du\,\frac{u}{1+ x^{-1}e^{u}}
\end{equation}
The steady-state void density $V(x)$ becomes cumbersome as $x$ increases. In the large $x$ limit, the decay of the void density is essentially factorial \cite{KR-birds22} as in the lattice case, viz. $V(x)=e^{-w(x)}$ with 
\begin{equation}
w= x[\ln x + \ln(\ln x) - 1 - \ln 2] + \ldots
\end{equation}

The cumulant generating function has not been computed so far. The variance can be deduced from the previous results. The corresponding Fano factor 
\begin{equation}
\label{Fano2:1d}
\Phi_2 = -1+4 \ln(4/3) = 0.150\,728\ldots
\end{equation}
slightly exceeds the Fano factor $\Phi_2=\frac{2}{15}$ for $\mathbb{Z}$. 

The desorption probabilities are known \cite{KR-birds22} only in one dimension. In the steady state, the desorption probabilities are given by \eqref{p012:cont}. In higher dimensions, the steady-state desorption probabilities can be expressed via certain multiple integrals as we show below.

\subsection{High dimensions}
\label{subsec:high}

The continuous DSP process in $d>1$ dimensions first appeared in 1960 in a Ph.D. of a forester Bertil Mat\'{e}rn,  see \cite{Matern}. This process admits an exact mapping into a model \cite{Viot97} of irreversible multilayer adsorption. The continuous DSP has been also studied in Refs.~\cite{TS06b,TS06a,TS06c}. 

The desorption probabilities are known only in one  dimension, Eq.~\eqref{p012:cont}. The desorption probabilities $p_n(d)$ vanish when $n>N_d$. In one dimension, $N_1=2$. Conjecturally, $N_d$ are related to the kissing numbers $K_d$, namely $N_d=K_d$ apart from special dimensions $d=2,8,24$ where $N_d=K_d-1$ is expected. Even if the relation between $N_d$ and $K_d$ is correct,  it sheds little light since the kissing numbers are known \cite{VDW,Leech,Levenshtein79,Odlyzko79,Ziegler,Musin08} only when $d=1,2,3,4,8,24$. Generalizing \eqref{ratio} to the continuous case, i.e., computing the ratios $p_{N_d-1}/p_{N_d}$  and $p_{N_d-2}/p_{N_d}$, seems impossible. In two dimensions, the calculation of the probabilities $p_n(2)$ with $n=0,\ldots, 5$ is in principle feasible as we show below. 

In the continuous DSP, the desorption probabilities $p_n(d)$ also greatly simplify in the $d\to\infty$ limit. To probe the behavior of $p_n(\infty)$, we employ an intuitively plausible assertion that the distribution of spheres is asymptotically uniform as $d\to\infty$.  Suppose we add a ball of radius $\frac{1}{2}$ to such uniform distribution. This ball overlaps with balls centered on distance $\leq 1$ from the center of the new ball. The probability to have $n$ such balls is $e^{-\rho_d V_d}/n!=e^{-1}/n!$ where in the last step we have used the relation \eqref{rho:cont} between the steady state density $\rho_d$ and the volume $V_d$ of the unit ball. Thus, we arrive at the same Poisson distribution \eqref{pn:inf} as for the DSP on $\mathbb{Z}^d$ in the $d\to\infty$ limit. 

We now present a more rigorous derivation of \eqref{pn:inf} for the continuous DSP in high dimensions extending the arguments for the DSP on $\mathbb{Z}^d$ given in Sec.~\ref{subsec:DP}.  Let $n(\br)$ be the probability density to have a ball centered at $\br$. The pair correlation function evolves  according to
\begin{eqnarray}
\label{nn:eq}
\frac{d}{dt}\langle n(\br_1) n(\br_2) \rangle &=& - V_2(\br_1,\br_2) \langle n(\br_1) n(\br_2)\rangle \nonumber\\ 
&+& \theta_{12}\,[\langle n(\br_1)\rangle+\langle n(\br_2)\rangle]
\end{eqnarray}
where $r_{12} = |\br_1 - \br_2|$, $\theta_{12}=\theta(r_{12} -1)$, and $V_2(\br_1,\br_2)$ is the total volume of the union of two balls of unit radius centered around $\br_1$ and $\br_2$. The theta function encodes the fact that deposition attempts at $\br_1$ should not evaporate a ball at $\br_2$ and vice versa. We emphasize that the one-body correlation function is spatially homogeneous: $\langle n(\br)\rangle = \rho_d$ with $ \rho_d$ given by \eqref{rho:cont}. The two-body correlation function $\langle n(\br_1) n(\br_2) \rangle$ depends only on the separation $r_{12} = |\br_1 - \br_2|$. 

In the steady state, Eq.~\eqref{nn:eq} becomes
\begin{equation}
\label{nn}
\langle n(\br_1) n(\br_2)\rangle =  \frac{\theta_{12}}{V_2(\br_1,\br_2)}\, [\langle n(\br_1)\rangle+\langle n(\br_2)\rangle]
\end{equation}
Similarly we find
\begin{eqnarray}
\label{nnn}
&&\langle n(\br_1) n(\br_2) n(\br_3)\rangle =  \frac{\theta_{12}\, \theta_{23}\, \theta_{31}}{V_3(\br_1,\br_2,\br_3)}\nonumber \\
&& [\langle n(\br_1)n(\br_2)\rangle+\langle n(\br_2) n(\br_3)\rangle +\langle n(\br_3)n(\br_1) \rangle]
\end{eqnarray}
where $V_3(\br_1,\br_2,\br_3)$ is the total volume of the union of balls centered around $\br_1, \br_2, \br_3$. Similar equations can be written for  higher correlation functions. Such equations were derived by Torquato and Stillinger \cite{TS06b} for a ghost random sequential adsorption model which  is isomorphic to the continuous DSP model. 

Proceeding as in the lattice case [Sec.~\ref{subsec:DP}], we express the generating function \eqref{GFP:def} through the correlation functions $\langle n(\br_1)\ldots n(\br_k)\rangle$:
\begin{equation}
\label{P:cont}
\mathcal{P}(x) = \sum_{k=0}^{\eta_d} \frac{(x-1)^k}{k!}\int \cdots\int\prod_{i=1}^k d\br_i \left\langle\prod_{i=1}^k n(\br_i)\right\rangle
\end{equation}
The integrals $\int  d\br_i$ are over unit ball. Equation \eqref{P:cont} is the continuous analog of Eq.~\eqref{GFP}  where  the sum over clusters is replaced by an integral over sets of $k$ points in the unit ball and the factor $k!$ corrects the overcounting.

The number of terms in the sum on the right-hand side of Eq.~\eqref{P:cont} diverges when $d\to\infty$, but the integrands greatly simplify in this limit. First, we recall that the volume of a $d$-dimensional ball concentrates near its surface \cite{Ball} when $d\gg 1$. Further, two unit vectors are asymptotically orthogonal, so the dominant contribution to the integral $\int d\br_1 \int d\br_2$ comes from the region where $r_{12} = |\br_1 - \br_2|=\sqrt{2}$, and hence $\theta_{12} = \theta(r_{12} -1)=1$  and $V_2(\br_1,\br_2) \rightarrow 2 V_{d}$ as $d \rightarrow \infty$, where $V_d$ is given by \eqref{Vd}. Thus \eqref{nn} simplifies to 
\begin{equation*}
\langle n(\br_1) n(\br_2)\rangle =  \frac{\langle n(\br_1)\rangle}{V_d}=  \frac{1}{V_d^2}
\end{equation*}
since $\langle n(\br_1)\rangle = \rho_d$.  Similarly $V_3(\br_1,\br_2, \br_3) \rightarrow 3 V_{d}$ in the $d \rightarrow \infty$ limit as all three vectors $\br_1, \br_2,  \br_3$ are asymptotically orthogonal. Thus \eqref{nnn} simplifies to 
\begin{equation*}
\label{nnn:inf}
\langle n(\br_1) n(\br_2) n(\br_3)\rangle =  \frac{\langle n(\br_1) n(\br_2)\rangle}{V_d} = \frac{1}{V_d^3}
\end{equation*}
and generally
\begin{equation}
\label{n-k:inf}
\left\langle\prod_{i=1}^k n(\br_i)\right\rangle = \frac{1}{V_d^k}
\end{equation}
when $d\to\infty$. Inserting \eqref{n-k:inf} into \eqref{P:cont} gives the same generating function \eqref{Poisson} as in the lattice case:
\begin{equation}
\label{P:cont-inf}
\mathcal{P}(x) = \sum_{k\geq 0}\frac{(x-1)^k}{k! V_d^k}\int \cdots\int\prod_{i=1}^k d\br_i=e^{x-1}
\end{equation}

In two dimensions, the generating function reads
\begin{eqnarray}
\label{P:cont-2}
\mathcal{P} & = & x + \frac{(x-1)^2}{2!} \int d\br_1 d\br_2\, \langle 1 2\rangle \nonumber \\
&+& \frac{(x-1)^3}{3!}\! \int d\br_1 d\br_2 d\br_3\, \langle 1 2 3\rangle \nonumber \\
&+& \frac{(x-1)^4}{4!}\! \int d\br_1 d\br_2 d\br_3\, d\br_4\, \langle 1 2 3 4\rangle \nonumber \\
&+& \frac{(x-1)^5}{5!}\! \int d\br_1 d\br_2 d\br_3\, d\br_4\,  d\br_5\,   \langle 1 2 3 4  5\rangle
\end{eqnarray}
where the integrals are over unit disks. Using Eq.~\eqref{nn}  we find the first integrand  in \eqref{P:cont-2}:
\begin{subequations}
\begin{equation}
\label{nn-2}
\langle 1 2\rangle  = \frac{2}{\pi}\,\frac{\theta_{12}}{V_2(\br_1,\br_2)}
\end{equation}
The second  integrand follows from \eqref{nnn}
\begin{equation}
\label{nnn-2}
\langle 1 2 3\rangle  = \frac{\langle 1 2\rangle +\langle 13\rangle +\langle 23 \rangle}{V_3(\br_1,\br_2,\br_3)} \prod_{1\leq a<b\leq 3}\theta_{ab}
\end{equation}
\end{subequations}
with $\langle 13\rangle$ and $\langle 13\rangle$ appearing in \eqref{nnn-2} obtained from $\langle 12\rangle$ by re-labeling. 
The last two integrands  in \eqref{P:cont-2} are
\begin{equation*}
\begin{split}
\langle 1 2 3 4\rangle  =&~ \frac{\langle 123\rangle +\langle 124\rangle +\langle 134\rangle +\langle 234 \rangle}{V_4(\br_1,\br_2,\br_3, \br_4)}
\prod_{1\leq a<b\leq 4}\theta_{ab} \\
\langle 1 2 3 4 5\rangle  =&~ \frac{\langle 1234\rangle +\langle 1235\rangle +\langle 1245\rangle +\langle 1345 \rangle +\langle 2345 \rangle}{V_5(\br_1,\br_2,\br_3, \br_4,\br_5)}\\
 &~\prod_{1\leq a<b\leq 5}\theta_{ab}
\end{split}
\end{equation*}

The generating function \eqref{P:cont-2} contains all desorption probabilities $p_0,\ldots,p_5$. For instance,
\begin{equation}
p_5 = \frac{1}{5!}\! \int d\br_1 d\br_2 d\br_3\, d\br_4\,  d\br_5\,   \langle 1 2 3 4  5\rangle
\end{equation}

\section{Concluding remarks}
\label{sec:remarks}

We analyzed the dynamic space packing (DSP), a random process with sequemtial deposition of identical objects and the removal of nearby objects after each deposition event. In the lattice DSP, particles land on single sites of $\mathbb{Z}^d$, and after each deposition event, particles occupying $2d$ neighboring sites leave the lattice. We also studied the DSP of balls into $\mathbb{R}^d$ in which balls overlapping with the newly added ball are removed. These two DSP processes are exactly solvable, particularly the steady state characteristics admit an analytical description as we showed in Sects.~\ref{sec:lattice}--\ref{sec:cont}. 

Amongst the remaining challenges, we mention the determination of the extremal probabilities in $d\geq 2$ dimensions. Conjecturally, they exhibit the asymptotic behaviors \eqref{EM-d}. For continuous DSP processes, the computations tend to be much more challenging. For instance, we have determined explicit expressions for desorption probabilities only in one dimension. The Mandel $Q$ parameter characterizing fluctuations of the occupation number of a large domain remains unknown for continuous DSP processes; for the DSP processes on $\mathbb{Z}^d$, we computed $Q$ in all dimensions, see \eqref{Q-d}.

The DSPs with amended rules also tend to be exactly solvable. In our lattice DSP after a deposition event, we remove particles from its von Neumann neighborhood containing $2d$ adjacent sites. Another natural lattice DSP postulates that particles from the Moore neighborhood, i.e., a $3\times \ldots\times 3$ hypercube centered around the site, are removed. The steady-state density is 
\begin{equation}
\label{DSP:Moore}
\rho_\text{dsp}=3^{-d}
\end{equation}

The lower and upper bounds for the densest sphere packings are exponentially separated, $2^{-d}$ and $2^{-0.5990 d}$, see  \cite{Minkowski05,Blichfeldt29,Rogers,KL78,Conway}. Thus, the celebrated Minkowski's lower bound for densest sphere packings coincides with the steady-state volume fraction for the continuous DSP. 

Let us compare the DSP with another random algorithm known as random sequential adsorption (RSA), where deposition attempts leading to the overlap with already present balls are discarded, see \cite{Evans93,Talbot00,KRB,Tor} and references therein. The system reaches a jammed state that is a saturated sphere packing (since it is impossible to add a sphere without overlap). The volume fraction of any saturated sphere packing exceeds $2^{-d}$, so 
\begin{equation}
\rho_\text{rsa} > \rho_\text{dsp} = 2^{-d}
\end{equation}

Consider the RSA on $\mathbb{Z}^d$ where a deposition event into an empty site is successful only if the von Neumann neighborhood of this site is empty \cite{PK20}. 
For the corresponding DSP, the steady state density is known in arbitrary dimension, $\rho_\text{dsp}=(2d+1)^{-1}$, while RSA with nearest-neighbor exclusion, the jamming density is known only in one dimension \cite{Flory39}: 
\begin{equation}
\label{RSA:2}
\rho_\text{rsa}  = \frac{1-e^{-2}}{2}
\end{equation}
Generally for saturated packings of $\mathbb{Z}^d$  we have
\begin{equation}
\label{sat-d}
\frac{1}{2d+1}\leq \rho\leq \frac{1}{2}
\end{equation}
The lower bound is understood by noting that the von Neumann neighborhood of each occupied site is empty. The upper bound follows from the bipartite nature of the hyper-cubic lattice: $\mathbb{Z}^d= \mathbb{Z}^d_\text{even}\cup \mathbb{Z}^d_\text{odd}$ with sub-lattices \eqref{Z+-}. In a densest packing, one of the two sub-lattices is fully occupied while another is empty. These packings give the upper bound in \eqref{sat-d}. 

The jamming density of the RSA lies strictly between the bounds \eqref{sat-d}. In one dimension, $\rho_\text{rsa}$ is closer to the upper bound; for $d\gg 1$, the jamming density is much closer to the lower bound:
\begin{equation}
\label{RSA:inf}
\rho_\text{rsa} \simeq \frac{\ln(2d+1)}{2d+1}
\end{equation}
when $d\gg 1$. Thus the lattice RSA algorithm yields only slightly denser packings than the DSP algorithm. 

The asymptotic \eqref{RSA:inf} has been established in \cite{Baram89,Baram21}. To appreciate \eqref{RSA:inf} one can rely on the exact jamming density on the infinite $q-$regular tree \cite{Pippenger,Percus91}
\begin{equation}
\label{RSA:q}
\rho_\text{rsa} =\frac{1}{2}\left[1-(q-1)^{-\frac{2}{q-2}}\right]
\end{equation}
The one-dimensional lattice is the infinite $2-$regular tree. In the $q\downarrow 2$ limit, we recover the one-dimensional jamming density \eqref{RSA:2} from \eqref{RSA:q}. 

The infinite graph $\mathbb{Z}^d$ is $(2d)-$regular, but it is not a tree when $d>1$. In the $d\to\infty$ limit, however, $\mathbb{Z}^d$ is tree-like. From \eqref{RSA:q} with $q=2d$, we deduce \eqref{RSA:inf} in the leading order. 

Consider the RSA on $\mathbb{Z}^d$ where a deposition event into an empty site is successful only if the Moore neighborhood of this site is empty. 
We have $\rho_\text{rsa}>\rho_\text{dsp}=3^{-d}$ since the Moore neighborhood of each occupied site is empty. To determine $\rho_\text{rsa}$ when $d\gg 1$, we use \eqref{RSA:q} with $q=3^d-1$ and obtain an asymptotic
\begin{equation}
\label{RSA:inf-Moore}
\rho_\text{rsa} \simeq d\,\frac{\ln 3}{3^d}
\end{equation}
Another derivation of \eqref{RSA:inf-Moore} is given in \cite{Baram21}. 

Thus for the lattice RSA and DSP models based on the Moore neighborhood, Eqs.~\eqref{DSP:Moore} and \eqref{RSA:inf-Moore} show that $\rho_\text{dsp}$ and $\rho_\text{rsa}$ exhibit the same exponential decay and differ only by linear in $d$ factor. In lattice RSA and DSP models based on the von Neumann neighborhood, the stationary densities exhibit the same algebraic decay and differ only by $\ln(d)$ factor. 

The decay laws for the densities $\rho_\text{rsa}$ and $\rho_\text{dsp}$ in the lattice examples suggest that in the continuous version the densities $\rho_\text{rsa}$ and $\rho_\text{dsp}$ also decay in a similar way. Noting that the asymptotic decay laws \eqref{RSA:inf} and \eqref{RSA:inf-Moore} for the lattice RSA models can be uniformly written as 
\begin{equation}
\label{RSA-DSP}
\rho_\text{rsa} \simeq \rho_\text{dsp}\ln(1/\rho_\text{dsp})
\end{equation}
and assuming that \eqref{RSA-DSP} is also applicable to the continuous RSA, we arrive at the conjectural decay law
\begin{equation}
\label{RSA:inf-cont}
\rho_\text{rsa} \simeq d\,\frac{\ln 2}{2^d}
\end{equation}

Simulations of the continuous RSA in up to eight dimensions support the $2^{-d}$ decay. The best fit of low-dimensional data is \cite{Torquato13}
\begin{equation}
\label{RSA-1}
\rho_\text{rsa} = \frac{a_0 + a_1 d + a_2 d^2}{2^d}
\end{equation}
with certain positive parameters $a_0, a_1, a_2$. It is unclear how accurate is \eqref{RSA-1} for $d\geq 9$. Other theoretical considerations \cite{Torquato10} suggest 
\begin{equation}
\label{RSA-2}
\rho_\text{rsa} = \frac{b_0 + b_1 d + b_2 d \ln d}{2^d}
\end{equation}

Deriving the decay law for the jammed density $\rho_\text{rsa}$ for the continuous RSA of spheres in high dimensions is an intriguing challenge.

\bigskip
\noindent{\bf Acknowledgments.}
We benefitted from discussions with D. Dhar, J.-M. Luck, K. Mallick, S. Redner and P. Urbani. We are thankful to O. Adelman, H. Hilhorst, D. S\'{e}nizergues and N. Smith for correspondence. PLK is grateful to the IPhT for hospitality and excellent working conditions. 

\appendix
\section{Algorithm for computing desorption probabilities in arbitrary dimension}
\label{ap:algorithm}

Using the recurrence \eqref{CSk}, one can compute all correlation functions $C(S_k)$ and hence determine the generating function \eqref{GFP} encapsulating the desorption probabilities $p_k$. Specifically, we employ the following algorithm
\begin{enumerate}
\item ${\bf 0}$ is taken as a central site. The neighbors differ from ${\bf 0}$ in one coordinate that is $\pm 1$ instead of $0$. 
\item The Mathematica function `Subsets' is used to construct all subsets of the set of neighboring sites.
\item Starting from $k=1$, the numbers $C(S_k)$ are recurrently calculated for all $S_k$.  The volume $V[S_k]$ is calculated 
    by enumerating the set of nearest neighbors of $S_k$.  
\item Performing the series expansion of the generating function $\mathcal{P}(x)$ given by \eqref{GFP} around $x=0$, 
    and comparing with the definition \eqref{GFP:def}, we extract the probabilities $p_k(d)$. 
\end{enumerate}

By definition, all $V[S_k]$ are rational numbers. The correlation functions $C(S_k)$ are also rational numbers---this is easy to prove using \eqref{CSk} and induction.  Hence the desorption probabilities $p_k(d)$ are also rational. The nominators and denominators of $p_k(d)$ rapidly grow with $d$. For instance, from Table~\ref{Table:345} we see that e.g. $p_0(5)$ is the ratio of integers exceeding $10^{21}$. The probability $p_0(10)$ is the ratio of integers exceeding $10^{103}$. 

Using Mathematica, we have computed exact desorption probabilities $p_k(d)$ in dimensions $d\leq 11$. The probabilities $p_1(d)$ appear to decrease when $d$ increases, while $p_0(d)$ is an increasing function of $d$. The convergence of $p_0(d)$ and $p_1(d)$ to $p_0(\infty)$ and $p_1(\infty)$ is slow:
\begin{equation}
\label{p0d:exp}
p_0(d) = p_0(\infty)+Q_{01}\,d^{-1}+Q_{02}\,d^{-2}+O(d^{-3})
\end{equation}
and similarly for $p_1(d)$. To extract a good estimate for $p_0(\infty)$ from a slowly convergent expansion like \eqref{p0d:exp}, we used the Richardson extrapolation \cite{Bender}. For instance, the second order extrapolation 
\begin{equation}
\label{Richardson}
p^{(2)}_0(d)=\tfrac{d^2 p_0(d) - 2(d-1)^2 p_0(d-1)+(d-2)^2 p_0(d-2)}{2}
\end{equation}
eliminates $d^{-1}$ and $d^{-2}$ terms from the series \eqref{p0d:exp}. Using the exact results from Table~\ref{Table:345} we obtain
\begin{equation*}
\label{p0-25}
p^{(2)}_0(5) = \tfrac{2009267067948311818944440303}{5463438864271106652565632000}=0.367\,766\ldots
\end{equation*}
The third order extrapolation additionally eliminates a term of order $d^{-3}$. One finds
\begin{equation*}
\label{p0-35}
p^{(3)}_0(5) = \tfrac{18088703513185851470120549681}{49170949778439959873090688000}=0.367\,873\ldots
\end{equation*}
The Richardson extrapolation usually quickly converges to the limit \cite{Bender}. The exact value  
\begin{equation}
\label{p0:inf}
p_0(\infty) = e^{-1} = 0.367\,879\ldots
\end{equation}
is an excellent agreement with above approximations extracted from exact results in $d\leq 5$. Applying similar procedure to $p_1(d)$ also gives approximations close to the exact value $p_1(\infty) = e^{-1}$. 

\section{Proof of \eqref{2:far}}
\label{ap:induction}

Here, we prove \eqref{2:far} which is the starting point of many subsequent results. Applying \eqref{CSk} to the union $S_1 \cup S_2$  of two arbitrary sets we obtain
\begin{equation}
\label{CS12-gen}
V[S_1 \cup S_2] C(S_1 \cup S_2) = \sum_{j \in S_1 \cup S_2} C(S_1 \cup S_2 - \{j\})
\end{equation}
In the case of two arbitrary well-separated sets, the neighborhood of the union is the sum of the individual neighborhoods:  $V[S_1 \cup S_2]=V[S_1]+V[S_2]$. Using this result and simplifying the right-hand side of \eqref{CS12-gen}, we get
\begin{eqnarray}
\label{CS12}
(V[S_1]+V[S_2])C(S_1 \cup S_2) &= & \sum_{j \in S_1} C(S_1^{(j)} \cup S_2)\nonumber \\
&+& \sum_{j \in S_2} C(S_1 \cup S_2^{(j)})
\end{eqnarray}
where $S_a^{(j)}=S_a-\{j\}$. We now employ induction in the size of sets. Namely we assume that for well-separated sets $S_1'$ and $S_2'$ with $|S_1' \cup S_2'|<|S_1 \cup S_2|$, de-correlation holds: $C(S_1' \cup S_2')=C(S_1') C(S_2')$. Hence the right-hand side (RHS) of \eqref{CS12} becomes
\begin{eqnarray}
\label{CS12-de}
\text{RHS}&= & C(S_2)\sum_{j \in S_1} C(S_1^{(j)})+ C(S_1) \sum_{j \in S_2} C(S_2^{(j)}) \nonumber\\
&= & (V[S_1]+V[S_2])\,C(S_1)C(S_2)
\end{eqnarray}
where in the second step we used 
\begin{equation*}
V[S_a] C(S_a) = \sum_{j \in S_a} C(S_a^{(j)})
\end{equation*}
implied by \eqref{CSk}. Comparing the left-hand side of \eqref{CS12} with the right-hand side which we reduced to \eqref{CS12-de} we arrive at the announced result \eqref{2:far}.

\section{Empty interval distribution}
\label{ap:empty}

Using the generating function \eqref{GF} we have recasted an infinite set of ordinary differential equations \eqref{Ek:eq} into a single partial differential equation \eqref{E:PDE} which is an inhomogeneous linear first order partial differential equation. To solve \eqref{E:PDE} we introduce auxiliary variables 
\begin{equation}
\label{uv:def}
u = \tfrac{1}{2}(t+z), \qquad v = \tfrac{1}{2}(t-z)
\end{equation}
allowing us to re-write \eqref{E:PDE} as
\begin{equation}
\label{E:uv}
\partial_u\mathcal{E} = 2 e^{u-v}\left[\mathcal{E}+ e^{2(u-v)}\right]
\end{equation}
Since $e^{2e^{u-v}}$ is the solution of the homogeneous version of Eq.~\eqref{E:uv}, we use the ansatz
\begin{equation}
\label{EF}
\mathcal{E} = \mathcal{F}\,e^{2e^{u-v}}
\end{equation}
and transform \eqref{E:uv} into  $\partial_u  \mathcal{F} = 2\, e^{3(u-v)-2\, e^{u-v}}$ which is integrated to yield
\begin{equation}
\label{F:uv}
\mathcal{F}=2\int_{-2v}^{u-v}dw\,\exp\!\big[3w-2e^w\big]+G(v)
\end{equation}
The integral can be computed, so we just need to fix an additive `constant', the function $G(v)$ in the present case. To this end we notice that $u=-v$ when $t=0$, and hence $\mathcal{F}=G(v)$. On the other hand, $E_k(0)=1$ and hence the definition \eqref{GF} gives $\mathcal{E}(0,z) = e^{3z}/(1-e^z)$. Plugging  this into \eqref{EF} and using $z=-2v$ and $\mathcal{F}=G(v)$ at $t=0$ gives
\begin{equation}
\label{G}
G(v) = \frac{V^3}{1-V}\, e^{-2V}, \qquad V=e^{-2v}
\end{equation}
Computing the integral in Eq.~\eqref{F:uv} and using \eqref{G} we obtain
\begin{equation}
\label{E:ZV}
\mathcal{E} =\left( \frac{V}{1-V} + \frac{1}{2}\right)e^{2Z-2V} - Z^2  - Z - \frac{1}{2}
\end{equation}
where $Z=e^{u-v}=e^z$. Using $V=e^{-2v}=Z  e^{-t}$ we re-write \eqref{E:ZV} as
\begin{eqnarray}
\label{E:Zt}
\mathcal{E} =\left(\frac{e^{-t} Z}{1-e^{-t} Z} + \frac{1}{2}\right)e^{2Z(1-e^{-t})} - Z^2  - Z - \tfrac{1}{2}
\end{eqnarray}
Expressing the generating function $\mathcal{E}(t, z)$ defined by \eqref{GF} through $t$ and $Z$ gives 
\begin{equation}
\label{EZ:def}
\mathcal{E}(t,Z)= \sum_{k\geq 1} E_k(t)\,Z^{k+2}
\end{equation}
Expanding \eqref{E:Zt} and comparing with \eqref{EZ:def} we arrive at the announced result \eqref{Ekt}.

\section{Statistics of maxima in permutations}
\label{ap:perm}

Here, we show a connection between the DSP on an arbitrary graph and the statistics of permutation maxima on the same graph: The probability of a configuration of particles in the DSP on a graph with $N$ vertices is the number of permutations on the graph, which have local maxima at the same places as the particles, divided by $N!$. Equivalently, one can consider a random surface on a graph with height assigned to each vertex. Assuming that heights are independent identically distributed random variables, a configuration of particles in the peaks on this graph gives a steady state in the DSP. More precisely, the statistics of the steady states coincides with statistics of the peaks with respect to the distribution of the heights. 

We emphasize that the isomorphism is between the steady-state characteristics. For instance, the cumulant generating function describing the number of particles for the DSP process on the one-dimensional lattice (Secs.~\ref{subsec:pnk}--\ref{subsec:cumulants}) is the same as the cumulant generating function describing the number of weak bonds in the random-bond Ising model \cite{Derrida-SG86}. The Glauber dynamics natural for the zero-temperature spin glass \cite{PLK91} differs from the dynamics of the DSP. Therefore, the one-dimensional zero-temperature Ising spin glass evolves differently than the DSP. 

The statistics of maxima of permutations is a rather well-studied subject \cite{Entringer66,Entringer69,perm-carlitz,perm-fewster,perm-Ma}. A similar problem is the enumeration of peaks in uncorrelated landscapes \cite{peaks-Billera,Majumdar06,perm-Oshanin,peaks-hivert,peaks-carmi,peaks-Dani}. This problem is also popular in biology, where it is known as a house of cards model \cite{Kingman78,Kauffman87,Gavrilets,Krug12}. 

For concreteness, we outline the connection between the DSP and permutations. Consider an arbitrary graph with $N$ vertices and an ensemble of permutations of the integers $\{1,2,\dots,N\}$. Thus, an integer from $1$ to $N$ is assigned to every vertex, and every permutation of these integers over the graph vertices is equally likely. The total number of permutations is $N!$. We want to study the correlations of maxima on the graph. A vertex on the graph is a local maximum if the integer on the vertex is larger than the integers on all the neighboring vertices. Let $m_i=1$ if the integer at $i$ is a local maximum and $m_i=0$ otherwise.

To study the statistics of maxima on a finite set of vertices, we can consider the set of permutations on the subgraph defined by these vertices and their neighbors. Let $\mathcal{V}_{i}$ be the neighborhood of vertex $i$, i.e., the vertex $i$ and its neighbors. It suffices to consider permutations of the integers $\{1,2,\dots,V_i\}$, where $V_i=|\mathcal{V}_{i}|$. Let $\mathcal{P}(V_i)$ be the total number of permutations and $\mathcal{P}(V_i|m_i=1)$ the number of permutations conditioned on the fact that the vertex $i$ is a local maximum. Averaging over the ensemble of all permutations one obtains 
\begin{equation}
\langle m_i \rangle = \frac{\mathcal{P}(V_i|m_i=1)}{\mathcal{P}(V_i)} = \frac{1}{V_i}
\end{equation}

For any two vertices $i$ and $j$ we denote again by $\mathcal{V}_{i,j}$ their neighborhood, and by  $V_{i,j}=|\mathcal{V}_{i,j}|$ the size of the neighborhood. We want to count the number of permutations such that $i$ and $j$ are both local maxima. Since the maximal element is $V_{i,j}$, it has to be a local maximum and hence has to be on either vertex $i$ or $j$.

If the maximal element $V_{i,j}$ is in $j$, other $\mathcal{V}_{i,j} - \{j\}$ vertices contain a permutation of $\{1,2,\dots,V_{i,j}-1\}$. When restricted to the neighbors of site $i$, this has to be a permutation with a maximum at $i$, and the $V_i$ elements of this permutation are be chosen from the total $V_{i,j}-1$ elements. The other $V_{i,j}-1-V_i$ elements are free to permute amongst themselves. The total number of such permutations is 
\begin{eqnarray}
{{V_{i,j}-1}\choose{V_i}} \times \mathcal{P}(V_i|m_i=1) \times (V_{i,j}-V_i-1)! \nonumber\\
= (V_{i,j}-1)! \langle m_i \rangle \nonumber
\end{eqnarray}
Similarly, the number of permutations where the maximal element $V_{i,j}$ is on site $i$ is $(V_{i,j}-1)! \langle m_j \rangle$. Adding the two contributions yields the pair correlation function
\begin{eqnarray}
\langle m_i m_j \rangle &=& \frac{(V_{i,j}-1)! \langle m_i \rangle + (V_{i,j}-1)! \langle m_j \rangle}{(V_{i,j})!} \nonumber\\
&=& \frac{\langle m_i \rangle + \langle m_j \rangle}{V_{i,j}}
\end{eqnarray}
Similarly one computes 
\begin{eqnarray}
&&\langle m_i m_j m_k \rangle  = \frac{1}{(V_{i,j,k})!}  \bigg[(V_{i,j,k}-1)! \langle m_i m_j \rangle  \nonumber\\
&&+ (V_{i,j,k}-1)! \langle m_j  m_k \rangle + (V_{i,j,k}-1)! \langle m_i m_k \rangle \bigg] \nonumber\\
	&=& \frac{\langle m_i m_j \rangle + \langle m_j m_k \rangle + \langle m_i m_k \rangle}{V_{i,j,k}} 
\end{eqnarray}
and higher-order correlation functions.

For a set $S_k$ of $k$ vertices, we denote by $\mathcal{V}[S_k]$ the neighborhood of $S_k$ and by $V[S_k]=|\mathcal{V}[S_k]|$ the volume of this neighborhood. The correlation function
\begin{equation}
	C_M(S_k) = \bigg\langle \prod_{i \in S_k} m_i \bigg\rangle
\end{equation}
satisfies
\begin{eqnarray}
\label{CMSk}
C_M(S_k) &=& \frac{1}{(V[S_k])!}\left(\sum_{j\in S_k} (V[S_k]-1)!~ C_M(S_k - \{j\})\right) \nonumber\\
	       &=& \frac{1}{V[S_k]}\sum_{j\in S_k} C_M(S_k - \{j\})
\end{eqnarray}
Each term in the top line on the right-hand side corresponds to the number of permutations on $V[S_k]$ vertices with the maximal element of value $V[S_k]$ on vertex $j$. Equation \eqref{CMSk} coincides with the formula for the DSP steady state correlation function $C(S_k)$, see Eq.~\eqref{CSk}, and thereby establishes the announced relation between the steady states in the DSP and permutations.

\section{Derivation of \eqref{v-d}}
\label{ap:var}

Consider the non-vanishing pair correlation functions involving the central site $\mathbf{0}$. Unit vectors along various directions are denoted as $\hat{i}$, $\hat{j}$, etc. We have
\begin{equation}
\label{eq:dcorr1}
\langle \eta_{\mathbf{0}}^2 \rangle = \rho =\frac{1}{2d+1} 
\end{equation}
in the steady state. 
Since two nearest neighbors cannot be occupied simultaneously 
\begin{equation}
\langle \eta_{\mathbf{0}} \eta_{\hat{i}} \rangle = 0 \label{eq:dcorr2}
\end{equation}

There are two types of sites separated by distance two from the central site. (Recall, that for the hyper-cubic lattice we use the norm $|k| = |k_1|+\ldots+|k_d|$.) For `diagonal' sites of the form $\hat{i}+\hat{j}$, we denote the correlation function by $D \equiv \langle \eta_{\mathbf{0}} \eta_{\hat{i} + \hat{j}} \rangle$. This correlation function satisfies an equation \begin{equation*}
\frac{dD}{dt} = -4 d D + 2 \rho
\end{equation*}
generalizing \eqref{D:eq} in two dimensions, from which
\begin{equation}
\label{eq:dcorr3}
D = \frac{\rho}{2d} = \frac{1}{2d(2d+1)} 
\end{equation}
in the steady state. There are also `horizontal' sites of the form $2\hat{i}$ separated by distance two from the central site. The corresponding correlation function $\langle \eta_{\mathbf{0}} \eta_{2\hat{i}} \rangle$ which we denote by $H$ satisfies an equation 
\begin{equation*}
\frac{dH}{dt} = - (4 d + 1)H + 2 \rho
\end{equation*}
generalizing \eqref{H:eq} in two dimensions, from which
\begin{equation}
 \label{eq:dcorr4}
H = \frac{2\rho}{4d+1} =  \frac{2}{(2d+1)(4d+1)}
\end{equation}
in the steady state. 

All other correlation functions involving site $\mathbf{0}$ are with sites that are not reachable in $2$ steps or less, and hence are well-separated from $\mathbf{0}$. For these sites the correlation function with the central site is equal to $\rho^2$. 

To calculate the variance of the total number of particles $k = \sum_{\mathbf{i}} \eta_{\mathbf{i}}$ in a large subsystem of $N$ sites, we use translational invariance to simplify the expressions. Neglecting boundary terms, we have
\begin{equation}
\label{var:sum}
\langle k^2 \rangle - \langle k \rangle^2 = N \sum_{\mathbf{i}} \left[\langle \eta_{\mathbf{0}} \eta_{\mathbf{i}} \rangle - \rho^2 \right] + o(N)
\end{equation}
In the sum on the right-hand side most terms vanish due to de-correlation. The exceptions are one term involving the site itself, $2d$ terms involving the nearest neighbors, $4 {{d}\choose{2}}$ terms of type $D$ and $2d$ terms of type $H$. We compute these non-vanishing terms using Eqs.~\eqref{eq:dcorr1}--\eqref{eq:dcorr4}, and arrive at the announced result \eqref{v-d}.

\bibliography{references-pack}

\begin{thebibliography}{122}%
\makeatletter
\providecommand \@ifxundefined [1]{%
 \@ifx{#1\undefined}
}%
\providecommand \@ifnum [1]{%
 \ifnum #1\expandafter \@firstoftwo
 \else \expandafter \@secondoftwo
 \fi
}%
\providecommand \@ifx [1]{%
 \ifx #1\expandafter \@firstoftwo
 \else \expandafter \@secondoftwo
 \fi
}%
\providecommand \natexlab [1]{#1}%
\providecommand \enquote  [1]{``#1''}%
\providecommand \bibnamefont  [1]{#1}%
\providecommand \bibfnamefont [1]{#1}%
\providecommand \citenamefont [1]{#1}%
\providecommand \href@noop [0]{\@secondoftwo}%
\providecommand \href [0]{\begingroup \@sanitize@url \@href}%
\providecommand \@href[1]{\@@startlink{#1}\@@href}%
\providecommand \@@href[1]{\endgroup#1\@@endlink}%
\providecommand \@sanitize@url [0]{\catcode `\\12\catcode `\$12\catcode
  `\&12\catcode `\#12\catcode `\^12\catcode `\_12\catcode `\%12\relax}%
\providecommand \@@startlink[1]{}%
\providecommand \@@endlink[0]{}%
\providecommand \url  [0]{\begingroup\@sanitize@url \@url }%
\providecommand \@url [1]{\endgroup\@href {#1}{\urlprefix }}%
\providecommand \urlprefix  [0]{URL }%
\providecommand \Eprint [0]{\href }%
\providecommand \doibase [0]{http://dx.doi.org/}%
\providecommand \selectlanguage [0]{\@gobble}%
\providecommand \bibinfo  [0]{\@secondoftwo}%
\providecommand \bibfield  [0]{\@secondoftwo}%
\providecommand \translation [1]{[#1]}%
\providecommand \BibitemOpen [0]{}%
\providecommand \bibitemStop [0]{}%
\providecommand \bibitemNoStop [0]{.\EOS\space}%
\providecommand \EOS [0]{\spacefactor3000\relax}%
\providecommand \BibitemShut  [1]{\csname bibitem#1\endcsname}%
\let\auto@bib@innerbib\@empty
\bibitem [{\citenamefont {Hales}(2005)}]{Hales05}%
  \BibitemOpen
  \bibfield  {author} {\bibinfo {author} {\bibfnamefont {T.~C.}\ \bibnamefont
  {Hales}},\ }\bibfield  {title} {\enquote {\bibinfo {title} {A proof of the
  {K}epler conjecture},}\ }\href {https://doi.org/10.4007/annals.2005.168.1065}
  {\bibfield  {journal} {\bibinfo  {journal} {Ann. Math.}\ }\textbf {\bibinfo
  {volume} {168}},\ \bibinfo {pages} {1065--1185} (\bibinfo {year}
  {2005})}\BibitemShut {NoStop}%
\bibitem [{\citenamefont {Conway}\ and\ \citenamefont {Sloane}(1999)}]{Conway}%
  \BibitemOpen
  \bibfield  {author} {\bibinfo {author} {\bibfnamefont {J.~H.}\ \bibnamefont
  {Conway}}\ and\ \bibinfo {author} {\bibfnamefont {N.~J.~A}\ \bibnamefont
  {Sloane}},\ }\href {https://doi.org/10.1007/978-1-4757-6568-7} {\emph
  {\bibinfo {title} {Sphere Packings, Lattices and Groups}}}\ (\bibinfo
  {publisher} {Springer-Verlag},\ \bibinfo {address} {New York},\ \bibinfo
  {year} {1999})\BibitemShut {NoStop}%
\bibitem [{\citenamefont {Viazovska}(2017)}]{Maryna17a}%
  \BibitemOpen
  \bibfield  {author} {\bibinfo {author} {\bibfnamefont {M.}~\bibnamefont
  {Viazovska}},\ }\bibfield  {title} {\enquote {\bibinfo {title} {The sphere
  packing problem in dimension 8},}\ }\href
  {https://doi.org/10.4007/annals.2017.185.3.7} {\bibfield  {journal} {\bibinfo
   {journal} {Ann. of Math.}\ }\textbf {\bibinfo {volume} {185}},\ \bibinfo
  {pages} {991--1015} (\bibinfo {year} {2017})}\BibitemShut {NoStop}%
\bibitem [{\citenamefont {Cohn}\ \emph {et~al.}(2017)\citenamefont {Cohn},
  \citenamefont {Kumar}, \citenamefont {Miller}, \citenamefont {Radchenko},\
  and\ \citenamefont {Viazovska}}]{Maryna17b}%
  \BibitemOpen
  \bibfield  {author} {\bibinfo {author} {\bibfnamefont {H.}~\bibnamefont
  {Cohn}}, \bibinfo {author} {\bibfnamefont {A.}~\bibnamefont {Kumar}},
  \bibinfo {author} {\bibfnamefont {S.}~\bibnamefont {Miller}}, \bibinfo
  {author} {\bibfnamefont {D.}~\bibnamefont {Radchenko}}, \ and\ \bibinfo
  {author} {\bibfnamefont {M.}~\bibnamefont {Viazovska}},\ }\bibfield  {title}
  {\enquote {\bibinfo {title} {The sphere packing problem in dimension 24},}\
  }\href {https://doi.org/10.4007/annals.2017.185.3.8} {\bibfield  {journal}
  {\bibinfo  {journal} {Ann. of Math.}\ }\textbf {\bibinfo {volume} {185}},\
  \bibinfo {pages} {1017--1033} (\bibinfo {year} {2017})}\BibitemShut {NoStop}%
\bibitem [{\citenamefont {Minkowski}(1905)}]{Minkowski05}%
  \BibitemOpen
  \bibfield  {author} {\bibinfo {author} {\bibfnamefont {H.}~\bibnamefont
  {Minkowski}},\ }\bibfield  {title} {\enquote {\bibinfo {title}
  {Diskontinuit\"{a}tsbereich f\"{u}r arithmetische \"{A}quivalenz},}\
  }\href@noop {} {\bibfield  {journal} {\bibinfo  {journal} {J. Reine Angew.
  Math.}\ }\textbf {\bibinfo {volume} {129}},\ \bibinfo {pages} {220--274}
  (\bibinfo {year} {1905})}\BibitemShut {NoStop}%
\bibitem [{\citenamefont {Blichfeldt}(1929)}]{Blichfeldt29}%
  \BibitemOpen
  \bibfield  {author} {\bibinfo {author} {\bibfnamefont {H.~F.}\ \bibnamefont
  {Blichfeldt}},\ }\bibfield  {title} {\enquote {\bibinfo {title} {The minimum
  value of quadratic forms, and the closest packing of spheres},}\ }\href
  {https://doi.org/10.1007/BF01454863} {\bibfield  {journal} {\bibinfo
  {journal} {Math. Ann.}\ }\textbf {\bibinfo {volume} {101}},\ \bibinfo {pages}
  {605--608} (\bibinfo {year} {1929})}\BibitemShut {NoStop}%
\bibitem [{\citenamefont {Rogers}(1964)}]{Rogers}%
  \BibitemOpen
  \bibfield  {author} {\bibinfo {author} {\bibfnamefont {C.~A.}\ \bibnamefont
  {Rogers}},\ }\href {\doibase 10.2307/3614728} {\emph {\bibinfo {title}
  {Packing and Covering}}}\ (\bibinfo  {publisher} {Cambridge University
  Press},\ \bibinfo {address} {Cambridge, UK},\ \bibinfo {year}
  {1964})\BibitemShut {NoStop}%
\bibitem [{\citenamefont {Kabatyansky}\ and\ \citenamefont
  {Levenshtein}(1978)}]{KL78}%
  \BibitemOpen
  \bibfield  {author} {\bibinfo {author} {\bibfnamefont {G.~A.}\ \bibnamefont
  {Kabatyansky}}\ and\ \bibinfo {author} {\bibfnamefont {V.~I.}\ \bibnamefont
  {Levenshtein}},\ }\bibfield  {title} {\enquote {\bibinfo {title} {On bounds
  for packing on a sphere and in space},}\ }\href
  {http://mi.mathnet.ru/eng/ppi/v14/i1/p3} {\bibfield  {journal} {\bibinfo
  {journal} {Probl. Inf. Transmission}\ }\textbf {\bibinfo {volume} {14}},\
  \bibinfo {pages} {1--17} (\bibinfo {year} {1978})}\BibitemShut {NoStop}%
\bibitem [{\citenamefont {Cohn}\ and\ \citenamefont {Elkies}(2003)}]{Cohn03}%
  \BibitemOpen
  \bibfield  {author} {\bibinfo {author} {\bibfnamefont {H.}~\bibnamefont
  {Cohn}}\ and\ \bibinfo {author} {\bibfnamefont {N.}~\bibnamefont {Elkies}},\
  }\bibfield  {title} {\enquote {\bibinfo {title} {New upper bounds on sphere
  packings {I}},}\ }\href {https://doi.org/10.4007/annals.2003.157.689}
  {\bibfield  {journal} {\bibinfo  {journal} {Ann. Math.}\ }\textbf {\bibinfo
  {volume} {157}},\ \bibinfo {pages} {689--714} (\bibinfo {year}
  {2003})}\BibitemShut {NoStop}%
\bibitem [{\citenamefont {Cohn}(2002)}]{Cohn02}%
  \BibitemOpen
  \bibfield  {author} {\bibinfo {author} {\bibfnamefont {H.}~\bibnamefont
  {Cohn}},\ }\bibfield  {title} {\enquote {\bibinfo {title} {New upper bounds
  on sphere packings {II}},}\ }\href
  {https://doi.org/10.4007/annals.2003.157.689} {\bibfield  {journal} {\bibinfo
   {journal} {Geom. Topol.}\ }\textbf {\bibinfo {volume} {6}},\ \bibinfo
  {pages} {329--353} (\bibinfo {year} {2002})}\BibitemShut {NoStop}%
\bibitem [{\citenamefont {Krivelevich}\ \emph {et~al.}(2004)\citenamefont
  {Krivelevich}, \citenamefont {Litsyn},\ and\ \citenamefont {Vardy}}]{Vardy}%
  \BibitemOpen
  \bibfield  {author} {\bibinfo {author} {\bibfnamefont {M.}~\bibnamefont
  {Krivelevich}}, \bibinfo {author} {\bibfnamefont {S.}~\bibnamefont {Litsyn}},
  \ and\ \bibinfo {author} {\bibfnamefont {A.}~\bibnamefont {Vardy}},\
  }\bibfield  {title} {\enquote {\bibinfo {title} {A lower bound on the density
  of sphere packings via graph theory},}\ }\href {\doibase
  10.1155/S1073792804140464} {\bibfield  {journal} {\bibinfo  {journal} {Int.
  Math. Res. Not.}\ }\textbf {\bibinfo {volume} {2004}},\ \bibinfo {pages}
  {2271--2279} (\bibinfo {year} {2004})}\BibitemShut {NoStop}%
\bibitem [{\citenamefont {Vance}(2011)}]{Vance11}%
  \BibitemOpen
  \bibfield  {author} {\bibinfo {author} {\bibfnamefont {S.}~\bibnamefont
  {Vance}},\ }\bibfield  {title} {\enquote {\bibinfo {title} {Improved sphere
  packing lower bounds from {H}urwitz lattices},}\ }\href {\doibase
  https://doi.org/10.1016/j.aim.2011.04.016} {\bibfield  {journal} {\bibinfo
  {journal} {Adv. Math.}\ }\textbf {\bibinfo {volume} {227}},\ \bibinfo {pages}
  {2144--2156} (\bibinfo {year} {2011})}\BibitemShut {NoStop}%
\bibitem [{\citenamefont {Venkatesh}(2013)}]{Venkatesh}%
  \BibitemOpen
  \bibfield  {author} {\bibinfo {author} {\bibfnamefont {A.}~\bibnamefont
  {Venkatesh}},\ }\bibfield  {title} {\enquote {\bibinfo {title} {A note on
  sphere packings in high dimension},}\ }\href
  {https://doi.org/10.1093/imrn/rns096} {\bibfield  {journal} {\bibinfo
  {journal} {Int. Math. Res. Not.}\ }\textbf {\bibinfo {volume} {2013}},\
  \bibinfo {pages} {1628--1642} (\bibinfo {year} {2013})}\BibitemShut {NoStop}%
\bibitem [{\citenamefont {Parisi}(2008)}]{Parisi08}%
  \BibitemOpen
  \bibfield  {author} {\bibinfo {author} {\bibfnamefont {G.}~\bibnamefont
  {Parisi}},\ }\bibfield  {title} {\enquote {\bibinfo {title} {On the most
  compact regular lattices in large dimensions: A statistical mechanical
  approach},}\ }\href {https://doi.org/10.1007/s10955-008-9539-6} {\bibfield
  {journal} {\bibinfo  {journal} {J. Stat. Phys.}\ }\textbf {\bibinfo {volume}
  {132}},\ \bibinfo {pages} {207--234} (\bibinfo {year} {2008})}\BibitemShut
  {NoStop}%
\bibitem [{\citenamefont {Cohn}\ and\ \citenamefont {Zhao}(2014)}]{Cohn14}%
  \BibitemOpen
  \bibfield  {author} {\bibinfo {author} {\bibfnamefont {H.}~\bibnamefont
  {Cohn}}\ and\ \bibinfo {author} {\bibfnamefont {Y.}~\bibnamefont {Zhao}},\
  }\bibfield  {title} {\enquote {\bibinfo {title} {Packing, coding, and ground
  states},}\ }\href {https://doi.org/10.1215/00127094-2738857} {\bibfield
  {journal} {\bibinfo  {journal} {Duke Math. J.}\ }\textbf {\bibinfo {volume}
  {163}},\ \bibinfo {pages} {1965--2002} (\bibinfo {year} {2014})}\BibitemShut
  {NoStop}%
\bibitem [{\citenamefont {Cohn}(2017{\natexlab{a}})}]{Cohn17}%
  \BibitemOpen
  \bibfield  {author} {\bibinfo {author} {\bibfnamefont {H.}~\bibnamefont
  {Cohn}},\ }\bibfield  {title} {\enquote {\bibinfo {title} {A conceptual
  breakthrough in sphere packing},}\ }\href {https://doi.org/10.1090/noti1474}
  {\bibfield  {journal} {\bibinfo  {journal} {Notices Amer. Math. Soc.}\
  }\textbf {\bibinfo {volume} {64}},\ \bibinfo {pages} {102--115} (\bibinfo
  {year} {2017}{\natexlab{a}})}\BibitemShut {NoStop}%
\bibitem [{\citenamefont {Cohn}(2017{\natexlab{b}})}]{Cohn16}%
  \BibitemOpen
  \bibfield  {author} {\bibinfo {author} {\bibfnamefont {H.}~\bibnamefont
  {Cohn}},\ }\bibfield  {title} {\enquote {\bibinfo {title} {Packing, coding,
  and ground states},}\ }in\ \href {https://bookstore.ams.org/pcms-23} {\emph
  {\bibinfo {booktitle} {Mathematics and materials}}},\ \bibinfo {editor}
  {edited by\ \bibinfo {editor} {\bibfnamefont {M.~J.}\ \bibnamefont {Bowick}},
  \bibinfo {editor} {\bibfnamefont {D.}~\bibnamefont {Kinderlehrer}}, \bibinfo
  {editor} {\bibfnamefont {G.}~\bibnamefont {Menon}}, \ and\ \bibinfo {editor}
  {\bibfnamefont {C.}~\bibnamefont {Radin}}}\ (\bibinfo  {publisher} {AMS},\
  \bibinfo {address} {Providence, RI},\ \bibinfo {year} {2017})\BibitemShut
  {NoStop}%
\bibitem [{\citenamefont {Jenssen}\ \emph {et~al.}(2019)\citenamefont
  {Jenssen}, \citenamefont {Joos},\ and\ \citenamefont {Perkins}}]{Perkins}%
  \BibitemOpen
  \bibfield  {author} {\bibinfo {author} {\bibfnamefont {M.}~\bibnamefont
  {Jenssen}}, \bibinfo {author} {\bibfnamefont {F.}~\bibnamefont {Joos}}, \
  and\ \bibinfo {author} {\bibfnamefont {W.}~\bibnamefont {Perkins}},\
  }\bibfield  {title} {\enquote {\bibinfo {title} {On the hard sphere model and
  sphere packings in high dimensions},}\ }\href {\doibase 10.1017/fms.2018.25}
  {\bibfield  {journal} {\bibinfo  {journal} {Forum Math. Sigma}\ }\textbf
  {\bibinfo {volume} {7}},\ \bibinfo {pages} {e1} (\bibinfo {year}
  {2019})}\BibitemShut {NoStop}%
\bibitem [{\citenamefont {Cohn}\ and\ \citenamefont {Salmon}(2022)}]{Cohn22a}%
  \BibitemOpen
  \bibfield  {author} {\bibinfo {author} {\bibfnamefont {H.}~\bibnamefont
  {Cohn}}\ and\ \bibinfo {author} {\bibfnamefont {A.}~\bibnamefont {Salmon}},\
  }\bibfield  {title} {\enquote {\bibinfo {title} {Sphere packing bounds via
  rescaling},}\ }\href {https://arxiv.org/abs/2108.10936} {\bibfield  {journal}
  {\bibinfo  {journal} {arXiv:2108.10936}\ } (\bibinfo {year}
  {2022})}\BibitemShut {NoStop}%
\bibitem [{\citenamefont {Cohn}\ \emph {et~al.}(2022)\citenamefont {Cohn},
  \citenamefont {de~Laat},\ and\ \citenamefont {Salmon}}]{Cohn23}%
  \BibitemOpen
  \bibfield  {author} {\bibinfo {author} {\bibfnamefont {H.}~\bibnamefont
  {Cohn}}, \bibinfo {author} {\bibfnamefont {D.}~\bibnamefont {de~Laat}}, \
  and\ \bibinfo {author} {\bibfnamefont {A.}~\bibnamefont {Salmon}},\
  }\bibfield  {title} {\enquote {\bibinfo {title} {Three-point bounds for
  sphere packing},}\ }\href {https://arxiv.org/abs/2206.15373} {\bibfield
  {journal} {\bibinfo  {journal} {arXiv:2206.15373}\ } (\bibinfo {year}
  {2022})}\BibitemShut {NoStop}%
\bibitem [{\citenamefont {de~Courcy-Ireland}\ \emph {et~al.}(2022)\citenamefont
  {de~Courcy-Ireland}, \citenamefont {Dostert},\ and\ \citenamefont
  {Viazovska}}]{Courcy23}%
  \BibitemOpen
  \bibfield  {author} {\bibinfo {author} {\bibfnamefont {M.}~\bibnamefont
  {de~Courcy-Ireland}}, \bibinfo {author} {\bibfnamefont {M.}~\bibnamefont
  {Dostert}}, \ and\ \bibinfo {author} {\bibfnamefont {M.}~\bibnamefont
  {Viazovska}},\ }\bibfield  {title} {\enquote {\bibinfo {title}
  {Six-dimensional sphere packing and linear programming},}\ }\href
  {https://arxiv.org/abs/2211.09044} {\bibfield  {journal} {\bibinfo  {journal}
  {arXiv:2211.09044}\ } (\bibinfo {year} {2022})}\BibitemShut {NoStop}%
\bibitem [{\citenamefont {Elser}(2023)}]{Elser23}%
  \BibitemOpen
  \bibfield  {author} {\bibinfo {author} {\bibfnamefont {V.}~\bibnamefont
  {Elser}},\ }\bibfield  {title} {\enquote {\bibinfo {title} {How densely can
  spheres be packed with moderate effort in high dimensions?}}\ }\href
  {https://arxiv.org/abs/2305.13492} {\bibfield  {journal} {\bibinfo  {journal}
  {arXiv:2305.13492}\ } (\bibinfo {year} {2023})}\BibitemShut {NoStop}%
\bibitem [{\citenamefont {Shannon}(1948{\natexlab{a}})}]{Shannon48a}%
  \BibitemOpen
  \bibfield  {author} {\bibinfo {author} {\bibfnamefont {C.~E.}\ \bibnamefont
  {Shannon}},\ }\bibfield  {title} {\enquote {\bibinfo {title} {A mathematical
  theory of communication},}\ }\href {\doibase
  10.1002/j.1538-7305.1948.tb01338.x} {\bibfield  {journal} {\bibinfo
  {journal} {Bell Syst. Tech. J.}\ }\textbf {\bibinfo {volume} {27}},\ \bibinfo
  {pages} {379--423} (\bibinfo {year} {1948}{\natexlab{a}})}\BibitemShut
  {NoStop}%
\bibitem [{\citenamefont {Shannon}(1948{\natexlab{b}})}]{Shannon48b}%
  \BibitemOpen
  \bibfield  {author} {\bibinfo {author} {\bibfnamefont {C.~E.}\ \bibnamefont
  {Shannon}},\ }\bibfield  {title} {\enquote {\bibinfo {title} {A mathematical
  theory of communication},}\ }\href {\doibase
  10.1002/j.1538-7305.1948.tb00917.x} {\bibfield  {journal} {\bibinfo
  {journal} {Bell Syst. Tech. J.}\ }\textbf {\bibinfo {volume} {27}},\ \bibinfo
  {pages} {623--656} (\bibinfo {year} {1948}{\natexlab{b}})}\BibitemShut
  {NoStop}%
\bibitem [{\citenamefont {Nemhauser}\ and\ \citenamefont
  {Wolsey}(1988)}]{Optimization}%
  \BibitemOpen
  \bibfield  {author} {\bibinfo {author} {\bibfnamefont {G.}~\bibnamefont
  {Nemhauser}}\ and\ \bibinfo {author} {\bibfnamefont {L.}~\bibnamefont
  {Wolsey}},\ }\href {\doibase 10.1002/9781118627372} {\emph {\bibinfo {title}
  {Integer and Combinatorial Optimization}}}\ (\bibinfo  {publisher} {John
  Wiley \& Sons, Inc.},\ \bibinfo {address} {Chichester},\ \bibinfo {year}
  {1988})\BibitemShut {NoStop}%
\bibitem [{\citenamefont {Belitz}\ and\ \citenamefont
  {Bewley}(2-13)}]{Belitz13}%
  \BibitemOpen
  \bibfield  {author} {\bibinfo {author} {\bibfnamefont {P.}~\bibnamefont
  {Belitz}}\ and\ \bibinfo {author} {\bibfnamefont {T.}~\bibnamefont
  {Bewley}},\ }\bibfield  {title} {\enquote {\bibinfo {title} {New horizons in
  sphere-packing theory, part {II}: {L}attice-based derivative-free
  optimization via global surrogates},}\ }\href
  {https://doi.org/10.1007/s10898-012-9866-7} {\bibfield  {journal} {\bibinfo
  {journal} {J. Glob. Optim.}\ }\textbf {\bibinfo {volume} {56}},\ \bibinfo
  {pages} {61--91} (\bibinfo {year} {2-13})}\BibitemShut {NoStop}%
\bibitem [{\citenamefont {Sch\"{u}tte}\ and\ \citenamefont {van~der
  Warden}(1953)}]{VDW}%
  \BibitemOpen
  \bibfield  {author} {\bibinfo {author} {\bibfnamefont {K.}~\bibnamefont
  {Sch\"{u}tte}}\ and\ \bibinfo {author} {\bibfnamefont {B.~L.}\ \bibnamefont
  {van~der Warden}},\ }\bibfield  {title} {\enquote {\bibinfo {title} {Das
  problem der dreizehn kugeln},}\ }\href {https://doi.org/10.1007/BF01343127}
  {\bibfield  {journal} {\bibinfo  {journal} {Math. Ann.}\ }\textbf {\bibinfo
  {volume} {125}},\ \bibinfo {pages} {325--334} (\bibinfo {year}
  {1953})}\BibitemShut {NoStop}%
\bibitem [{\citenamefont {Leech}(1956)}]{Leech}%
  \BibitemOpen
  \bibfield  {author} {\bibinfo {author} {\bibfnamefont {J.}~\bibnamefont
  {Leech}},\ }\bibfield  {title} {\enquote {\bibinfo {title} {The problem of
  the thirteen spheres},}\ }\href {https://doi.org/10.2307/3610264} {\bibfield
  {journal} {\bibinfo  {journal} {Math. Gazette}\ }\textbf {\bibinfo {volume}
  {41}},\ \bibinfo {pages} {22--23} (\bibinfo {year} {1956})}\BibitemShut
  {NoStop}%
\bibitem [{\citenamefont {Levenshtein}(1979)}]{Levenshtein79}%
  \BibitemOpen
  \bibfield  {author} {\bibinfo {author} {\bibfnamefont {V.~I.}\ \bibnamefont
  {Levenshtein}},\ }\bibfield  {title} {\enquote {\bibinfo {title} {On bounds
  for packing in $n$-dimensional euclidean space},}\ }\href
  {http://mi.mathnet.ru/eng/dan/v245/i6/p1299} {\bibfield  {journal} {\bibinfo
  {journal} {Dokl. Akad. Nauk {SSSR}}\ }\textbf {\bibinfo {volume} {245}},\
  \bibinfo {pages} {1299--1303} (\bibinfo {year} {1979})}\BibitemShut {NoStop}%
\bibitem [{\citenamefont {Odlyzko}\ and\ \citenamefont
  {Sloane}(1979)}]{Odlyzko79}%
  \BibitemOpen
  \bibfield  {author} {\bibinfo {author} {\bibfnamefont {A.~M.}\ \bibnamefont
  {Odlyzko}}\ and\ \bibinfo {author} {\bibfnamefont {N.~J.~A.}\ \bibnamefont
  {Sloane}},\ }\bibfield  {title} {\enquote {\bibinfo {title} {New bounds on
  the number of unit spheres that can touch a unit sphere in $n$ dimensions},}\
  }\href {\doibase https://doi.org/10.1016/0097-3165(79)90074-8} {\bibfield
  {journal} {\bibinfo  {journal} {J. Combin. Theory {A}}\ }\textbf {\bibinfo
  {volume} {26}},\ \bibinfo {pages} {210--214} (\bibinfo {year}
  {1979})}\BibitemShut {NoStop}%
\bibitem [{\citenamefont {Pfender}\ and\ \citenamefont
  {Ziegler}(2004)}]{Ziegler}%
  \BibitemOpen
  \bibfield  {author} {\bibinfo {author} {\bibfnamefont {F.}~\bibnamefont
  {Pfender}}\ and\ \bibinfo {author} {\bibfnamefont {G.~M.}\ \bibnamefont
  {Ziegler}},\ }\bibfield  {title} {\enquote {\bibinfo {title} {Kissing
  numbers, sphere packings, and some unexpected proofs},}\ }\href
  {http://www.ams.org/notices/200408/fea-pfender.pdf} {\bibfield  {journal}
  {\bibinfo  {journal} {Notices Amer. Math. Soc.}\ }\textbf {\bibinfo {volume}
  {51}},\ \bibinfo {pages} {873--883} (\bibinfo {year} {2004})}\BibitemShut
  {NoStop}%
\bibitem [{\citenamefont {Musin}(2008)}]{Musin08}%
  \BibitemOpen
  \bibfield  {author} {\bibinfo {author} {\bibfnamefont {O.~R.}\ \bibnamefont
  {Musin}},\ }\bibfield  {title} {\enquote {\bibinfo {title} {The kissing
  number in four dimensions},}\ }\href {https://www.jstor.org/stable/40345407}
  {\bibfield  {journal} {\bibinfo  {journal} {Ann. Math.}\ }\textbf {\bibinfo
  {volume} {168}},\ \bibinfo {pages} {1--32} (\bibinfo {year}
  {2008})}\BibitemShut {NoStop}%
\bibitem [{\citenamefont {Brauchart}\ and\ \citenamefont
  {Grabner}(2015)}]{Min-energy15}%
  \BibitemOpen
  \bibfield  {author} {\bibinfo {author} {\bibfnamefont {J.~S.}\ \bibnamefont
  {Brauchart}}\ and\ \bibinfo {author} {\bibfnamefont {P.~J.}\ \bibnamefont
  {Grabner}},\ }\bibfield  {title} {\enquote {\bibinfo {title} {Distributing
  many points on spheres: Minimal energy and designs},}\ }\href {\doibase
  https://doi.org/10.1016/j.jco.2015.02.003} {\bibfield  {journal} {\bibinfo
  {journal} {J. Complexity}\ }\textbf {\bibinfo {volume} {31}},\ \bibinfo
  {pages} {293--326} (\bibinfo {year} {2015})}\BibitemShut {NoStop}%
\bibitem [{\citenamefont {Jenssen}\ \emph {et~al.}(2018)\citenamefont
  {Jenssen}, \citenamefont {Joos},\ and\ \citenamefont {Perkins}}]{Kiss18}%
  \BibitemOpen
  \bibfield  {author} {\bibinfo {author} {\bibfnamefont {M.}~\bibnamefont
  {Jenssen}}, \bibinfo {author} {\bibfnamefont {F.}~\bibnamefont {Joos}}, \
  and\ \bibinfo {author} {\bibfnamefont {W.}~\bibnamefont {Perkins}},\
  }\bibfield  {title} {\enquote {\bibinfo {title} {On kissing numbers and
  spherical codes in high dimensions},}\ }\href
  {https://doi.org/10.1016/j.aim.2018.07.001} {\bibfield  {journal} {\bibinfo
  {journal} {Adv. Math.}\ }\textbf {\bibinfo {volume} {335}},\ \bibinfo {pages}
  {307--321} (\bibinfo {year} {2018})}\BibitemShut {NoStop}%
\bibitem [{\citenamefont {Borodachov}\ \emph {et~al.}(2019)\citenamefont
  {Borodachov}, \citenamefont {Hardin},\ and\ \citenamefont {Saff}}]{Hardin}%
  \BibitemOpen
  \bibfield  {author} {\bibinfo {author} {\bibfnamefont {S.~V.}\ \bibnamefont
  {Borodachov}}, \bibinfo {author} {\bibfnamefont {D.~P.}\ \bibnamefont
  {Hardin}}, \ and\ \bibinfo {author} {\bibfnamefont {E.~B.}\ \bibnamefont
  {Saff}},\ }\href {https://doi.org/10.1007/978-0-387-84808-2} {\emph {\bibinfo
  {title} {Discrete Energy on Rectifiable Sets}}}\ (\bibinfo  {publisher}
  {Springer},\ \bibinfo {address} {New York, NY},\ \bibinfo {year}
  {2019})\BibitemShut {NoStop}%
\bibitem [{\citenamefont {Shannon}(1959)}]{Shannon59}%
  \BibitemOpen
  \bibfield  {author} {\bibinfo {author} {\bibfnamefont {C.~E.}\ \bibnamefont
  {Shannon}},\ }\bibfield  {title} {\enquote {\bibinfo {title} {Probability of
  error for optimal codes in a {G}aussian channel},}\ }\href {\doibase
  10.1002/j.1538-7305.1959.tb03905.x} {\bibfield  {journal} {\bibinfo
  {journal} {Bell Syst. Tech. J.}\ }\textbf {\bibinfo {volume} {38}},\ \bibinfo
  {pages} {611--656} (\bibinfo {year} {1959})}\BibitemShut {NoStop}%
\bibitem [{\citenamefont {Wyner}(1965)}]{Wyner}%
  \BibitemOpen
  \bibfield  {author} {\bibinfo {author} {\bibfnamefont {A.~D.}\ \bibnamefont
  {Wyner}},\ }\bibfield  {title} {\enquote {\bibinfo {title} {Capabilities of
  bounded discrepancy decoding},}\ }\href {\doibase
  10.1002/j.1538-7305.1965.tb04170.x} {\bibfield  {journal} {\bibinfo
  {journal} {Bell Syst. Tech. J.}\ }\textbf {\bibinfo {volume} {44}},\ \bibinfo
  {pages} {1061--1122} (\bibinfo {year} {1965})}\BibitemShut {NoStop}%
\bibitem [{\citenamefont {Frisch}\ \emph {et~al.}(1985)\citenamefont {Frisch},
  \citenamefont {Rivier},\ and\ \citenamefont {Wyler}}]{Frisch85}%
  \BibitemOpen
  \bibfield  {author} {\bibinfo {author} {\bibfnamefont {H.~L.}\ \bibnamefont
  {Frisch}}, \bibinfo {author} {\bibfnamefont {N.}~\bibnamefont {Rivier}}, \
  and\ \bibinfo {author} {\bibfnamefont {D.}~\bibnamefont {Wyler}},\ }\bibfield
   {title} {\enquote {\bibinfo {title} {Classical hard-sphere fluid in
  infinitely many dimensions},}\ }\href {\doibase 10.1103/PhysRevLett.54.2061}
  {\bibfield  {journal} {\bibinfo  {journal} {Phys. Rev. Lett.}\ }\textbf
  {\bibinfo {volume} {54}},\ \bibinfo {pages} {2061--2063} (\bibinfo {year}
  {1985})}\BibitemShut {NoStop}%
\bibitem [{\citenamefont {Klein}\ and\ \citenamefont
  {Frisch}(1986)}]{Frisch86}%
  \BibitemOpen
  \bibfield  {author} {\bibinfo {author} {\bibfnamefont {W.}~\bibnamefont
  {Klein}}\ and\ \bibinfo {author} {\bibfnamefont {H.~L.}\ \bibnamefont
  {Frisch}},\ }\bibfield  {title} {\enquote {\bibinfo {title} {Instability in
  the infinite dimensional hard sphere fluid},}\ }\href
  {https://doi.org/10.1063/1.450544} {\bibfield  {journal} {\bibinfo  {journal}
  {J. Chem. Phys.}\ }\textbf {\bibinfo {volume} {84}},\ \bibinfo {pages}
  {968--970} (\bibinfo {year} {1986})}\BibitemShut {NoStop}%
\bibitem [{\citenamefont {Wyler}\ \emph {et~al.}(1987)\citenamefont {Wyler},
  \citenamefont {Rivier},\ and\ \citenamefont {Frisch}}]{Frisch87}%
  \BibitemOpen
  \bibfield  {author} {\bibinfo {author} {\bibfnamefont {D.}~\bibnamefont
  {Wyler}}, \bibinfo {author} {\bibfnamefont {N.}~\bibnamefont {Rivier}}, \
  and\ \bibinfo {author} {\bibfnamefont {H.~L.}\ \bibnamefont {Frisch}},\
  }\bibfield  {title} {\enquote {\bibinfo {title} {Hard-sphere fluid in
  infinite dimensions},}\ }\href {\doibase 10.1103/PhysRevA.36.2422} {\bibfield
   {journal} {\bibinfo  {journal} {Phys. Rev. A}\ }\textbf {\bibinfo {volume}
  {36}},\ \bibinfo {pages} {2422--2431} (\bibinfo {year} {1987})}\BibitemShut
  {NoStop}%
\bibitem [{\citenamefont {Elskens}\ and\ \citenamefont
  {Frisch}(1988)}]{Frisch88}%
  \BibitemOpen
  \bibfield  {author} {\bibinfo {author} {\bibfnamefont {Y.}~\bibnamefont
  {Elskens}}\ and\ \bibinfo {author} {\bibfnamefont {H.~L.}\ \bibnamefont
  {Frisch}},\ }\bibfield  {title} {\enquote {\bibinfo {title} {Kinetic theory
  of hard spheres in infinite dimensions},}\ }\href {\doibase
  10.1103/PhysRevA.37.4351} {\bibfield  {journal} {\bibinfo  {journal} {Phys.
  Rev. A}\ }\textbf {\bibinfo {volume} {37}},\ \bibinfo {pages} {4351--4353}
  (\bibinfo {year} {1988})}\BibitemShut {NoStop}%
\bibitem [{\citenamefont {Carmesin}\ \emph {et~al.}(1989)\citenamefont
  {Carmesin}, \citenamefont {Frisch},\ and\ \citenamefont {Percus}}]{Frisch89}%
  \BibitemOpen
  \bibfield  {author} {\bibinfo {author} {\bibfnamefont {H.-O.}\ \bibnamefont
  {Carmesin}}, \bibinfo {author} {\bibfnamefont {H.~L.}\ \bibnamefont
  {Frisch}}, \ and\ \bibinfo {author} {\bibfnamefont {J.~K.}\ \bibnamefont
  {Percus}},\ }\bibfield  {title} {\enquote {\bibinfo {title} {Liquid crystals
  at high dimensionality},}\ }\href {\doibase 10.1103/PhysRevB.40.9416}
  {\bibfield  {journal} {\bibinfo  {journal} {Phys. Rev. B}\ }\textbf {\bibinfo
  {volume} {40}},\ \bibinfo {pages} {9416--9418} (\bibinfo {year}
  {1989})}\BibitemShut {NoStop}%
\bibitem [{\citenamefont {Mehta}(1994)}]{Mehta}%
  \BibitemOpen
  \bibfield  {author} {\bibinfo {author} {\bibfnamefont {A.}~\bibnamefont
  {Mehta}},\ }\href {https://doi.org/10.1007/978-1-4612-4290-1} {\emph
  {\bibinfo {title} {Granular Matter: An Interdisciplinary Approach}}}\
  (\bibinfo  {publisher} {Springer-Verlag},\ \bibinfo {address} {New York},\
  \bibinfo {year} {1994})\BibitemShut {NoStop}%
\bibitem [{\citenamefont {Georges}\ \emph {et~al.}(1996)\citenamefont
  {Georges}, \citenamefont {Kotliar}, \citenamefont {Krauth},\ and\
  \citenamefont {Rozenberg}}]{Georges96}%
  \BibitemOpen
  \bibfield  {author} {\bibinfo {author} {\bibfnamefont {A.}~\bibnamefont
  {Georges}}, \bibinfo {author} {\bibfnamefont {G.}~\bibnamefont {Kotliar}},
  \bibinfo {author} {\bibfnamefont {W.}~\bibnamefont {Krauth}}, \ and\ \bibinfo
  {author} {\bibfnamefont {M.~J.}\ \bibnamefont {Rozenberg}},\ }\bibfield
  {title} {\enquote {\bibinfo {title} {Dynamical mean-field theory of strongly
  correlated fermion systems and the limit of infinite dimensions},}\ }\href
  {\doibase 10.1103/RevModPhys.68.13} {\bibfield  {journal} {\bibinfo
  {journal} {Rev. Mod. Phys.}\ }\textbf {\bibinfo {volume} {68}},\ \bibinfo
  {pages} {13--125} (\bibinfo {year} {1996})}\BibitemShut {NoStop}%
\bibitem [{\citenamefont {Frisch}\ and\ \citenamefont
  {Percus}(1999)}]{Frisch99}%
  \BibitemOpen
  \bibfield  {author} {\bibinfo {author} {\bibfnamefont {H.~L.}\ \bibnamefont
  {Frisch}}\ and\ \bibinfo {author} {\bibfnamefont {J.~K.}\ \bibnamefont
  {Percus}},\ }\bibfield  {title} {\enquote {\bibinfo {title} {High
  dimensionality as an organizing device for classical fluids},}\ }\href
  {\doibase 10.1103/PhysRevE.60.2942} {\bibfield  {journal} {\bibinfo
  {journal} {Phys. Rev. E}\ }\textbf {\bibinfo {volume} {60}},\ \bibinfo
  {pages} {2942--2948} (\bibinfo {year} {1999})}\BibitemShut {NoStop}%
\bibitem [{\citenamefont {Parisi}\ and\ \citenamefont
  {Slanina}(2000)}]{Parisi00}%
  \BibitemOpen
  \bibfield  {author} {\bibinfo {author} {\bibfnamefont {G.}~\bibnamefont
  {Parisi}}\ and\ \bibinfo {author} {\bibfnamefont {F.}~\bibnamefont
  {Slanina}},\ }\bibfield  {title} {\enquote {\bibinfo {title} {Toy model for
  the mean-field theory of hard-sphere liquids},}\ }\href {\doibase
  10.1103/PhysRevE.62.6554} {\bibfield  {journal} {\bibinfo  {journal} {Phys.
  Rev. E}\ }\textbf {\bibinfo {volume} {62}},\ \bibinfo {pages} {6554--6559}
  (\bibinfo {year} {2000})}\BibitemShut {NoStop}%
\bibitem [{\citenamefont {Torquato}\ and\ \citenamefont
  {Stillinger}(2010)}]{TS10}%
  \BibitemOpen
  \bibfield  {author} {\bibinfo {author} {\bibfnamefont {S.}~\bibnamefont
  {Torquato}}\ and\ \bibinfo {author} {\bibfnamefont {F.~H.}\ \bibnamefont
  {Stillinger}},\ }\bibfield  {title} {\enquote {\bibinfo {title} {Jammed
  hard-particle packings: From {K}epler to {B}ernal and beyond},}\ }\href
  {\doibase 10.1103/RevModPhys.82.2633} {\bibfield  {journal} {\bibinfo
  {journal} {Rev. Mod. Phys.}\ }\textbf {\bibinfo {volume} {82}},\ \bibinfo
  {pages} {2633--2672} (\bibinfo {year} {2010})}\BibitemShut {NoStop}%
\bibitem [{\citenamefont {Parisi}\ and\ \citenamefont
  {Zamponi}(2010)}]{Parisi10}%
  \BibitemOpen
  \bibfield  {author} {\bibinfo {author} {\bibfnamefont {G.}~\bibnamefont
  {Parisi}}\ and\ \bibinfo {author} {\bibfnamefont {F.}~\bibnamefont
  {Zamponi}},\ }\bibfield  {title} {\enquote {\bibinfo {title} {Mean-field
  theory of hard sphere glasses and jamming},}\ }\href {\doibase
  10.1103/RevModPhys.82.789} {\bibfield  {journal} {\bibinfo  {journal} {Rev.
  Mod. Phys.}\ }\textbf {\bibinfo {volume} {82}},\ \bibinfo {pages} {789--845}
  (\bibinfo {year} {2010})}\BibitemShut {NoStop}%
\bibitem [{\citenamefont {Schmid}\ and\ \citenamefont
  {Schilling}(2010)}]{Schilling10}%
  \BibitemOpen
  \bibfield  {author} {\bibinfo {author} {\bibfnamefont {B.}~\bibnamefont
  {Schmid}}\ and\ \bibinfo {author} {\bibfnamefont {R.}~\bibnamefont
  {Schilling}},\ }\bibfield  {title} {\enquote {\bibinfo {title} {Glass
  transition of hard spheres in high dimensions},}\ }\href {\doibase
  10.1103/PhysRevE.81.041502} {\bibfield  {journal} {\bibinfo  {journal} {Phys.
  Rev. E}\ }\textbf {\bibinfo {volume} {81}},\ \bibinfo {pages} {041502}
  (\bibinfo {year} {2010})}\BibitemShut {NoStop}%
\bibitem [{\citenamefont {Charbonneau}\ \emph {et~al.}(2011)\citenamefont
  {Charbonneau}, \citenamefont {Ikeda}, \citenamefont {Parisi},\ and\
  \citenamefont {Zamponi}}]{Charbonneau11}%
  \BibitemOpen
  \bibfield  {author} {\bibinfo {author} {\bibfnamefont {P.}~\bibnamefont
  {Charbonneau}}, \bibinfo {author} {\bibfnamefont {A.}~\bibnamefont {Ikeda}},
  \bibinfo {author} {\bibfnamefont {G.}~\bibnamefont {Parisi}}, \ and\ \bibinfo
  {author} {\bibfnamefont {F.}~\bibnamefont {Zamponi}},\ }\bibfield  {title}
  {\enquote {\bibinfo {title} {Glass transition and random close packing above
  three dimensions},}\ }\href {\doibase 10.1103/PhysRevLett.107.185702}
  {\bibfield  {journal} {\bibinfo  {journal} {Phys. Rev. Lett.}\ }\textbf
  {\bibinfo {volume} {107}},\ \bibinfo {pages} {185702} (\bibinfo {year}
  {2011})}\BibitemShut {NoStop}%
\bibitem [{\citenamefont {Kallus}(2013)}]{Kallus13}%
  \BibitemOpen
  \bibfield  {author} {\bibinfo {author} {\bibfnamefont {Y.}~\bibnamefont
  {Kallus}},\ }\bibfield  {title} {\enquote {\bibinfo {title} {Statistical
  mechanics of the lattice sphere packing problem},}\ }\href {\doibase
  10.1103/PhysRevE.87.063307} {\bibfield  {journal} {\bibinfo  {journal} {Phys.
  Rev. E}\ }\textbf {\bibinfo {volume} {87}},\ \bibinfo {pages} {063307}
  (\bibinfo {year} {2013})}\BibitemShut {NoStop}%
\bibitem [{\citenamefont {Maimbourg}\ \emph {et~al.}(2016)\citenamefont
  {Maimbourg}, \citenamefont {Kurchan},\ and\ \citenamefont
  {Zamponi}}]{Kurchan16a}%
  \BibitemOpen
  \bibfield  {author} {\bibinfo {author} {\bibfnamefont {T.}~\bibnamefont
  {Maimbourg}}, \bibinfo {author} {\bibfnamefont {J.}~\bibnamefont {Kurchan}},
  \ and\ \bibinfo {author} {\bibfnamefont {F.}~\bibnamefont {Zamponi}},\
  }\bibfield  {title} {\enquote {\bibinfo {title} {Solution of the dynamics of
  liquids in the large-dimensional limit},}\ }\href {\doibase
  10.1103/PhysRevLett.116.015902} {\bibfield  {journal} {\bibinfo  {journal}
  {Phys. Rev. Lett.}\ }\textbf {\bibinfo {volume} {116}},\ \bibinfo {pages}
  {015902} (\bibinfo {year} {2016})}\BibitemShut {NoStop}%
\bibitem [{\citenamefont {Kurchan}\ \emph {et~al.}(2016)\citenamefont
  {Kurchan}, \citenamefont {Maimbourg},\ and\ \citenamefont
  {Zamponi}}]{Kurchan16b}%
  \BibitemOpen
  \bibfield  {author} {\bibinfo {author} {\bibfnamefont {J.}~\bibnamefont
  {Kurchan}}, \bibinfo {author} {\bibfnamefont {T.}~\bibnamefont {Maimbourg}},
  \ and\ \bibinfo {author} {\bibfnamefont {F.}~\bibnamefont {Zamponi}},\
  }\bibfield  {title} {\enquote {\bibinfo {title} {Statics and dynamics of
  infinite-dimensional liquids and glasses: a parallel and compact
  derivation},}\ }\href {\doibase 10.1088/1742-5468/2016/03/033210} {\bibfield
  {journal} {\bibinfo  {journal} {J. Stat. Mech.}\ }\textbf {\bibinfo {volume}
  {2016}},\ \bibinfo {pages} {033210} (\bibinfo {year} {2016})}\BibitemShut
  {NoStop}%
\bibitem [{\citenamefont {Szamel}(2017)}]{Szamel17}%
  \BibitemOpen
  \bibfield  {author} {\bibinfo {author} {\bibfnamefont {G.}~\bibnamefont
  {Szamel}},\ }\bibfield  {title} {\enquote {\bibinfo {title} {Simple theory
  for the dynamics of mean-field-like models of glass-forming fluids},}\ }\href
  {\doibase 10.1103/PhysRevLett.119.155502} {\bibfield  {journal} {\bibinfo
  {journal} {Phys. Rev. Lett.}\ }\textbf {\bibinfo {volume} {119}},\ \bibinfo
  {pages} {155502} (\bibinfo {year} {2017})}\BibitemShut {NoStop}%
\bibitem [{\citenamefont {Charbonneau}\ \emph {et~al.}(2017)\citenamefont
  {Charbonneau}, \citenamefont {Kurchan}, \citenamefont {Parisi}, \citenamefont
  {Urbani},\ and\ \citenamefont {Zamponi}}]{Charbonneau17}%
  \BibitemOpen
  \bibfield  {author} {\bibinfo {author} {\bibfnamefont {P.}~\bibnamefont
  {Charbonneau}}, \bibinfo {author} {\bibfnamefont {J.}~\bibnamefont
  {Kurchan}}, \bibinfo {author} {\bibfnamefont {G.}~\bibnamefont {Parisi}},
  \bibinfo {author} {\bibfnamefont {P.}~\bibnamefont {Urbani}}, \ and\ \bibinfo
  {author} {\bibfnamefont {F.}~\bibnamefont {Zamponi}},\ }\bibfield  {title}
  {\enquote {\bibinfo {title} {Glass and jamming transitions: From exact
  results to finite-dimensional descriptions},}\ }\href {\doibase
  10.1146/annurev-conmatphys-031016-025334} {\bibfield  {journal} {\bibinfo
  {journal} {Annual Rev. Cond. Matter Phys.}\ }\textbf {\bibinfo {volume}
  {8}},\ \bibinfo {pages} {265--288} (\bibinfo {year} {2017})}\BibitemShut
  {NoStop}%
\bibitem [{\citenamefont {Biroli}\ and\ \citenamefont
  {Urbani}(2018)}]{Biroli18}%
  \BibitemOpen
  \bibfield  {author} {\bibinfo {author} {\bibfnamefont {G.}~\bibnamefont
  {Biroli}}\ and\ \bibinfo {author} {\bibfnamefont {P.}~\bibnamefont
  {Urbani}},\ }\bibfield  {title} {\enquote {\bibinfo {title} {{Liu-{N}agel
  phase diagrams in infinite dimension}},}\ }\href {\doibase
  10.21468/SciPostPhys.4.4.020} {\bibfield  {journal} {\bibinfo  {journal}
  {SciPost Phys.}\ }\textbf {\bibinfo {volume} {4}},\ \bibinfo {pages} {020}
  (\bibinfo {year} {2018})}\BibitemShut {NoStop}%
\bibitem [{\citenamefont {Torquato}(2018)}]{Torquato18}%
  \BibitemOpen
  \bibfield  {author} {\bibinfo {author} {\bibfnamefont {S.}~\bibnamefont
  {Torquato}},\ }\bibfield  {title} {\enquote {\bibinfo {title} {Perspective:
  Basic understanding of condensed phases of matter via packing models},}\
  }\href {\doibase 10.1063/1.5036657} {\bibfield  {journal} {\bibinfo
  {journal} {J. Chem. Phys.}\ }\textbf {\bibinfo {volume} {149}},\ \bibinfo
  {pages} {020901} (\bibinfo {year} {2018})}\BibitemShut {NoStop}%
\bibitem [{\citenamefont {Parisi}\ \emph {et~al.}(2020)\citenamefont {Parisi},
  \citenamefont {Urbani},\ and\ \citenamefont {Zamponi}}]{Parisi20}%
  \BibitemOpen
  \bibfield  {author} {\bibinfo {author} {\bibfnamefont {G.}~\bibnamefont
  {Parisi}}, \bibinfo {author} {\bibfnamefont {P.}~\bibnamefont {Urbani}}, \
  and\ \bibinfo {author} {\bibfnamefont {F.}~\bibnamefont {Zamponi}},\ }\href
  {https://doi.org/10.1017/9781108120494} {\emph {\bibinfo {title} {Theory of
  simple glasses: Exact solutions in infinite dimensions}}}\ (\bibinfo
  {publisher} {Cambridge University Press},\ \bibinfo {address} {Cambridge,
  UK},\ \bibinfo {year} {2020})\BibitemShut {NoStop}%
\bibitem [{\citenamefont {Arnoulx~de Pirey}\ \emph {et~al.}(2019)\citenamefont
  {Arnoulx~de Pirey}, \citenamefont {Lozano},\ and\ \citenamefont {van
  Wijland}}]{Fred19}%
  \BibitemOpen
  \bibfield  {author} {\bibinfo {author} {\bibfnamefont {T.}~\bibnamefont
  {Arnoulx~de Pirey}}, \bibinfo {author} {\bibfnamefont {G.}~\bibnamefont
  {Lozano}}, \ and\ \bibinfo {author} {\bibfnamefont {F.}~\bibnamefont {van
  Wijland}},\ }\bibfield  {title} {\enquote {\bibinfo {title} {Active hard
  spheres in infinitely many dimensions},}\ }\href {\doibase
  10.1103/PhysRevLett.123.260602} {\bibfield  {journal} {\bibinfo  {journal}
  {Phys. Rev. Lett.}\ }\textbf {\bibinfo {volume} {123}},\ \bibinfo {pages}
  {260602} (\bibinfo {year} {2019})}\BibitemShut {NoStop}%
\bibitem [{\citenamefont {Arnoulx~de Pirey}\ \emph {et~al.}(2021)\citenamefont
  {Arnoulx~de Pirey}, \citenamefont {Manacorda}, \citenamefont {van Wijland},\
  and\ \citenamefont {Zamponi}}]{Fred21}%
  \BibitemOpen
  \bibfield  {author} {\bibinfo {author} {\bibfnamefont {T.}~\bibnamefont
  {Arnoulx~de Pirey}}, \bibinfo {author} {\bibfnamefont {A.}~\bibnamefont
  {Manacorda}}, \bibinfo {author} {\bibfnamefont {F.}~\bibnamefont {van
  Wijland}}, \ and\ \bibinfo {author} {\bibfnamefont {F.}~\bibnamefont
  {Zamponi}},\ }\bibfield  {title} {\enquote {\bibinfo {title} {Active matter
  in infinite dimensions: {F}okker-{P}lanck equation and dynamical mean-field
  theory at low density},}\ }\href {https://doi.org/10.1063/5.0065893}
  {\bibfield  {journal} {\bibinfo  {journal} {J. Chem. Phys.}\ }\textbf
  {\bibinfo {volume} {155}},\ \bibinfo {pages} {174106} (\bibinfo {year}
  {2021})}\BibitemShut {NoStop}%
\bibitem [{\citenamefont {Charbonneau}\ \emph {et~al.}(2021)\citenamefont
  {Charbonneau}, \citenamefont {Morse}, \citenamefont {Perkins},\ and\
  \citenamefont {Zamponi}}]{Charbonneau21}%
  \BibitemOpen
  \bibfield  {author} {\bibinfo {author} {\bibfnamefont {P.}~\bibnamefont
  {Charbonneau}}, \bibinfo {author} {\bibfnamefont {P.~K.}\ \bibnamefont
  {Morse}}, \bibinfo {author} {\bibfnamefont {W.}~\bibnamefont {Perkins}}, \
  and\ \bibinfo {author} {\bibfnamefont {F.}~\bibnamefont {Zamponi}},\
  }\bibfield  {title} {\enquote {\bibinfo {title} {Three simple scenarios for
  high-dimensional sphere packings},}\ }\href
  {https://link.aps.org/doi/10.1103/PhysRevE.104.064612} {\bibfield  {journal}
  {\bibinfo  {journal} {Phys. Rev. E}\ }\textbf {\bibinfo {volume} {104}},\
  \bibinfo {pages} {064612} (\bibinfo {year} {2021})}\BibitemShut {NoStop}%
\bibitem [{\citenamefont {Hartman}\ \emph {et~al.}(2019)\citenamefont
  {Hartman}, \citenamefont {Maz\'{a}\u{c}},\ and\ \citenamefont
  {Rastelli}}]{Rastelli19}%
  \BibitemOpen
  \bibfield  {author} {\bibinfo {author} {\bibfnamefont {T.}~\bibnamefont
  {Hartman}}, \bibinfo {author} {\bibfnamefont {D.}~\bibnamefont
  {Maz\'{a}\u{c}}}, \ and\ \bibinfo {author} {\bibfnamefont {L.}~\bibnamefont
  {Rastelli}},\ }\bibfield  {title} {\enquote {\bibinfo {title} {Sphere packing
  and quantum gravity},}\ }\href {https://doi.org/10.1007/JHEP12(2019)048}
  {\bibfield  {journal} {\bibinfo  {journal} {J. High Energ. Phys.}\ }\textbf
  {\bibinfo {volume} {2019}},\ \bibinfo {pages} {48} (\bibinfo {year}
  {2019})}\BibitemShut {NoStop}%
\bibitem [{\citenamefont {Afkhami-Jeddi}\ \emph {et~al.}(2020)\citenamefont
  {Afkhami-Jeddi}, \citenamefont {Cohn}, \citenamefont {Hartman}, \citenamefont
  {de~Laat},\ and\ \citenamefont {Tajdini}}]{Cohn20}%
  \BibitemOpen
  \bibfield  {author} {\bibinfo {author} {\bibfnamefont {N.}~\bibnamefont
  {Afkhami-Jeddi}}, \bibinfo {author} {\bibfnamefont {H.}~\bibnamefont {Cohn}},
  \bibinfo {author} {\bibfnamefont {T.}~\bibnamefont {Hartman}}, \bibinfo
  {author} {\bibfnamefont {D.}~\bibnamefont {de~Laat}}, \ and\ \bibinfo
  {author} {\bibfnamefont {A.}~\bibnamefont {Tajdini}},\ }\bibfield  {title}
  {\enquote {\bibinfo {title} {High-dimensional sphere packing and the modular
  bootstrap},}\ }\href {https://doi.org/10.1007/JHEP12(2020)066} {\bibfield
  {journal} {\bibinfo  {journal} {J. High Energ. Phys.}\ }\textbf {\bibinfo
  {volume} {2020}},\ \bibinfo {pages} {66} (\bibinfo {year}
  {2020})}\BibitemShut {NoStop}%
\bibitem [{\citenamefont {Liu}\ \emph {et~al.}(2021)\citenamefont {Liu},
  \citenamefont {Biroli}, \citenamefont {Reichman},\ and\ \citenamefont
  {Szamel}}]{Biroli22}%
  \BibitemOpen
  \bibfield  {author} {\bibinfo {author} {\bibfnamefont {C.}~\bibnamefont
  {Liu}}, \bibinfo {author} {\bibfnamefont {G.}~\bibnamefont {Biroli}},
  \bibinfo {author} {\bibfnamefont {D.~R.}\ \bibnamefont {Reichman}}, \ and\
  \bibinfo {author} {\bibfnamefont {G.}~\bibnamefont {Szamel}},\ }\bibfield
  {title} {\enquote {\bibinfo {title} {Dynamics of liquids in the
  large-dimensional limit},}\ }\href {\doibase 10.1103/PhysRevE.104.054606}
  {\bibfield  {journal} {\bibinfo  {journal} {Phys. Rev. E}\ }\textbf {\bibinfo
  {volume} {104}},\ \bibinfo {pages} {054606} (\bibinfo {year}
  {2021})}\BibitemShut {NoStop}%
\bibitem [{\citenamefont {Krapivsky}\ and\ \citenamefont
  {Redner}(2022)}]{KR-birds22}%
  \BibitemOpen
  \bibfield  {author} {\bibinfo {author} {\bibfnamefont {P.~L.}\ \bibnamefont
  {Krapivsky}}\ and\ \bibinfo {author} {\bibfnamefont {S.}~\bibnamefont
  {Redner}},\ }\bibfield  {title} {\enquote {\bibinfo {title} {Birds on a
  wire},}\ }\href {\doibase 10.1088/1742-5468/ac98bf} {\bibfield  {journal}
  {\bibinfo  {journal} {J. Stat. Mech.}\ }\textbf {\bibinfo {volume} {2022}},\
  \bibinfo {pages} {103405} (\bibinfo {year} {2022})}\BibitemShut {NoStop}%
\bibitem [{Note1()}]{Note1}%
  \BibitemOpen
  \bibinfo {note} {For $\protect \mathbb {Z}^d$, we use the Manhattan norm
  $|i|=|i_1|+\protect \ldots +|i_d|$.}\BibitemShut {Stop}%
\bibitem [{\citenamefont {Jin}\ \emph {et~al.}(2010)\citenamefont {Jin},
  \citenamefont {Charbonneau}, \citenamefont {Meyer}, \citenamefont {Song},\
  and\ \citenamefont {Zamponi}}]{Charbonneau10}%
  \BibitemOpen
  \bibfield  {author} {\bibinfo {author} {\bibfnamefont {Y.}~\bibnamefont
  {Jin}}, \bibinfo {author} {\bibfnamefont {P.}~\bibnamefont {Charbonneau}},
  \bibinfo {author} {\bibfnamefont {S.}~\bibnamefont {Meyer}}, \bibinfo
  {author} {\bibfnamefont {C.}~\bibnamefont {Song}}, \ and\ \bibinfo {author}
  {\bibfnamefont {F.}~\bibnamefont {Zamponi}},\ }\bibfield  {title} {\enquote
  {\bibinfo {title} {Application of {E}dwards' statistical mechanics to
  high-dimensional jammed sphere packings},}\ }\href {\doibase
  10.1103/PhysRevE.82.051126} {\bibfield  {journal} {\bibinfo  {journal} {Phys.
  Rev. E}\ }\textbf {\bibinfo {volume} {82}},\ \bibinfo {pages} {051126}
  (\bibinfo {year} {2010})}\BibitemShut {NoStop}%
\bibitem [{\citenamefont {Andreanov}\ \emph {et~al.}(2016)\citenamefont
  {Andreanov}, \citenamefont {Scardicchio},\ and\ \citenamefont
  {Torquato}}]{Torquato16}%
  \BibitemOpen
  \bibfield  {author} {\bibinfo {author} {\bibfnamefont {A.}~\bibnamefont
  {Andreanov}}, \bibinfo {author} {\bibfnamefont {A.}~\bibnamefont
  {Scardicchio}}, \ and\ \bibinfo {author} {\bibfnamefont {S.}~\bibnamefont
  {Torquato}},\ }\bibfield  {title} {\enquote {\bibinfo {title} {Extreme
  lattices: symmetries and decorrelation},}\ }\href {\doibase
  10.1088/1742-5468/2016/11/113301} {\bibfield  {journal} {\bibinfo  {journal}
  {J. Stat. Mech.}\ }\textbf {\bibinfo {volume} {2016}},\ \bibinfo {pages}
  {113301} (\bibinfo {year} {2016})}\BibitemShut {NoStop}%
\bibitem [{\citenamefont {Rohrmann}\ and\ \citenamefont
  {Santos}(2007)}]{Santos07}%
  \BibitemOpen
  \bibfield  {author} {\bibinfo {author} {\bibfnamefont {R.~D.}\ \bibnamefont
  {Rohrmann}}\ and\ \bibinfo {author} {\bibfnamefont {A.}~\bibnamefont
  {Santos}},\ }\bibfield  {title} {\enquote {\bibinfo {title} {Structure of
  hard-hypersphere fluids in odd dimensions},}\ }\href {\doibase
  10.1103/PhysRevE.76.051202} {\bibfield  {journal} {\bibinfo  {journal} {Phys.
  Rev. E}\ }\textbf {\bibinfo {volume} {76}},\ \bibinfo {pages} {051202}
  (\bibinfo {year} {2007})}\BibitemShut {NoStop}%
\bibitem [{\citenamefont {Cohn}\ and\ \citenamefont {Kumar}(2008)}]{Cohn08}%
  \BibitemOpen
  \bibfield  {author} {\bibinfo {author} {\bibfnamefont {H.}~\bibnamefont
  {Cohn}}\ and\ \bibinfo {author} {\bibfnamefont {A.}~\bibnamefont {Kumar}},\
  }\bibfield  {title} {\enquote {\bibinfo {title} {Counterintuitive ground
  states in soft-core models},}\ }\href {\doibase 10.1103/PhysRevE.78.061113}
  {\bibfield  {journal} {\bibinfo  {journal} {Phys. Rev. E}\ }\textbf {\bibinfo
  {volume} {78}},\ \bibinfo {pages} {061113} (\bibinfo {year}
  {2008})}\BibitemShut {NoStop}%
\bibitem [{\citenamefont {Zahary}\ \emph {et~al.}(2008)\citenamefont {Zahary},
  \citenamefont {Stillinger},\ and\ \citenamefont {Torquato}}]{Torquato08a}%
  \BibitemOpen
  \bibfield  {author} {\bibinfo {author} {\bibfnamefont {C.~E.}\ \bibnamefont
  {Zahary}}, \bibinfo {author} {\bibfnamefont {F.~H.}\ \bibnamefont
  {Stillinger}}, \ and\ \bibinfo {author} {\bibfnamefont {S.}~\bibnamefont
  {Torquato}},\ }\bibfield  {title} {\enquote {\bibinfo {title} {Gaussian core
  model phase diagram and pair correlations in high {E}uclidean dimensions},}\
  }\href {https://doi.org/10.1063/1.2928843} {\bibfield  {journal} {\bibinfo
  {journal} {J. Chem. Phys.}\ }\textbf {\bibinfo {volume} {128}},\ \bibinfo
  {pages} {224505} (\bibinfo {year} {2008})}\BibitemShut {NoStop}%
\bibitem [{\citenamefont {Cohn}\ and\ \citenamefont
  {de~Courcy-Ireland}(2019)}]{Cohn18}%
  \BibitemOpen
  \bibfield  {author} {\bibinfo {author} {\bibfnamefont {H.}~\bibnamefont
  {Cohn}}\ and\ \bibinfo {author} {\bibfnamefont {M.}~\bibnamefont
  {de~Courcy-Ireland}},\ }\bibfield  {title} {\enquote {\bibinfo {title} {The
  {G}aussian core model in high dimensions},}\ }\href
  {https://doi.org/10.1215/00127094-2018-0018} {\bibfield  {journal} {\bibinfo
  {journal} {Duke Math. J.}\ }\textbf {\bibinfo {volume} {167}},\ \bibinfo
  {pages} {2417--2455} (\bibinfo {year} {2019})}\BibitemShut {NoStop}%
\bibitem [{\citenamefont {Torquato}\ \emph {et~al.}(2008)\citenamefont
  {Torquato}, \citenamefont {Scardicchio},\ and\ \citenamefont
  {Zahary}}]{Torquato08b}%
  \BibitemOpen
  \bibfield  {author} {\bibinfo {author} {\bibfnamefont {S.}~\bibnamefont
  {Torquato}}, \bibinfo {author} {\bibfnamefont {A.}~\bibnamefont
  {Scardicchio}}, \ and\ \bibinfo {author} {\bibfnamefont {C.~E.}\ \bibnamefont
  {Zahary}},\ }\bibfield  {title} {\enquote {\bibinfo {title} {Point processes
  in arbitrary dimension from fermionic gases, random matrix theory, and number
  theory},}\ }\href {\doibase 10.1088/1742-5468/2008/11/P11019} {\bibfield
  {journal} {\bibinfo  {journal} {J. Stat. Mech.}\ }\textbf {\bibinfo {volume}
  {2008}},\ \bibinfo {pages} {P11019} (\bibinfo {year} {2008})}\BibitemShut
  {NoStop}%
\bibitem [{\citenamefont {Bender}\ and\ \citenamefont {Orszag}(1999)}]{Bender}%
  \BibitemOpen
  \bibfield  {author} {\bibinfo {author} {\bibfnamefont {C.~M.}\ \bibnamefont
  {Bender}}\ and\ \bibinfo {author} {\bibfnamefont {S.~A.}\ \bibnamefont
  {Orszag}},\ }\href {https://doi.org/10.1007/978-1-4757-3069-2} {\emph
  {\bibinfo {title} {Advanced Mathematical Methods for Scientists and Engineers
  {I}. Asymptotic Methods and Perturbation Theory}}}\ (\bibinfo  {publisher}
  {Springer},\ \bibinfo {address} {New York},\ \bibinfo {year}
  {1999})\BibitemShut {NoStop}%
\bibitem [{Note2()}]{Note2}%
  \BibitemOpen
  \bibinfo {note} {A sphere packing is saturated if no other sphere can be
  added without overlap.}\BibitemShut {Stop}%
\bibitem [{\citenamefont {Van~Tassel}\ and\ \citenamefont
  {Viot}(1997)}]{Viot97}%
  \BibitemOpen
  \bibfield  {author} {\bibinfo {author} {\bibfnamefont {P.~R.}\ \bibnamefont
  {Van~Tassel}}\ and\ \bibinfo {author} {\bibfnamefont {P.}~\bibnamefont
  {Viot}},\ }\bibfield  {title} {\enquote {\bibinfo {title} {An exactly
  solvable continuum model of multilayer irreversible adsorption},}\ }\href
  {\doibase 10.1209/epl/i1997-00463-3} {\bibfield  {journal} {\bibinfo
  {journal} {EPL}\ }\textbf {\bibinfo {volume} {40}},\ \bibinfo {pages}
  {293--298} (\bibinfo {year} {1997})}\BibitemShut {NoStop}%
\bibitem [{\citenamefont {Torquato}\ and\ \citenamefont
  {Stillinger}(2006{\natexlab{a}})}]{TS06b}%
  \BibitemOpen
  \bibfield  {author} {\bibinfo {author} {\bibfnamefont {S.}~\bibnamefont
  {Torquato}}\ and\ \bibinfo {author} {\bibfnamefont {F.~H.}\ \bibnamefont
  {Stillinger}},\ }\bibfield  {title} {\enquote {\bibinfo {title} {Exactly
  solvable disordered sphere-packing model in arbitrary-dimensional euclidean
  spaces},}\ }\href {\doibase 10.1103/PhysRevE.73.031106} {\bibfield  {journal}
  {\bibinfo  {journal} {Phys. Rev. E}\ }\textbf {\bibinfo {volume} {73}},\
  \bibinfo {pages} {031106} (\bibinfo {year} {2006}{\natexlab{a}})}\BibitemShut
  {NoStop}%
\bibitem [{\citenamefont {Torquato}\ and\ \citenamefont
  {Stillinger}(2006{\natexlab{b}})}]{TS06a}%
  \BibitemOpen
  \bibfield  {author} {\bibinfo {author} {\bibfnamefont {S.}~\bibnamefont
  {Torquato}}\ and\ \bibinfo {author} {\bibfnamefont {F.~H.}\ \bibnamefont
  {Stillinger}},\ }\bibfield  {title} {\enquote {\bibinfo {title} {New
  conjectural lower bounds on the optimal density of sphere packings},}\ }\href
  {https://doi.org/10.1080/10586458.2006.10128964} {\bibfield  {journal}
  {\bibinfo  {journal} {Exp. Math.}\ }\textbf {\bibinfo {volume} {15}},\
  \bibinfo {pages} {307--331} (\bibinfo {year}
  {2006}{\natexlab{b}})}\BibitemShut {NoStop}%
\bibitem [{\citenamefont {Torquato}\ \emph {et~al.}(2006)\citenamefont
  {Torquato}, \citenamefont {Uche},\ and\ \citenamefont {Stillinger}}]{TS06c}%
  \BibitemOpen
  \bibfield  {author} {\bibinfo {author} {\bibfnamefont {S.}~\bibnamefont
  {Torquato}}, \bibinfo {author} {\bibfnamefont {O.~U.}\ \bibnamefont {Uche}},
  \ and\ \bibinfo {author} {\bibfnamefont {F.~H.}\ \bibnamefont {Stillinger}},\
  }\bibfield  {title} {\enquote {\bibinfo {title} {Random sequential addition
  of hard spheres in high euclidean dimensions},}\ }\href
  {https://link.aps.org/doi/10.1103/PhysRevE.74.061308} {\bibfield  {journal}
  {\bibinfo  {journal} {Phys. Rev. E}\ }\textbf {\bibinfo {volume} {74}},\
  \bibinfo {pages} {061308} (\bibinfo {year} {2006})}\BibitemShut {NoStop}%
\bibitem [{\citenamefont {Kubrusly}\ and\ \citenamefont
  {Conci}(2018)}]{pseudo}%
  \BibitemOpen
  \bibfield  {author} {\bibinfo {author} {\bibfnamefont {C.}~\bibnamefont
  {Kubrusly}}\ and\ \bibinfo {author} {\bibfnamefont {A.}~\bibnamefont
  {Conci}},\ }\bibfield  {title} {\enquote {\bibinfo {title} {Distance between
  sets --- a survey},}\ }\href {https://arxiv.org/abs/1808.02574} {\bibfield
  {journal} {\bibinfo  {journal} {arXiv:1808.02574}\ } (\bibinfo {year}
  {2018})}\BibitemShut {NoStop}%
\bibitem [{\citenamefont {Krapivsky}\ \emph {et~al.}(2010)\citenamefont
  {Krapivsky}, \citenamefont {Redner},\ and\ \citenamefont {Ben-Naim}}]{KRB}%
  \BibitemOpen
  \bibfield  {author} {\bibinfo {author} {\bibfnamefont {P.~L.}\ \bibnamefont
  {Krapivsky}}, \bibinfo {author} {\bibfnamefont {S.}~\bibnamefont {Redner}}, \
  and\ \bibinfo {author} {\bibfnamefont {E.}~\bibnamefont {Ben-Naim}},\ }\href
  {https://doi.org/10.1017/CBO9780511780516} {\emph {\bibinfo {title} {A
  Kinetic View of Statistical Physics}}}\ (\bibinfo  {publisher} {Cambridge
  University Press},\ \bibinfo {address} {Cambridge, UK},\ \bibinfo {year}
  {2010})\BibitemShut {NoStop}%
\bibitem [{\citenamefont {Olver}\ \emph {et~al.}(2010)\citenamefont {Olver},
  \citenamefont {Lozier}, \citenamefont {Boisvert},\ and\ \citenamefont
  {Clark}}]{NIST}%
  \BibitemOpen
  \bibfield  {author} {\bibinfo {author} {\bibfnamefont {F.~W.~J.}\
  \bibnamefont {Olver}}, \bibinfo {author} {\bibfnamefont {D.~W.}\ \bibnamefont
  {Lozier}}, \bibinfo {author} {\bibfnamefont {R.~F.}\ \bibnamefont
  {Boisvert}}, \ and\ \bibinfo {author} {\bibfnamefont {C.~W.}\ \bibnamefont
  {Clark}},\ }\href {https://dlmf.nist.gov} {\emph {\bibinfo {title} {NIST
  Handbook of Mathematical Functions}}}\ (\bibinfo  {publisher} {Cambridge
  University Press},\ \bibinfo {address} {Cambridge, UK},\ \bibinfo {year}
  {2010})\BibitemShut {NoStop}%
\bibitem [{\citenamefont {Mandel}(1979)}]{Mandel79}%
  \BibitemOpen
  \bibfield  {author} {\bibinfo {author} {\bibfnamefont {L.}~\bibnamefont
  {Mandel}},\ }\bibfield  {title} {\enquote {\bibinfo {title} {Sub-{P}oissonian
  photon statistics in resonance fluorescence},}\ }\href {\doibase
  10.1364/OL.4.000205} {\bibfield  {journal} {\bibinfo  {journal} {Opt. Lett.}\
  }\textbf {\bibinfo {volume} {4}},\ \bibinfo {pages} {205--207} (\bibinfo
  {year} {1979})}\BibitemShut {NoStop}%
\bibitem [{\citenamefont {Fano}(1947)}]{Fano}%
  \BibitemOpen
  \bibfield  {author} {\bibinfo {author} {\bibfnamefont {U.}~\bibnamefont
  {Fano}},\ }\bibfield  {title} {\enquote {\bibinfo {title} {Ionization yield
  of radiations. {II}. {T}he fluctuations of the number of ions},}\ }\href
  {\doibase 10.1103/PhysRev.72.26} {\bibfield  {journal} {\bibinfo  {journal}
  {Phys. Rev.}\ }\textbf {\bibinfo {volume} {72}},\ \bibinfo {pages} {26--29}
  (\bibinfo {year} {1947})}\BibitemShut {NoStop}%
\bibitem [{\citenamefont {Krapivsky}(2023)}]{PK23}%
  \BibitemOpen
  \bibfield  {author} {\bibinfo {author} {\bibfnamefont {P.~L.}\ \bibnamefont
  {Krapivsky}},\ }\bibfield  {title} {\enquote {\bibinfo {title} {Random
  sequential covering},}\ }\href {\doibase 10.1088/1742-5468/acbc20} {\bibfield
   {journal} {\bibinfo  {journal} {J. Stat. Mech.}\ }\textbf {\bibinfo {volume}
  {2023}},\ \bibinfo {pages} {033202} (\bibinfo {year} {2023})}\BibitemShut
  {NoStop}%
\bibitem [{\citenamefont {Derrida}\ and\ \citenamefont
  {Gardner}(1986)}]{Derrida-SG86}%
  \BibitemOpen
  \bibfield  {author} {\bibinfo {author} {\bibfnamefont {B.}~\bibnamefont
  {Derrida}}\ and\ \bibinfo {author} {\bibfnamefont {E.}~\bibnamefont
  {Gardner}},\ }\bibfield  {title} {\enquote {\bibinfo {title} {Metastable
  states of a spin glass chain at 0 temperature},}\ }\href
  {https://doi.org/10.1051/jphys:01986004706095900} {\bibfield  {journal}
  {\bibinfo  {journal} {J. Phys. France}\ }\textbf {\bibinfo {volume} {47}},\
  \bibinfo {pages} {959--965} (\bibinfo {year} {1986})}\BibitemShut {NoStop}%
\bibitem [{\citenamefont {Hivert}\ \emph {et~al.}(2007)\citenamefont {Hivert},
  \citenamefont {Nechaev}, \citenamefont {Oshanin},\ and\ \citenamefont
  {Vasilyev}}]{peaks-hivert}%
  \BibitemOpen
  \bibfield  {author} {\bibinfo {author} {\bibfnamefont {F.}~\bibnamefont
  {Hivert}}, \bibinfo {author} {\bibfnamefont {S}~\bibnamefont {Nechaev}},
  \bibinfo {author} {\bibfnamefont {G}~\bibnamefont {Oshanin}}, \ and\ \bibinfo
  {author} {\bibfnamefont {O}~\bibnamefont {Vasilyev}},\ }\bibfield  {title}
  {\enquote {\bibinfo {title} {On the distribution of surface extrema in
  several one- and two-dimensional random landscapes},}\ }\href
  {https://doi.org/10.1007/s10955-006-9231-7} {\bibfield  {journal} {\bibinfo
  {journal} {J. Stat. Phys.}\ }\textbf {\bibinfo {volume} {126}},\ \bibinfo
  {pages} {243--279} (\bibinfo {year} {2007})}\BibitemShut {NoStop}%
\bibitem [{\citenamefont {Billera}\ \emph {et~al.}(2003)\citenamefont
  {Billera}, \citenamefont {Hsiao},\ and\ \citenamefont {{van
  Willigenburg}}}]{peaks-Billera}%
  \BibitemOpen
  \bibfield  {author} {\bibinfo {author} {\bibfnamefont {L.~J.}\ \bibnamefont
  {Billera}}, \bibinfo {author} {\bibfnamefont {S.~K.}\ \bibnamefont {Hsiao}},
  \ and\ \bibinfo {author} {\bibfnamefont {S.}~\bibnamefont {{van
  Willigenburg}}},\ }\bibfield  {title} {\enquote {\bibinfo {title} {Peak
  quasisymmetric functions and {E}ulerian enumeration},}\ }\href {\doibase
  https://doi.org/10.1016/S0001-8708(02)00067-1} {\bibfield  {journal}
  {\bibinfo  {journal} {Adv. Math.}\ }\textbf {\bibinfo {volume} {176}},\
  \bibinfo {pages} {248--276} (\bibinfo {year} {2003})}\BibitemShut {NoStop}%
\bibitem [{\citenamefont {Majumdar}\ and\ \citenamefont
  {Martin}(2006)}]{Majumdar06}%
  \BibitemOpen
  \bibfield  {author} {\bibinfo {author} {\bibfnamefont {S.~N.}\ \bibnamefont
  {Majumdar}}\ and\ \bibinfo {author} {\bibfnamefont {O.~C.}\ \bibnamefont
  {Martin}},\ }\bibfield  {title} {\enquote {\bibinfo {title} {Statistics of
  the number of minima in a random energy landscape},}\ }\href {\doibase
  10.1103/PhysRevE.74.061112} {\bibfield  {journal} {\bibinfo  {journal} {Phys.
  Rev. E}\ }\textbf {\bibinfo {volume} {74}},\ \bibinfo {pages} {061112}
  (\bibinfo {year} {2006})}\BibitemShut {NoStop}%
\bibitem [{\citenamefont {Oshanin}\ and\ \citenamefont
  {Voituriez}(2004)}]{perm-Oshanin}%
  \BibitemOpen
  \bibfield  {author} {\bibinfo {author} {\bibfnamefont {G.}~\bibnamefont
  {Oshanin}}\ and\ \bibinfo {author} {\bibfnamefont {R.}~\bibnamefont
  {Voituriez}},\ }\bibfield  {title} {\enquote {\bibinfo {title} {Random walk
  generated by random permutations of (1, 2, 3, ... , n + 1)},}\ }\href
  {\doibase 10.1088/0305-4470/37/24/002} {\bibfield  {journal} {\bibinfo
  {journal} {J. Phys. A}\ }\textbf {\bibinfo {volume} {37}},\ \bibinfo {pages}
  {6221} (\bibinfo {year} {2004})}\BibitemShut {NoStop}%
\bibitem [{\citenamefont {Carmi}\ \emph {et~al.}(2008)\citenamefont {Carmi},
  \citenamefont {Krapivsky},\ and\ \citenamefont {ben Avraham}}]{peaks-carmi}%
  \BibitemOpen
  \bibfield  {author} {\bibinfo {author} {\bibfnamefont {S.}~\bibnamefont
  {Carmi}}, \bibinfo {author} {\bibfnamefont {P.~L.}\ \bibnamefont
  {Krapivsky}}, \ and\ \bibinfo {author} {\bibfnamefont {D.}~\bibnamefont {ben
  Avraham}},\ }\bibfield  {title} {\enquote {\bibinfo {title} {Partition of
  networks into basins of attraction},}\ }\href {\doibase
  10.1103/PhysRevE.78.066111} {\bibfield  {journal} {\bibinfo  {journal} {Phys.
  Rev. E}\ }\textbf {\bibinfo {volume} {78}},\ \bibinfo {pages} {066111}
  (\bibinfo {year} {2008})}\BibitemShut {NoStop}%
\bibitem [{\citenamefont {Knecht}\ \emph {et~al.}(2012)\citenamefont {Knecht},
  \citenamefont {Trump}, \citenamefont {ben Avraham},\ and\ \citenamefont
  {Ziff}}]{peaks-Dani}%
  \BibitemOpen
  \bibfield  {author} {\bibinfo {author} {\bibfnamefont {C.~L.}\ \bibnamefont
  {Knecht}}, \bibinfo {author} {\bibfnamefont {W.}~\bibnamefont {Trump}},
  \bibinfo {author} {\bibfnamefont {D.}~\bibnamefont {ben Avraham}}, \ and\
  \bibinfo {author} {\bibfnamefont {R.~M.}\ \bibnamefont {Ziff}},\ }\bibfield
  {title} {\enquote {\bibinfo {title} {Retention capacity of random
  surfaces},}\ }\href {\doibase 10.1103/PhysRevLett.108.045703} {\bibfield
  {journal} {\bibinfo  {journal} {Phys. Rev. Lett.}\ }\textbf {\bibinfo
  {volume} {108}},\ \bibinfo {pages} {045703} (\bibinfo {year}
  {2012})}\BibitemShut {NoStop}%
\bibitem [{\citenamefont {Luck}(2014)}]{peaks-JML}%
  \BibitemOpen
  \bibfield  {author} {\bibinfo {author} {\bibfnamefont {J.~M.}\ \bibnamefont
  {Luck}},\ }\bibfield  {title} {\enquote {\bibinfo {title} {On the frequencies
  of patterns of rises and falls},}\ }\href
  {https://doi.org/10.1016/j.physa.2014.04.010} {\bibfield  {journal} {\bibinfo
   {journal} {Physica A}\ }\textbf {\bibinfo {volume} {407}},\ \bibinfo {pages}
  {252--275} (\bibinfo {year} {2014})}\BibitemShut {NoStop}%
\bibitem [{\citenamefont {Andr\'{e}}(1879)}]{Andre1879}%
  \BibitemOpen
  \bibfield  {author} {\bibinfo {author} {\bibfnamefont {D.}~\bibnamefont
  {Andr\'{e}}},\ }\bibfield  {title} {\enquote {\bibinfo {title}
  {D\'{e}veloppements de s\'{e}c x et de tang x},}\ }\href@noop {} {\bibfield
  {journal} {\bibinfo  {journal} {C. R. Acad. Sci. Paris}\ }\textbf {\bibinfo
  {volume} {88}},\ \bibinfo {pages} {965--967} (\bibinfo {year}
  {1879})}\BibitemShut {NoStop}%
\bibitem [{\citenamefont {Andr\'{e}}(1881)}]{Andre1881}%
  \BibitemOpen
  \bibfield  {author} {\bibinfo {author} {\bibfnamefont {D.}~\bibnamefont
  {Andr\'{e}}},\ }\bibfield  {title} {\enquote {\bibinfo {title} {M\'{e}moire
  sur les permutations altern\'{e}es},}\ }\href
  {http://www.numdam.org/item/JMPA_1881_3_7__167_0} {\bibfield  {journal}
  {\bibinfo  {journal} {J. Math. Pures Appl.}\ }\textbf {\bibinfo {volume}
  {7}},\ \bibinfo {pages} {167--184} (\bibinfo {year} {1881})}\BibitemShut
  {NoStop}%
\bibitem [{\citenamefont {Arnol'd}(1992)}]{Arnold92}%
  \BibitemOpen
  \bibfield  {author} {\bibinfo {author} {\bibfnamefont {V.~I.}\ \bibnamefont
  {Arnol'd}},\ }\bibfield  {title} {\enquote {\bibinfo {title} {The calculus of
  snakes and the combinatorics of {B}ernoulli, {E}uler and {S}pringer numbers
  of {C}oxeter groups},}\ }\href {\doibase 10.1070/RM1992v047n01ABEH000861}
  {\bibfield  {journal} {\bibinfo  {journal} {Russian Math. Surveys}\ }\textbf
  {\bibinfo {volume} {47}},\ \bibinfo {pages} {1--51} (\bibinfo {year}
  {1992})}\BibitemShut {NoStop}%
\bibitem [{\citenamefont {Stanley}(2011)}]{Stanley}%
  \BibitemOpen
  \bibfield  {author} {\bibinfo {author} {\bibfnamefont {R.~P.}\ \bibnamefont
  {Stanley}},\ }\href {https://doi.org/10.1017/CBO9781139058520} {\emph
  {\bibinfo {title} {Enumerative Combinatorics. Vol. I}}}\ (\bibinfo
  {publisher} {Cambridge University Press},\ \bibinfo {address} {Cambridge,
  UK},\ \bibinfo {year} {2011})\BibitemShut {NoStop}%
\bibitem [{\citenamefont {Majumdar}(2002)}]{Majumdar02}%
  \BibitemOpen
  \bibfield  {author} {\bibinfo {author} {\bibfnamefont {S.~N.}\ \bibnamefont
  {Majumdar}},\ }\bibfield  {title} {\enquote {\bibinfo {title} {Statistics of
  multiple sign changes in a discrete non-markovian sequence},}\ }\href
  {\doibase 10.1103/PhysRevE.65.035104} {\bibfield  {journal} {\bibinfo
  {journal} {Phys. Rev. E}\ }\textbf {\bibinfo {volume} {65}},\ \bibinfo
  {pages} {035104} (\bibinfo {year} {2002})}\BibitemShut {NoStop}%
\bibitem [{\citenamefont {Zagier}(2007)}]{Zagier}%
  \BibitemOpen
  \bibfield  {author} {\bibinfo {author} {\bibfnamefont {D.}~\bibnamefont
  {Zagier}},\ }\bibfield  {title} {\enquote {\bibinfo {title} {The dilogarithm
  function},}\ }in\ \href {https://doi.org/10.1007/978-3-540-30308-4_1} {\emph
  {\bibinfo {booktitle} {Frontiers in Number Theory, Physics, and Geometry
  {II}}}},\ \bibinfo {editor} {edited by\ \bibinfo {editor} {\bibfnamefont
  {P.}~\bibnamefont {Cartier}}, \bibinfo {editor} {\bibfnamefont
  {P.}~\bibnamefont {Moussa}}, \bibinfo {editor} {\bibfnamefont
  {B.}~\bibnamefont {Julia}}, \ and\ \bibinfo {editor} {\bibfnamefont
  {P.}~\bibnamefont {Vanhove}}}\ (\bibinfo  {publisher} {Springer},\ \bibinfo
  {address} {Berlin},\ \bibinfo {year} {2007})\ pp.\ \bibinfo {pages}
  {3--65}\BibitemShut {NoStop}%
\bibitem [{\citenamefont {Mat\'{e}rn}(1986)}]{Matern}%
  \BibitemOpen
  \bibfield  {author} {\bibinfo {author} {\bibfnamefont {B.}~\bibnamefont
  {Mat\'{e}rn}},\ }\href {https://doi.org/10.1007/978-1-4615-7892-5} {\emph
  {\bibinfo {title} {Spatial Variation}}}\ (\bibinfo  {publisher} {Springer},\
  \bibinfo {address} {New York, NY},\ \bibinfo {year} {1986})\BibitemShut
  {NoStop}%
\bibitem [{\citenamefont {Ball}(1997)}]{Ball}%
  \BibitemOpen
  \bibfield  {author} {\bibinfo {author} {\bibfnamefont {K.}~\bibnamefont
  {Ball}},\ }\bibfield  {title} {\enquote {\bibinfo {title} {An elementary
  introduction to modern convex geometry},}\ }in\ \href@noop {} {\emph
  {\bibinfo {booktitle} {Flavors of Geometry. {MSRI} Lecture Notes}}},\
  Vol.~\bibinfo {volume} {31},\ \bibinfo {editor} {edited by\ \bibinfo {editor}
  {\bibfnamefont {S.}~\bibnamefont {Levy}}}\ (\bibinfo  {publisher} {Cambridge
  University Press},\ \bibinfo {address} {Cambridge, UK},\ \bibinfo {year}
  {1997})\ pp.\ \bibinfo {pages} {1--58}\BibitemShut {NoStop}%
\bibitem [{\citenamefont {Evans}(1993)}]{Evans93}%
  \BibitemOpen
  \bibfield  {author} {\bibinfo {author} {\bibfnamefont {J.~W.}\ \bibnamefont
  {Evans}},\ }\bibfield  {title} {\enquote {\bibinfo {title} {Random and
  cooperative sequential adsorption},}\ }\href {\doibase
  10.1103/RevModPhys.65.1281} {\bibfield  {journal} {\bibinfo  {journal} {Rev.
  Mod. Phys.}\ }\textbf {\bibinfo {volume} {65}},\ \bibinfo {pages}
  {1281--1329} (\bibinfo {year} {1993})}\BibitemShut {NoStop}%
\bibitem [{\citenamefont {Talbot}\ \emph {et~al.}(2000)\citenamefont {Talbot},
  \citenamefont {Tarjus}, \citenamefont {{Van Tassel}},\ and\ \citenamefont
  {Viot}}]{Talbot00}%
  \BibitemOpen
  \bibfield  {author} {\bibinfo {author} {\bibfnamefont {J.}~\bibnamefont
  {Talbot}}, \bibinfo {author} {\bibfnamefont {G.}~\bibnamefont {Tarjus}},
  \bibinfo {author} {\bibfnamefont {P.R.}\ \bibnamefont {{Van Tassel}}}, \ and\
  \bibinfo {author} {\bibfnamefont {P.}~\bibnamefont {Viot}},\ }\bibfield
  {title} {\enquote {\bibinfo {title} {From car parking to protein adsorption:
  an overview of sequential adsorption processes},}\ }\href {\doibase
  https://doi.org/10.1016/S0927-7757(99)00409-4} {\bibfield  {journal}
  {\bibinfo  {journal} {Colloids Surfaces A}\ }\textbf {\bibinfo {volume}
  {165}},\ \bibinfo {pages} {287--324} (\bibinfo {year} {2000})}\BibitemShut
  {NoStop}%
\bibitem [{\citenamefont {Torquato}(2002)}]{Tor}%
  \BibitemOpen
  \bibfield  {author} {\bibinfo {author} {\bibfnamefont {S.}~\bibnamefont
  {Torquato}},\ }\href {https://doi.org/10.1007/978-1-4757-6355-3} {\emph
  {\bibinfo {title} {Random Heterogeneous Materials: Microstructure and
  Macroscopic Properties}}}\ (\bibinfo  {publisher} {Springer-Verlag},\
  \bibinfo {address} {New York},\ \bibinfo {year} {2002})\BibitemShut {NoStop}%
\bibitem [{\citenamefont {Krapivsky}(2020)}]{PK20}%
  \BibitemOpen
  \bibfield  {author} {\bibinfo {author} {\bibfnamefont {P.~L.}\ \bibnamefont
  {Krapivsky}},\ }\bibfield  {title} {\enquote {\bibinfo {title} {Large
  deviations in one-dimensional random sequential adsorption},}\ }\href
  {\doibase 10.1103/PhysRevE.102.062108} {\bibfield  {journal} {\bibinfo
  {journal} {Phys. Rev. E}\ }\textbf {\bibinfo {volume} {102}},\ \bibinfo
  {pages} {062108} (\bibinfo {year} {2020})}\BibitemShut {NoStop}%
\bibitem [{\citenamefont {Flory}(1939)}]{Flory39}%
  \BibitemOpen
  \bibfield  {author} {\bibinfo {author} {\bibfnamefont {P.~J.}\ \bibnamefont
  {Flory}},\ }\bibfield  {title} {\enquote {\bibinfo {title} {Intramolecular
  reaction between neighboring substituents of vinyl polymers},}\ }\href
  {\doibase 10.1021/ja01875a053} {\bibfield  {journal} {\bibinfo  {journal} {J.
  Amer. Chem. Soc.}\ }\textbf {\bibinfo {volume} {61}},\ \bibinfo {pages}
  {1518--1521} (\bibinfo {year} {1939})}\BibitemShut {NoStop}%
\bibitem [{\citenamefont {Baram}\ and\ \citenamefont
  {Kutasov}(1989)}]{Baram89}%
  \BibitemOpen
  \bibfield  {author} {\bibinfo {author} {\bibfnamefont {A.}~\bibnamefont
  {Baram}}\ and\ \bibinfo {author} {\bibfnamefont {D.}~\bibnamefont
  {Kutasov}},\ }\bibfield  {title} {\enquote {\bibinfo {title} {On the dynamics
  of random sequential absorption},}\ }\href {\doibase
  10.1088/0305-4470/22/6/011} {\bibfield  {journal} {\bibinfo  {journal} {J.
  Phys. A}\ }\textbf {\bibinfo {volume} {22}},\ \bibinfo {pages} {L251--L254}
  (\bibinfo {year} {1989})}\BibitemShut {NoStop}%
\bibitem [{\citenamefont {Baram}\ and\ \citenamefont
  {Lipshtat}(2021)}]{Baram21}%
  \BibitemOpen
  \bibfield  {author} {\bibinfo {author} {\bibfnamefont {A.}~\bibnamefont
  {Baram}}\ and\ \bibinfo {author} {\bibfnamefont {A.}~\bibnamefont
  {Lipshtat}},\ }\bibfield  {title} {\enquote {\bibinfo {title} {Jamming
  densities of random sequential adsorption on $d$-dimensional cubic
  lattices},}\ }\href {\doibase 10.1103/PhysRevE.104.014104} {\bibfield
  {journal} {\bibinfo  {journal} {Phys. Rev. E}\ }\textbf {\bibinfo {volume}
  {104}},\ \bibinfo {pages} {014104} (\bibinfo {year} {2021})}\BibitemShut
  {NoStop}%
\bibitem [{\citenamefont {Pippenger}(1989)}]{Pippenger}%
  \BibitemOpen
  \bibfield  {author} {\bibinfo {author} {\bibfnamefont {N.}~\bibnamefont
  {Pippenger}},\ }\bibfield  {title} {\enquote {\bibinfo {title} {Random
  sequential adsorption on graphs},}\ }\href {\doibase 10.1137/0402034}
  {\bibfield  {journal} {\bibinfo  {journal} {SIAM J. Disc. Math.}\ }\textbf
  {\bibinfo {volume} {2}},\ \bibinfo {pages} {393--401} (\bibinfo {year}
  {1989})}\BibitemShut {NoStop}%
\bibitem [{\citenamefont {Fan}\ and\ \citenamefont {Percus}(1991)}]{Percus91}%
  \BibitemOpen
  \bibfield  {author} {\bibinfo {author} {\bibfnamefont {Y.}~\bibnamefont
  {Fan}}\ and\ \bibinfo {author} {\bibfnamefont {J.~K.}\ \bibnamefont
  {Percus}},\ }\bibfield  {title} {\enquote {\bibinfo {title} {Asymptotic
  coverage in random sequential adsorption on a lattice},}\ }\href {\doibase
  10.1103/PhysRevA.44.5099} {\bibfield  {journal} {\bibinfo  {journal} {Phys.
  Rev. A}\ }\textbf {\bibinfo {volume} {44}},\ \bibinfo {pages} {5099--5103}
  (\bibinfo {year} {1991})}\BibitemShut {NoStop}%
\bibitem [{\citenamefont {Zhang}\ and\ \citenamefont
  {Torquato}(2013)}]{Torquato13}%
  \BibitemOpen
  \bibfield  {author} {\bibinfo {author} {\bibfnamefont {G.}~\bibnamefont
  {Zhang}}\ and\ \bibinfo {author} {\bibfnamefont {S.}~\bibnamefont
  {Torquato}},\ }\bibfield  {title} {\enquote {\bibinfo {title} {Precise
  algorithm to generate random sequential addition of hard hyperspheres at
  saturation},}\ }\href {\doibase 10.1103/PhysRevE.88.053312} {\bibfield
  {journal} {\bibinfo  {journal} {Phys. Rev. E}\ }\textbf {\bibinfo {volume}
  {88}},\ \bibinfo {pages} {053312} (\bibinfo {year} {2013})}\BibitemShut
  {NoStop}%
\bibitem [{\citenamefont {Torquato}(2010)}]{Torquato10}%
  \BibitemOpen
  \bibfield  {author} {\bibinfo {author} {\bibfnamefont {S.}~\bibnamefont
  {Torquato}},\ }\bibfield  {title} {\enquote {\bibinfo {title} {Reformulation
  of the covering and quantizer problems as ground states of interacting
  particles},}\ }\href {\doibase 10.1103/PhysRevE.82.056109} {\bibfield
  {journal} {\bibinfo  {journal} {Phys. Rev. E}\ }\textbf {\bibinfo {volume}
  {82}},\ \bibinfo {pages} {056109} (\bibinfo {year} {2010})}\BibitemShut
  {NoStop}%
\bibitem [{\citenamefont {Krapivsky}(1991)}]{PLK91}%
  \BibitemOpen
  \bibfield  {author} {\bibinfo {author} {\bibfnamefont {P.~L.}\ \bibnamefont
  {Krapivsky}},\ }\bibfield  {title} {\enquote {\bibinfo {title} {Zero
  temperature dynamics of a spin glass chain},}\ }\href
  {https://doi.org/10.1051/jp1:1991185} {\bibfield  {journal} {\bibinfo
  {journal} {J. Phys. France}\ }\textbf {\bibinfo {volume} {1}},\ \bibinfo
  {pages} {1013--1018} (\bibinfo {year} {1991})}\BibitemShut {NoStop}%
\bibitem [{\citenamefont {Entringer}(1966)}]{Entringer66}%
  \BibitemOpen
  \bibfield  {author} {\bibinfo {author} {\bibfnamefont {R.~C.}\ \bibnamefont
  {Entringer}},\ }\bibfield  {title} {\enquote {\bibinfo {title} {A
  combinatorial interpretation of the {E}uler and {B}ernoulli numbers},}\
  }\href@noop {} {\bibfield  {journal} {\bibinfo  {journal} {Nieuw Arch.
  Wisk.}\ }\textbf {\bibinfo {volume} {14}},\ \bibinfo {pages} {241--246}
  (\bibinfo {year} {1966})}\BibitemShut {NoStop}%
\bibitem [{\citenamefont {Entringer}(1969)}]{Entringer69}%
  \BibitemOpen
  \bibfield  {author} {\bibinfo {author} {\bibfnamefont {R.~C.}\ \bibnamefont
  {Entringer}},\ }\bibfield  {title} {\enquote {\bibinfo {title} {Enumeration
  of permutations of (1,.., n) by number of maxima},}\ }\href {\doibase
  10.1215/S0012-7094-69-03669-2} {\bibfield  {journal} {\bibinfo  {journal}
  {Duke Math. J.}\ }\textbf {\bibinfo {volume} {36}},\ \bibinfo {pages}
  {575--579} (\bibinfo {year} {1969})}\BibitemShut {NoStop}%
\bibitem [{\citenamefont {Carlitz}(1974)}]{perm-carlitz}%
  \BibitemOpen
  \bibfield  {author} {\bibinfo {author} {\bibfnamefont {L.}~\bibnamefont
  {Carlitz}},\ }\bibfield  {title} {\enquote {\bibinfo {title} {Permutations
  and sequences},}\ }\href {https://doi.org/10.1016/0001-8708(74)90025-5}
  {\bibfield  {journal} {\bibinfo  {journal} {Adv. Math.}\ }\textbf {\bibinfo
  {volume} {14}},\ \bibinfo {pages} {92--120} (\bibinfo {year}
  {1974})}\BibitemShut {NoStop}%
\bibitem [{\citenamefont {Fewster}\ and\ \citenamefont
  {Siemssen}(2014)}]{perm-fewster}%
  \BibitemOpen
  \bibfield  {author} {\bibinfo {author} {\bibfnamefont {C.~J.}\ \bibnamefont
  {Fewster}}\ and\ \bibinfo {author} {\bibfnamefont {D.}~\bibnamefont
  {Siemssen}},\ }\bibfield  {title} {\enquote {\bibinfo {title} {Enumerating
  permutations by their run structure},}\ }\href
  {https://doi.org/10.37236/4235} {\bibfield  {journal} {\bibinfo  {journal}
  {Electronic J. Combin.}\ }\textbf {\bibinfo {volume} {21}},\ \bibinfo {pages}
  {P4.18} (\bibinfo {year} {2014})}\BibitemShut {NoStop}%
\bibitem [{\citenamefont {Ma}\ and\ \citenamefont {Yeh}(2016)}]{perm-Ma}%
  \BibitemOpen
  \bibfield  {author} {\bibinfo {author} {\bibfnamefont {S.-M.}\ \bibnamefont
  {Ma}}\ and\ \bibinfo {author} {\bibfnamefont {Y.-N.}\ \bibnamefont {Yeh}},\
  }\bibfield  {title} {\enquote {\bibinfo {title} {Enumeration of permutations
  by number of alternating descents},}\ }\href
  {https://doi.org/10.1016/j.disc.2015.12.007} {\bibfield  {journal} {\bibinfo
  {journal} {Discrete Math.}\ }\textbf {\bibinfo {volume} {339}},\ \bibinfo
  {pages} {1362--1367} (\bibinfo {year} {2016})}\BibitemShut {NoStop}%
\bibitem [{\citenamefont {Kingman}(1978)}]{Kingman78}%
  \BibitemOpen
  \bibfield  {author} {\bibinfo {author} {\bibfnamefont {J.~F.~C.}\
  \bibnamefont {Kingman}},\ }\bibfield  {title} {\enquote {\bibinfo {title} {A
  simple model for the balance between selection and mutation},}\ }\href
  {https://doi.org/10.2307/3213231} {\bibfield  {journal} {\bibinfo  {journal}
  {J. Appl. Probab.}\ }\textbf {\bibinfo {volume} {15}},\ \bibinfo {pages}
  {1--12} (\bibinfo {year} {1978})}\BibitemShut {NoStop}%
\bibitem [{\citenamefont {Kauffman}\ and\ \citenamefont
  {Levin}(1987)}]{Kauffman87}%
  \BibitemOpen
  \bibfield  {author} {\bibinfo {author} {\bibfnamefont {S.}~\bibnamefont
  {Kauffman}}\ and\ \bibinfo {author} {\bibfnamefont {S.}~\bibnamefont
  {Levin}},\ }\bibfield  {title} {\enquote {\bibinfo {title} {Towards a general
  theory of adaptive walks on rugged landscapes},}\ }\href
  {https://doi.org/10.1016/S0022-5193(87)80029-2} {\bibfield  {journal}
  {\bibinfo  {journal} {J. Theor. Biol.}\ }\textbf {\bibinfo {volume} {128}},\
  \bibinfo {pages} {11--45} (\bibinfo {year} {1987})}\BibitemShut {NoStop}%
\bibitem [{\citenamefont {Gavrilets}(2004)}]{Gavrilets}%
  \BibitemOpen
  \bibfield  {author} {\bibinfo {author} {\bibfnamefont {S.}~\bibnamefont
  {Gavrilets}},\ }\href@noop {} {\emph {\bibinfo {title} {Fitness Landscapes
  and the Origin of Species}}}\ (\bibinfo  {publisher} {Princeton University
  Press},\ \bibinfo {address} {Princeton, NJ},\ \bibinfo {year}
  {2004})\BibitemShut {NoStop}%
\bibitem [{\citenamefont {Franke}\ and\ \citenamefont {Krug}(2012)}]{Krug12}%
  \BibitemOpen
  \bibfield  {author} {\bibinfo {author} {\bibfnamefont {J.}~\bibnamefont
  {Franke}}\ and\ \bibinfo {author} {\bibfnamefont {J.}~\bibnamefont {Krug}},\
  }\bibfield  {title} {\enquote {\bibinfo {title} {Evolutionary accessibility
  in tunably rugged fitness landscapes},}\ }\href
  {https://doi.org/10.1007/s10955-012-0507-9} {\bibfield  {journal} {\bibinfo
  {journal} {J. Stat. Phys.}\ }\textbf {\bibinfo {volume} {148}},\ \bibinfo
  {pages} {706--723} (\bibinfo {year} {2012})}\BibitemShut {NoStop}%
\end{thebibliography}%

\end{document}